\RequirePackage[latest]{latexrelease}
\documentclass[11pt,final,onecolumn]{IEEEtran}

\usepackage[colorlinks,tables=false,bibtex=true]{preamble}
\include{auxiliary-commands}
\DeclareGraphicsExtensions{.pdf,.png,.jpg}
\graphicspath{{Figures/}}

\title{Active Target Localization using Low-Rank Matrix Completion and Unimodal Regression}
\author{Sunav~Choudhary,~\IEEEmembership{Student~Member,~IEEE,} Naveen~Kumar,~\IEEEmembership{Student~Member,~IEEE,} Srikanth~Narayanan,~\IEEEmembership{Fellow,~IEEE} and~Urbashi~Mitra,~\IEEEmembership{Fellow,~IEEE}%
		\thanks{This work has been funded in part by the following grants and organizations:~ONR~N00014-09-1-0700, AFOSR~FA9550-12-1-0215, DOT~CA-26-7084-00, NSF~CPS-1446901, NSF~CNS-0832186, NSF~CNS-1213128, NSF~CCF-1410009 and NSF~CCF-1117896.
		Parts of this paper were presented at the IEEE International Conference on Acoustic, Speech, and Signal Processing (ICASSP), Florence, Italy, May 4-9, 2014~\cite{choudhary2014activetargetdetection}, at the 52nd Annual Allerton Conference on Communication, Control, and Computing (Allerton), Monticello, Illinois, Sep.~30 - Oct.~3, 2014~\cite{choudhary2014ATDwithNav} and at the IEEE International Conference on Acoustic, Speech, and Signal Processing (ICASSP), Brisbane, Australia, April 19-24, 2015~\cite{choudhary2015AnalysisTargetDetection}.}%
		\thanks{S.~Choudhary, N.~Kumar, S.~Narayanan and U.~Mitra are with the Ming Hsieh Department of Electrical Engineering, Viterbi School of Engineering, University of Southern California, Los Angeles CA 90089, USA (email: \href{mailto:sunavcho@usc.edu}{\protect\nolinkurl{sunavcho@usc.edu}}, \href{mailto:komathnk@usc.edu}{\protect\nolinkurl{komathnk@usc.edu}}, \href{mailto:shri@sipi.usc.edu}{\protect\nolinkurl{shri@sipi.usc.edu}}, \href{mailto:ubli@usc.edu}{\protect\nolinkurl{ubli@usc.edu}}).}}

\begin{document}
	\maketitle

	\begin{abstract}
		The detection and localization of a target from samples of its generated field is a problem of interest in a broad range of applications.
		Often, the target field admits structural properties that enable the design of lower sample detection strategies with good performance.
		This paper designs a sampling and localization strategy which exploits separability and unimodality in target fields and theoretically analyzes the trade-off achieved between sampling density, noise level and convergence rate of localization.
		In particular, the strategy adopts an \emph{exploration-exploitation} approach to target detection and utilizes the theory of low-rank matrix completion, coupled with unimodal regression, on decaying and approximately separable target fields.
		The assumptions on the field are fairly generic and are applicable to many decay profiles since no specific knowledge of the field is necessary, besides its admittance of an approximately rank-one representation.
		Extensive numerical experiments and comparisons are performed to test the efficacy and robustness of the presented approach.
		Numerical results suggest that the proposed strategy outperforms algorithms based on mean-shift clustering, surface interpolation and naive low-rank matrix completion with peak detection, under low sampling density.
	\end{abstract}

	\IEEEpeerreviewmaketitle

	\section{Introduction}
		\label{sec:intro}
\begin{figure}[t]
			\centering
			\makeatletter
				\if@twocolumn
					\includegraphics[width=0.45\figwidth]{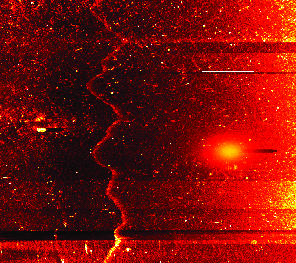}
				\else
					\includegraphics[width=0.6\figwidth]{spread_target-cropped}
				\fi
			\makeatother
			\caption{An underwater side-scan sonar image with a synthetic target signature.
			The background noise and artifacts are due to reflections from the seabed.}
			\label{fig:spread target signature}
		\end{figure}

		\IEEEPARstart{D}{etecting} and localizing a target, from samples of its induced field, is an important problem of interest with manifestations in a wide variety of applications like environmental monitoring, cyber-security, medical diagnosis and military surveillance.
		Because of its ubiquity, a rich literature has evolved around this problem and its application specific variations utilizing ideas from statistics, signal processing, information theory, machine learning and data mining.
		In this paper, we study a variation of the target detection and localization problem with sampling constraints on the induced target field.
		In particular, we consider the scenario where localization is desired from a set of samples that is \emph{information theoretically} insufficient to reconstruct the complete target field, and construct a localization algorithm with accompanying theoretical performance analysis.
		The possibility of reducing the number of samples required for target detection/localization is of interest for time critical applications where speed of acquisition is a bottleneck, like in magnetic resonance imaging due to the slow sampling process and in underwater sonar imaging due to large search spaces.
		As a simple illustrative example, consider the side-scan sonar image in \figurename~\ref{fig:spread target signature}, acquired by an autonomous underwater vehicle (AUV) with the goal of locating the position of the target (marked by a region of high intensity reflection) amongst background clutter (reflections from the sea bed).
		Examining the \emph{complete} image, it is easy to identify the location of the object of interest.
		However, we note that the target field in \figurename~\ref{fig:spread target signature} is highly structured and recalling the philosophy of compressed sensing~\cite{donoho2006compressed}, good detection/localization may be possible from very few samples of the complete field in \figurename~\ref{fig:spread target signature} at the expense of using a more sophisticated (but computationally tractable) algorithm.

		\subsection{Contributions and Organization}
			\label{sec:contributions}
			We consider a static \emph{separable} target field whose magnitude decays monotonically with increasing distance from the true location of the target.
			We employ an approach based on low-rank matrix completion~\cite{candes2010noisematrixcompletion} that allows us to derive a localization algorithm that \emph{does not} need the knowledge of the target field decay profile; the only requirement is that the target field should be separable along some \emph{known} directions.
			In particular, the algorithm can be viewed as a solution to the exploration-exploitation problem wherein the possible location of the target is unknown \emph{a priori} and the sampling strategy enables the coarse learning of the location and presence of target, resulting in subsequent sampling in more informed locations.
			We prove correctness and convergence of the proposed algorithm and further develop an analytical trade-off between the number of collected samples and the target localization error in the presence of noise when employing a uniformly random spatial pixel sampling strategy.
			In contrast to our results, most prior literature on noisy low-rank matrix completion investigates bounds on mean squared estimation error, and very little is known about the performance of matrix completion for other tasks (like detection or localization).
			Our approach is fairly general and as such does not exploit specialized models for the background clutter, beyond that of reduced sharpness of the separability assumption.
			Thus, further improvement in performance may be possible by taking this information into consideration.
			For example, the sonar images of the form in \figurename~\ref{fig:spread target signature} suffer from certain position dependent imaging artifacts that may be removed by intermediate processing.
			We perform extensive numerical experiments on synthetic and real datasets to validate the efficacy and robustness of the proposed approach.

			The rest of the paper is organized as follows.
			In the remainder of this section, we explore related prior art and define the mathematical notation used in the paper.
			\sectionname~\ref{sec:model} describes and justifies our assumptions on the target field and introduces the lifted reformulation of the underlying field.
			\sectionname~\ref{sec:strategy} describes our localization algorithm and states theoretical results to prove its correctness.
			\sectionname~\ref{sec:methods} reviews other methods that we use to compare against our algorithm for the purpose of numerical simulations.
			\sectionsname~\ref{sec:synthetic simulations} and~\ref{sec:real simulations} respectively describe our simulation results on synthetic and real data sets.
			\sectionname~\ref{sec:conclusions} concludes the paper.
			Detailed proofs of all results in the paper and useful supplementary material can be found in \appendicesname~\ref{sec:inner product bound proof}-\ref{sec:coherence computation}.

		\subsection{Related Work}
			\label{sec:prior art}
			For an early survey of active target detection, we refer the reader to~\cite{bhanu1986automatictargetrecognition} consisting of statistical and signal processing approaches that assume availability of the \emph{full} target field/signature (see also \cite{aridgides1995adaptive,hyland1995sea}).
			The field of anomaly detection~\cite{chandola2009AnomalyDetectionSurvey} further generalizes the scope of target detection and employs tools from machine learning, \eg~\cite{schweizer1989automatictargetdetection,dura2005active,reed2003automatic,mukherjee2011symbolic,kumar2012object,hollinger2012underwaterdatacollection} perform window based target detection in \emph{full} sonar images.
			General theoretical analysis on either of these problems is plagued by the lack of good models for experimental scenarios that are amenable to tractable analysis.
			In \cite{yilmaz2005pathplanningautonomous,bourgault2006optimalsearchlost,mihaylova2003comparisondecisionmaking,hummel2011missiondesigncompressive,englot2012samplingbasedcoverage,hollinger2012uncertaintydrivenview,hollinger2012underwaterdatacollection} there is a focus on path planning for active sensing of structured fields (in particular, \cite{hummel2011missiondesigncompressive} uses compressed sensing) with an explicit consideration of the navigation cost and stopping time.
			In contrast, the goal of this paper is to explore theoretical properties of adaptive sensing for structured fields stemming from the exploration-exploitation trade-off.
			Early work \cite{bekkerman2006targetdetectionand} focusing on target detection in multiple-in-multiple-out (MIMO) radar used a statistical approach, which was refined in \cite{kalogerias2013sparsesensingin,kalogerias2014matrixcompletionin,sun2013targetestimationin} using a combination of joint sparse sensing and low-rank matrix completion ideas, relying on the strong theoretical guarantees of low-rank matrix completion from random samples~\cite{candes2010noisematrixcompletion,gross2011recovering,negahban2012restrictedstrongconvexity}.
			The focus in the papers~\cite{kalogerias2013sparsesensingin,kalogerias2014matrixcompletionin,sun2013targetestimationin} is to adapt the design of the MIMO radar array to optimize coherence, which is also very different from our goal here of studying the detection and localization error performance of low-rank matrix completion.
			Finally, we note that distilled sensing~\cite{HauptIT11,HauptSSP12,Malloy2012Nearoptimaladaptive} has a somewhat similar algorithmic philosophy as ours for target detection, but therein the field is assumed to be sparse rather than low-rank, thus facing basis mismatch challenges~\cite{chi2011Sensitivitytobasis} that we can avoid completely.

		\subsection{Notation}
			\label{sec:notation}
			We use lowercase boldface alphabets to denote column vectors~(\eg~$\vec{z}$) and uppercase boldface alphabets to denote matrices~(\eg~$\mat{A}$).
			The MATLAB\textsuperscript{\circledR} indexing rules will be used to denote parts of a vector/matrix~(\eg~$\mat{A}\bb{2:3,4:6}$ denotes the sub-matrix of $\mat{A}$ formed by the rows $\cc{2,3}$ and columns $\cc{4,5,6}$).
			The all zero, all one and identity matrices shall be respectively denoted by $\vec{0}$, $\vec{1}$ and $\eye$ with dimensions dictated by context.
			$\tpose{\bb{\cdot}}$ denotes the transpose operation and $\ip{\cdot}{\cdot}$ denotes the standard inner product on $\setR^{n}$.
			The functions $\fronorm{\cdot}$ and $\nucnorm{\cdot}$ respectively return the Frobenius and nuclear norms of their matrix argument.
			The function $\abs{\cdot}$ applied to a scalar (respectively a set) returns its absolute value (respectively cardinality).
			Vector inequalities are assumed to hold element-wise, \eg~if $\vec{z} \in \setR^{n}$ then $\vec{z} \leq \vec{0}$ is shorthand for the $n$ inequality relations $\vec{z}\bb{j} \leq 0$, $\forall \, 1 \leq j \leq n$.
			$\setR$ and $\setZ_{+}$ respectively denote the set of real numbers and the set of positive integers.
			We shall use the $\BigOh{\cdot}$ notation, to upper bound the order of growth of any function $f \fcolon \setR \to \setR$ of $h \in \setR$ \wrt~its argument, \ie~$f\bb{h} = \BigOh{h} \iff \lim_{h \to \infty} \frac{f\bb{h}}{h} < \infty$.

	\section{System Model}
		\label{sec:model}
		\subsection{Target Field Assumptions}
			\label{sec:assumptions on the field}
			Let the search region (see \figurename~\ref{fig:spread target signature}) be the two dimensional unit square $\BB{0,1}^{2} \subset \setR^{2}$, and $\vec{y} = \bb{\yc,\yr} \in \BB{0,1}^{2}$ denote an arbitrary location in the search space.
			Let $H \colon \setR^{2} \to \setR$ denote the \textit{scalar valued} field induced by the target, \ie~the target signature.
			Thus, a mobile agent measuring the field value at location $\vec{y} \in \setR^{2}$ would record the value $H\bb{\vec{y}} = H\bb{\yc, \yr} \in \setR$.
			We shall make the following key (physically motivated) assumptions on the field $H\bb{\vec{y}}$:
			\begin{enumerate}[({A}1)]
				\item	\label{assume:separable}
						$H\bb{\vec{y}}$ is separable in some known basis of $\setR^{2}$, independent of the true location of the target.
				\item	\label{assume:monotonic}
						The magnitude of the field, $\abs{H\bb{\vec{y}}}$ is a monotonically non-increasing function of the distance from the target in every direction.
				\item	\label{assume:invariant}
						$H\bb{\vec{y}}$ is spatially invariant relative to the target's position.
			\end{enumerate}
			Without loss of generality (\WLOG), we assume separability of $H\bb{\vec{y}}$ in the $\yc$ and $\yr$ directions (\ie~in the canonical basis $\cc{\BB{1, 0}, \BB{0, 1}}$) as per \aref{assume:separable}.
			This means that there exist functions $F \colon \setR \to \setR$ and $G \colon \setR \to \setR$ such that $H\bb{\vec{y}} = F\bb{\yc} G\bb{\yr}$, $\forall \bb{\yc, \yr} \in \setR^{2}$.
			Notice that if $H\bb{\vec{y}}$ is instead separable in the rotated directions $\mat{\Sigma} \tpose{\bb{1, 0}}$ and $\mat{\Sigma} \tpose{\bb{0, 1}}$ for some \textit{known} $\mat{\Sigma} \in \setR^{2 \times 2}$, then we can work in this rotated coordinate system.
			Assumption~\aref{assume:monotonic} is intuitively clear and can be mathematically described by the inequality:
			\begin{equation}
				\abs{H\bb{t_{1}\bb{\vec{y} - \vec{y}_{0}}}} \geq \abs{H\bb{t_{2}\bb{\vec{y} - \vec{y}_{0}}}},
			\end{equation}
			holding $\forall \vec{y} \in \setR^{2}, t_{2} > t_{1} > 0$, where $\vec{y}_{0}$ represents the unknown location of the target.
			Assumption~\aref{assume:invariant} implies that if the target were moved from $\vec{y}_{0}$ to a new position $\vec{y}'_{0}$, then the new field at location $\vec{y}$ would be given by $H\bb{\vec{y} - \vec{y}'_{0} + \vec{y}_{0}}$, thus ensuring that~\aref{assume:separable} holds in the canonical basis, regardless of the target's position $\vec{y}_{0}$.
			In this sense~\aref{assume:invariant} is stricter than necessary for our purposes, but we retain it for intuitive clarity.

			Scalar fields commonly correspond to intensity measurements (like the sonar image in \figurename~\ref{fig:spread target signature}).
			The following types of commonly assumed intensity fields satisfy our assumptions:
			\begin{enumerate}
				\item	Exponential fields: $H\bb{\vec{y}} = H_{0}\exp\bb{-\pnorm{\mat{\Sigma}\vec{y}}^{p}}$, for any $2 \times 2$ diagonal matrix $\mat{\Sigma} \in \setRn{2}{2}$ and constants $p, H_{0} > 0$.
						For $p = 1$, we get two dimensional Laplacian fields
						\begin{subequations}
							\begin{align}
								H\bb{\vec{y}} & = H_{0}\exp\bb{-\onenorm{\mat{\Sigma}\vec{y}}},	\\
								\intertext{and for $p = 2$, we get two dimensional Gaussian fields}
								H\bb{\vec{y}} & = H_{0}\exp\bb{-\twonorm{\mat{\Sigma}\vec{y}}^{2}}.
							\end{align}
						\end{subequations}
				
				\item	Power Law fields:
						\begin{equation}
							H\bb{\vec{y}} = \frac{H_{0}}{\bb{a_{1} + \abs{\yc}^{p_{1}}}^{r_{1}} \bb{a_{2} + \abs{\yr}^{p_{2}}}^{r_{2}}}
						\end{equation}
						for constants $H_{0}, p_{1}, p_{2}, a_{1}, a_{2}, r_{1}, r_{2} > 0$.
						With $p_{1} = p_{2} = 2$ and $r_{1} = r_{2} = 1$, we get a field that is separable as a product of two Cauchy fields
						\begin{equation}
							H\bb{\vec{y}} = \frac{H_{0}}{\bb{a_{1}+\yc^{2}}\bb{a_{2}+\yr^{2}}}.
						\end{equation}
				
				\item	Any \emph{multiplicative} combination of fields satisfying our assumptions, \eg
						\begin{equation}
							H\bb{\vec{y}} = H_{0} \frac{\exp\bb{-c_{1}\yc^{2} -c_{2}\abs{\yr}}}{\bb{1 + \abs{\yc}} \bb{1 + \yr^{2}}}
						\end{equation}
						for some constants $H_{0}, c_{1}, c_{2} > 0$.
						In particular, the set of separable fields is closed under multiplication.
			\end{enumerate}
			In some cases, the target field may not strictly satisfy the separability assumption~\aref{assume:separable}.
			However, the algorithm we develop in the sequel also works with \emph{approximate} separability (measured by how well can the field be approximated by a rank-1 matrix; see \sectionname~\ref{sec:lifting}).
			For example:
			\begin{enumerate}
				\item	The first ten singular values $\set{\sigma_{j}}{1 \leq j \leq 10}$ of the sonar image in \figurename~\ref{fig:spread target signature} are shown in \figurename~\ref{fig:spread target svd}, relative to the first singular value $\sigma_{1}$.
				Clearly, $\sigma_{2}$ is about 7dB below $\sigma_{1}$ which implies that the sonar image is approximately rank-1 (hence approximately separable).
				\item	The commonly occurring inverse square law field, $H\bb{\vec{y}} = H_{0}/\bb{\yc^{2} + \yr^{2}}$ for some constant $H_{0} > 0$, is not separable in the sense of assumption~\aref{assume:separable}.
				Let $\mat{H} \in \setRn{20}{20}$ denote the discretized field matrix formed by sampling $H\bb{\vec{y}}$ on the grid $\vec{y} \in \set{\bb{\yc,\yr}}{\yc,\yr \in \cc{-9.5,-8.5,\dotsc,9.5}}$ with $H_{0} = 1$.
				Plotting the first seven singular values $\set{\sigma_{j}}{1 \leq j \leq 7}$ of $\mat{H}$, relative to the first singular value $\sigma_{1}$, in \figurename~\ref{fig:inv sq law svd} shows that $\sigma_{2}$ is about 9dB below $\sigma_{1}$.
				Thus, inverse square law fields can be \emph{approximately separable}.
				Further, it can be computationally verified that the approximately 9dB attenuation from $\sigma_{1}$ to $\sigma_{2}$, for discretized inverse square law fields, also holds for few other sampling grids, \eg~$\set{\bb{\yc,\yr}}{\yc,\yr \in \cc{-99.5,-98.5,\dotsc,99.5}}$ and~$\set{\bb{\yc,\yr}}{\yc,\yr \in \cc{-19,-17,\dotsc,19}}$.
			\end{enumerate}

			\begin{figure}[t]
				\centering
				\makeatletter
					\if@twocolumn
						\subfloat[Plot of the first ten singular values of the sonar image in \figurename~\ref{fig:spread target signature}, normalized \wrt~the first singular value.]{\includegraphics[width=0.7\figwidth]{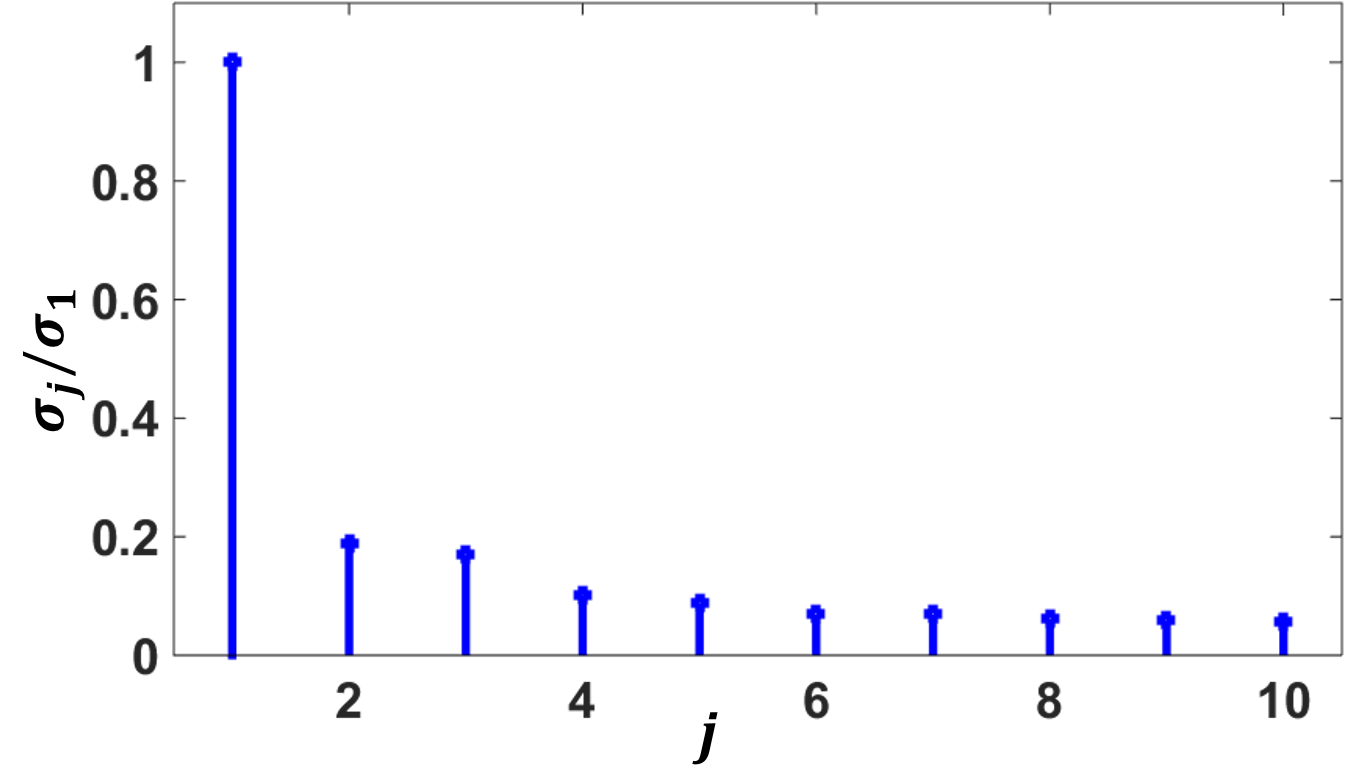}	\label{fig:spread target svd}}
						\hfill
						\subfloat[Plot of the first seven singular values of the discretized inverse square law field, normalized \wrt~the first singular value.]{\includegraphics[width=0.7\figwidth]{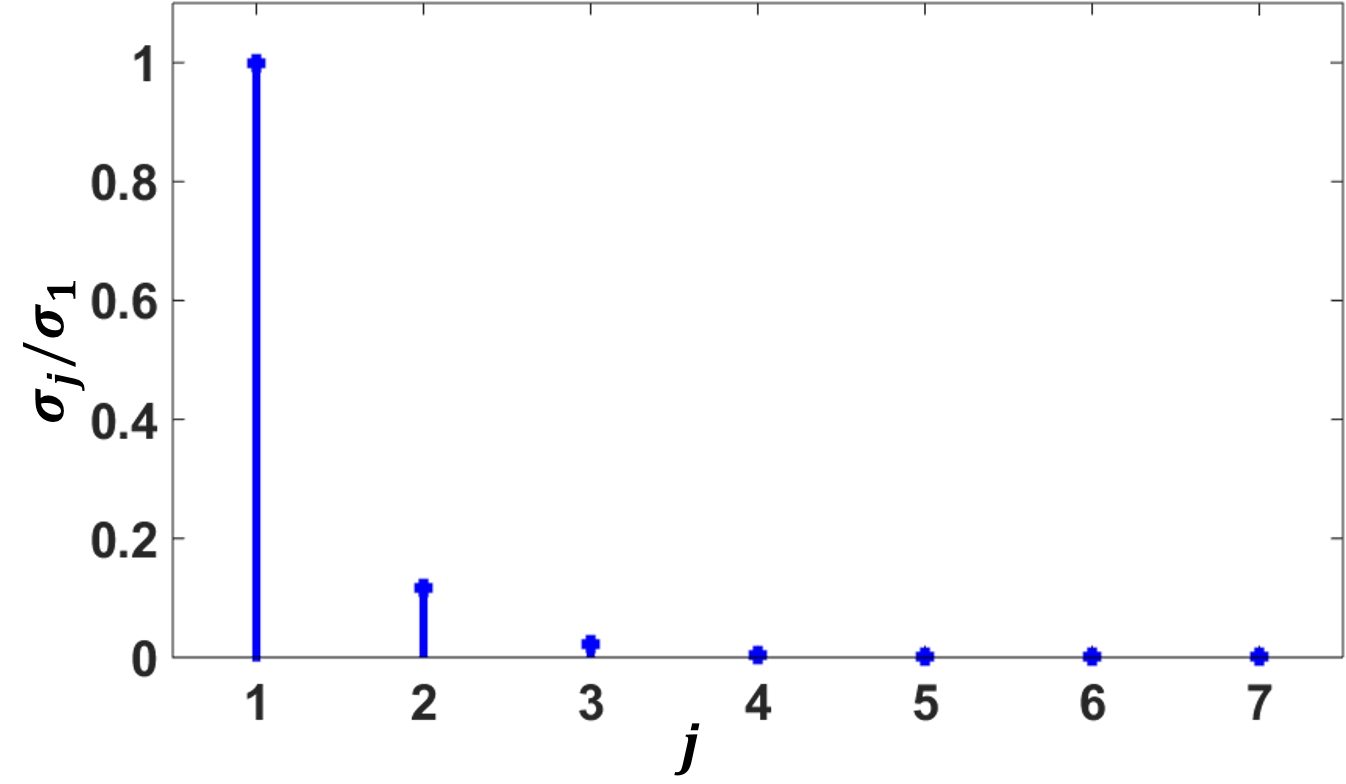}	\label{fig:inv sq law svd}}
					\else
						\subfloat[Plot of the first ten singular values of the sonar image in \figurename~\ref{fig:spread target signature}, normalized \wrt~the first singular value.]{\includegraphics[width=0.85\figwidth]{svd_spread_cropped}	\label{fig:spread target svd}}
						\hspace{0.1\linewidth}
						\subfloat[Plot of the first seven singular values of the discretized inverse square law field, normalized \wrt~the first singular value.]{\includegraphics[width=0.85\figwidth]{svd_inv_square_law}	\label{fig:inv sq law svd}}
					\fi
				\makeatother
				\caption{Plot of dominant singular values for approximately separable fields (approximately rank-1 fields).}
			\end{figure}

		\subsection{Lifted Formulation}
			\label{sec:lifting}
			By virtue of assumption~\aref{assume:monotonic}, localizing the target is synonymous with locating the peak of the induced field.
			In light of our assumptions, we can state the target detection problem as the following task: \textit{To determine the location of the peak in the field $H\bb{\vec{y}}$ from its values in only a few locations $\vec{y} \in \BB{0,1}^{2}$.}
			We use the \textit{lifting} technique from optimization~\cite{balas2005projection} to demonstrate that the separability assumption \aref{assume:separable} implies a rank one structure on the field.
			This key observation allows large reductions in both number of collected samples as well as the computational effort necessary for target detection by utilizing existing theoretical results for high-dimensional low-rank matrix completion algorithms~\cite{negahban2012restrictedstrongconvexity,candes2010noisematrixcompletion}.
			
			Let $H\bb{\vec{y}} = F\bb{\yc} G\bb{\yr}$ be the canonical separable representation of the target field and let $\mat{H}$ denote a high resolution discretized version of $H\bb{\vec{y}}$ on a $\nr \times \nc$ rectangular (not necessarily uniform) grid $\mathcal{V} \subset \BB{0,1}^{2}$.
			Let $\mathcal{V} = \cc{\yr^{1}, \yr^{2}, \dotsc, \yr^{\nr}} \times \cc{\yc^{1}, \yc^{2}, \dotsc, \yc^{\nc}}$ be the representation of the grid for $\yr^{1}, \yr^{2}, \dotsc, \yr^{\nr}, \yc^{1}, \yc^{2}, \dotsc, \yc^{\nc} \in \BB{0,1}$.
			The set of all possible sampled values of the field on the set $\mathcal{V}$ is given by $\set{H\bb[\big]{\yc^{i}, \yr^{j}}}{\bb[\big]{\yc^{i}, \yr^{j}} \in \mathcal{V}}$ and can be arranged in the form of the rank one matrix $\mat{H} \in \setRn{\nr}{\nc}$, whose $\bb{i, j}^{\thp}$ entry $\mat{H}\bb{i,j}$ is
			\begin{equation}
				\mat{H}\bb{i,j} = H\bb[\big]{\yc^{i}, \yr^{j}} = F\bb[\big]{\yc^{i}} G\bb[\big]{\yr^{j}}.
			\end{equation}
			where $\bb[\big]{\yc^{i}, \yr^{j}}$ is the physical location of the $\bb{i, j}^{\thp}$ point in $\mathcal{V}$.
			The matrix $\mat{H}$ is clearly of rank one since we can express it as the outer product $\mat{H} = \vec{f} \tpose{\vec{g}}$ where $\tpose{\vec{f}} = \BB{F\bb[]{\yc^{1}}, F\bb[]{\yc^{2}}, \dotsc, F\bb[]{\yc^{\nc}}}$ and $\tpose{\vec{g}} = \BB{G\bb[]{\yr^{1}}, G\bb[]{\yr^{2}}, \dotsc, G\bb[]{\yr^{\nr}}}$.
			Without loss of generality, we assume that both $\yr^{1}, \yr^{2}, \dotsc, \yr^{\nr}$ and $\yc^{1}, \yc^{2}, \dotsc, \yc^{\nc}$ are sorted in ascending order, corresponding respectively to traversing the grid from top to bottom and from left to right.
			Because of the preceding derivation, we can refer to $\mat{H}$ as the target field with a slight abuse of terminology.
			Consequently, we can consider $\mathcal{V}$ in a rescaled sense to refer to the set of index pairs $\cc{1,2,\dotsc,\nr} \times \cc{1,2,\dotsc,\nc}$ for the matrix $\mat{H}$.

	\section{Sampling and Reconstruction Approach}
		\label{sec:strategy}
		To convey the main aspects of our approach, we shall assume that $H\bb{\cdot}$ is a \emph{positive} scalar field and the sampling grid $\mathcal{V}$ is \emph{square} with $\nr = \nc = n$.
		These assumptions can be somewhat relaxed as described in \appendixname~\ref{sec:relax-positivity}.

		\subsection{The \texttt{PAMCUR} Algorithm}
			We use standard low-rank noisy matrix completion followed by peak localization along each axis.
			The algorithm starts with $n^{2}$ possible locations of the peak and after execution, returns a smaller set of index pairs that are guaranteed to contain the peak, provided that the error form the matrix completion step is sufficiently small.
			This can be considered as the ``first pass'' over the search region, giving us a coarse segmentation of the region into an area of interest that contains the peak, and its complement region which can be discarded.
			The algorithmic procedure can be repeated on this smaller region of interest, giving rise to the \emph{exploration-exploitation} interpretation of our hierarchical approach.
			The key steps for the first pass are described in \algorithmname~\ref{alg:PAMC-UMR} with $\proj<\mathcal{V}'>{\cdot}$ denoting the projection operator on the set of index pairs in $\mathcal{V}'$.

			\begin{algorithm}
				\caption{\texttt{PAMCUR}: Partial Adaptive Matrix Completion with Unimodal Regression}
				\textbf{Inputs:}
				\begin{enumerate}
					\item The regular grid $\mathcal{V} = \cc{1,2,\dotsc,n}^{2}$
					\item Upper bound on noise power per sample (averaged across samples), $\epsilon^{2}$
				\end{enumerate}
				\textbf{Output:} Localization index bounds $\lcL, \lcR, \lrL, \lrR \in \setZ_{+}$ such that the target is located within the rectangular region formed by the index pairs in $\cc{\lcL,\lcL+1,\dotsc,\lcR} \times \cc{\lrL,\lrL+1,\dotsc,\lrR}$.	\\
				\textbf{Steps:}
				\begin{enumerate}[({S}1)]
					\item	\label{itm:random sampling}
							Select a subset $\mathcal{V}' \subset \mathcal{V}$ of $\BigOh{n \log^{2} n}$ points uniformly and independently at random from the $n^{2}$ points in $\mathcal{V}$, and measure the (possibly noisy) samples $\mat{H}\bb{i,j}$ for every index pair $\bb{i,j} \in \mathcal{V}'$, \ie~record the projection $\proj<\mathcal{V}'>{\mat{H}}$.
					\item	\label{itm:reconstruction}
							Solve the convex nuclear norm heuristic to stable low-rank matrix completion~\cite{candes2010noisematrixcompletion}
							\minimize{\mat{Q}}
							{\nucnorm{\mat{Q}}}
							{\fronorm{\proj<\mathcal{V}'>{\mat{Q}} - \proj<\mathcal{V}'>{\mat{H}}} \leq \epsilon\card{\mathcal{V}'},}
							{\label{prob:stableMC}}
							to obtain the solution $\widehat{\mat{H}} \in \setRn{n}{n}$.
					\item	\label{itm:SVD}
							Compute the largest singular value and corresponding singular vectors of $\widehat{\mat{H}}$ as the triplet $\bb{\vec{u}, \sigma, \vec{v}}$.
					\item	\label{itm:localization LB}
							Compute the lower bounding index $\lrL \in \cc{1,2,\dotsc,n}$ for localization as the solution to the optimization problem
							\makeatletter
								\if@twocolumn
									\minimize{\vec{z},l}
									{l}
									{\vec{z} \geq \vec{0}, \, l \in \cc{1,2,\dotsc,n},
									\sep \vec{z}\bb{j+1} \leq \vec{z}\bb{j}, \quad j = l, l+1, \dotsc, n-1,
									\sep \vec{z}\bb{j-1} \leq \vec{z}\bb{j}, \quad j = 2, 3, \dotsc, l,
									\sep \ip{\vec{z}}{\vec{u}} \geq \twonorm{\vec{z}}\sqrt{1-\zeta^{2}/\sigma^{2}},}
									{\label{prob:localization}}
								\else
									\minimize{\vec{z},l}
									{l}
									{l \in \cc{1,2,\dotsc,n},
									\sep \vec{z}\bb{j+1} \leq \vec{z}\bb{j}, \quad j = l, l+1, \dotsc, n-1,
									\sep \vec{z}\bb{j-1} \leq \vec{z}\bb{j}, \quad j = 2, 3, \dotsc, l,
									\sep \ip{\vec{z}}{\vec{u}} \geq \twonorm{\vec{z}}\sqrt{1-\zeta^{2}/\sigma^{2}},
									\sep \vec{z} \geq \vec{0},}
									{\label{prob:localization}}
								\fi
							\makeatother
							where $\zeta$ is an upper bound on $\fronorm[\big]{\mat{H} - \widehat{\mat{H}}}$ from the theory of low-rank matrix completion and depends only on $n$, $\epsilon$ and $\card{\mathcal{V}'}$.
							If necessary, replace $\vec{u}$ by $-\vec{u}$ in \problemname~\pref{prob:localization} to make it feasible.	\\
							\textbf{Note:}	Since \problemname~\pref{prob:localization} is parametrized by the known parameters $n$, $\vec{u}$, $\zeta$ and $\sigma$, we shall refer to it as \problemname~\pref{prob:localization}[n,\zeta,\sigma,\vec{u}] if it is necessary to make the dependence explicit.
					
					\item	\label{itm:localization UB}
							Compute the upper bounding index $\lrR \in \cc{1,2,\dotsc,n}$ for localization as the solution to \problemname~\pref{prob:localization} with the objective function changed from $l$ to $-l$.
					
					\item	\label{itm:localization second set}
							Repeat steps~\sref{itm:localization LB} and~\sref{itm:localization UB} with \problemname~\pref{prob:localization}[n,\zeta,\sigma,\vec{v}] to respectively obtain the remaining two indices $\lcL$ and $\lcR$ (both in $\cc{1,2,\dotsc,n}$).
							If \problemname~\pref{prob:localization}[n,\zeta,\sigma,\vec{v}] is not feasible, solve \problemname~\pref{prob:localization}[n,\zeta,\sigma,-\vec{v}] instead.
				\end{enumerate}
				\label{alg:PAMC-UMR}
			\end{algorithm}
			
			We remark that for every fixed value of $l \in \cc{1,2,\dotsc,n}$, \problemname~\pref{prob:localization} reduces to a \emph{convex feasibility problem}~\cite{boyd2004convex}.
			Given that $l$ admits at most $n$ distinct values, \problemname~\pref{prob:localization} is efficiently solvable.
			Further, results from low-rank matrix completion~\cite{gross2011recovering,candes2010noisematrixcompletion} guarantee that a sample complexity of $\card{\mathcal{V}'} = \BigOh{n \log^{2} n}$ is sufficient for the solution to \problemname~\pref{prob:stableMC} to be a good reconstruction of $\mat{H}$ with high probability (\whp) over the realizations of $\mathcal{V}'$, and that the hidden constant depends on the \emph{coherence}~\cite{gross2011recovering} of $\mat{H}$ with the canonical basis for matrices in $\setRn{n}{n}$ (coherence parameters for decaying exponential and power-law fields are analytically computed in \appendixname~\ref{sec:coherence computation}).
			It is intuitive to reason that good mean-squared error (MSE) leads to good peak localization in $\mat{H}$.
			\theoremsname~\ref{thm:inner product bound} and~\ref{thm:convergence} below, give precise results to the same effect.

		\subsection{Correctness and Localization-Accuracy Trade-off}
			\label{sec:loc-acc tradeoff}
			For a quantitative comparison of the trade-offs involved, we present the following analysis, that holds \whp~over realizations of $\mathcal{V}'$.
			Suppose that the fraction $q = \card{\mathcal{V}'}/n^{2}$ of the total number of elements in $\mat{H}$ are sampled at random, where the sampling budget $\card{\mathcal{V}'} = \BigOm{n \log^{2}n}$ is \emph{sufficiently high} with the right constants as given by~\cite{negahban2012restrictedstrongconvexity} or~\cite{candes2010noisematrixcompletion}.
			Using \theoremname~7 from~\cite{candes2010noisematrixcompletion} on mean-squared-error performance of the low-rank matrix completion subproblem~\pref{prob:stableMC}, we get bounds on the Frobenius norm of the reconstruction error matrix $\mat{Z} = \mat{H} - \widehat{\mat{H}}$ as
			\begin{equation}
				\fronorm{\mat{Z}} \leq \bb{2 + 4 \sqrt{n\bb{1 + 2/q}}} \epsilon \card{\mathcal{V}'}
				= C\bb{q,n} \epsilon \card{\mathcal{V}'} = \zeta,
				\label{eqn:mse result}
			\end{equation}
			where we have used the shorthand notation $C\bb{q,n} = 2 + 4 \sqrt{n\bb{1 + 2/q}}$.
			In particular, we have a bound on the reconstruction error of the form $\fronorm[\big]{\mat{H} - \widehat{\mat{H}}} \leq \zeta$ for $\zeta$ depending only on $n$, $\epsilon$ and $\card{\mathcal{V}'}$, where $\widehat{\mat{H}}$ is the solution to \problemname~\pref{prob:stableMC}.
			Let $\mat{H} = \sigma_{0} \vec{u}_{0} \tpose{\vec{v}_{0}}$ and $\widehat{\mat{H}} = \sigma \vec{u} \tpose{\vec{v}} + \sum_{j=2}^{n} \sigma_{j} \vec{u}_{j} \tpose{\vec{v}}_{j}$ respectively denote singular value decompositions (SVDs), where $\sigma$ is the largest singular value of the matrix $\widehat{\mat{H}}$ (in agreement with step~\sref{itm:SVD} of the algorithm).
			Then, the SNR is (note that the noise power $\epsilon^{2}\card{\mathcal{V}'}^{2}$ is computed only over the observed entries)
			\begin{equation}
				\text{SNR} = \fronorm{\mat{H}}^{2}/\bb{\epsilon\card{\mathcal{V}'}}^{2}
				= \sigma_{0}^{2}/\bb{\epsilon\card{\mathcal{V}'}}^{2},
			\end{equation}
			and we can rewrite \eqref{eqn:mse result} as
			\makeatletter
				\if@twocolumn
					\begin{equation}
						\begin{split}
							\fronorm{\mat{Z}} & = \fronorm[\bigg]{\sigma_{0} \vec{u}_{0} \tpose{\vec{v}_{0}} - \sigma \vec{u} \tpose{\vec{v}} - \sum_{j=2}^{n} \sigma_{j} \vec{u}_{j} \tpose{\vec{v}}_{j}} \\
							& \leq C\bb{q,n} \sigma_{0}/\sqrt{\text{SNR}}
							= \zeta.
						\end{split}
						\label{eqn:mse error bound}
					\end{equation}
				\else
					\begin{equation}
						\fronorm{\mat{Z}} = \fronorm[\bigg]{\sigma_{0} \vec{u}_{0} \tpose{\vec{v}_{0}} - \sigma \vec{u} \tpose{\vec{v}} - \sum_{j=2}^{n} \sigma_{j} \vec{u}_{j} \tpose{\vec{v}}_{j}}
						\leq C\bb{q,n} \sigma_{0}/\sqrt{\text{SNR}} = \zeta.
						\label{eqn:mse error bound}
					\end{equation}
				\fi
			\makeatother
			
			Both the correctness of the proposed algorithm and the localization-accuracy trade-off characterization follow from the theorem below, which lower bounds the magnitudes of the inner products $\ip{\vec{v}_{0}}{\vec{v}}$ and $\ip{\vec{u}_{0}}{\vec{u}}$.
			
			\begin{theorem}
				\label{thm:inner product bound}
				Let $\mat{H} = \sigma_{0} \vec{u}_{0} \tpose{\vec{v}_{0}}$ and $\widehat{\mat{H}} = \sigma \vec{u} \tpose{\vec{v}} + \sum_{j=2}^{n} \sigma_{j} \vec{u}_{j} \tpose{\vec{v}}_{j}$ respectively denote SVDs, where $\sigma$ is the largest singular value of $\widehat{\mat{H}}$ and let the bound $\fronorm[\big]{\mat{H} - \widehat{\mat{H}}} \leq \zeta \leq \sigma$ be satisfied.
				Then $\ip{\vec{u}_{0}}{\vec{u}} \ip{\vec{v}_{0}}{\vec{v}} \geq \eta\bb{\sigma,\sigma_{0},\zeta}$ where
				\begin{equation}
					\eta\bb{\sigma,\sigma_{0},\zeta} \triangleq \bb{1-\frac{\sigma}{\sigma_{0}}} + \sqrt{\bb{1-\frac{\sigma}{\sigma_{0}}}^{2} + \bb{\frac{\sigma}{\sigma_{0}}}^{2} - \bb{\frac{\zeta}{\sigma_{0}}}^{2}}.
					\label{eqn:eta-defn}
				\end{equation}
			\end{theorem}
			
			\begin{IEEEproof}
				\appendixname~\ref{sec:inner product bound proof}.
			\end{IEEEproof}
			
			In the moderate to high SNR regimes, we expect good reconstruction so that the relative error $\zeta/\sigma$ is much smaller than 1.
			We also expect $\sigma/\sigma_{0}$ to be very close to 1, but slightly less than 1 since the native formulation in \problemname~\pref{prob:stableMC} is known to bias solutions towards zero~\cite{candes2010noisematrixcompletion}.
			Assuming $\sigma = \sigma_{0}$ we have the approximate bound $\alpha\beta \geq \sqrt{1-\zeta^{2}/\sigma_{0}^{2}}$ which implies $\max\cc{\alpha,\beta} \geq \sqrt[4]{1-\zeta^{2}/\sigma_{0}^{2}}$ and $\min\cc{\alpha,\beta} \geq \sqrt{1-\zeta^{2}/\sigma_{0}^{2}}$.
			Assuming $\alpha = \ip{\vec{u}_{0}}{\vec{u}} > \ip{\vec{v}_{0}}{\vec{v}} = \beta$ and comparing the lower bound expressions with \eqref{eqn:mse error bound}, we have the following SNR dependencies:
			\begin{enumerate}[a)]
				\item	$\ip{\vec{u}_{0}}{\vec{u}}$ scales as $\sqrt[4]{1-C^{2}\bb{q,n}/\text{SNR}}$, and
				\item	$\ip{\vec{v}_{0}}{\vec{v}}$ scales as $\sqrt{1-C^{2}\bb{q,n}/\text{SNR}}$.
			\end{enumerate}
			
			\begin{lemma}
				\label{lem:bound reduction}
				For $0 \leq \zeta \leq \sigma \leq \sigma_{0}$, $\eta\bb{\sigma,\sigma_{0},\zeta} \geq \sqrt{1-\zeta^{2}/\sigma^{2}}$.
			\end{lemma}

			\begin{IEEEproof}
				\appendixname~\ref{sec:bound reduction proof}.
			\end{IEEEproof}
			
			\begin{figure}[t]
				\centering
				\makeatletter
					\if@twocolumn
						\includegraphics[width=0.7\figwidth]{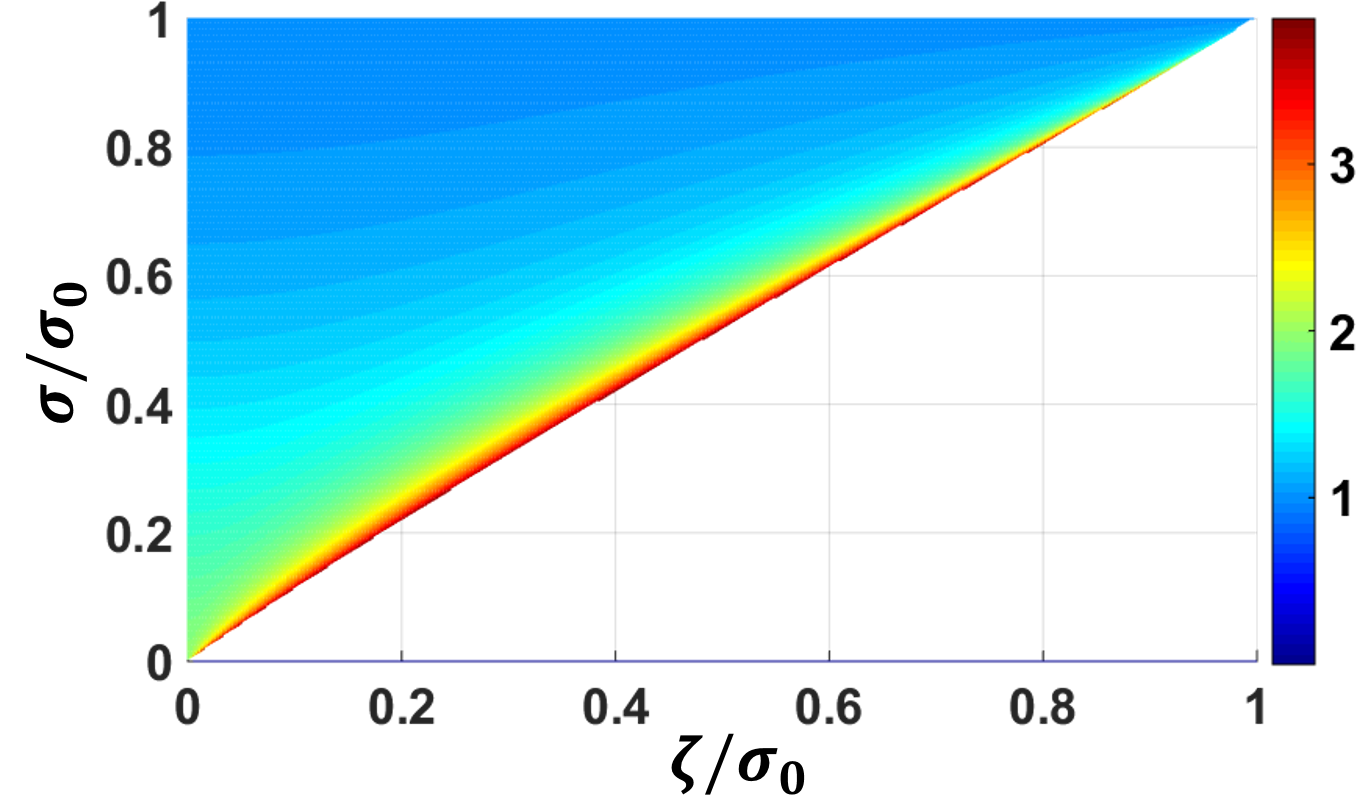}
					\else
						\includegraphics[width=0.95\figwidth]{Bound-Ratio}
					\fi
				\makeatother
				\caption{A heatmap of the ratio $\displaystyle \frac{\eta\bb{\sigma,\sigma_{0},\zeta}}{\sqrt{1-\zeta^{2}/\sigma^{2}}}$ over the domain $0 \leq \zeta \leq \sigma \leq \sigma_{0}$.
				In the high SNR regime, $\sigma/\sigma_{0} \approx 1$ and the bound $\eta\bb{\sigma,\sigma_{0},\zeta} \geq \sqrt{1-\zeta^{2}/\sigma^{2}}$ is very tight irrespective of the value of $\zeta$.}
				\label{fig:bound tightness}
			\end{figure}
			
			The main purpose of \lemmaname~\ref{lem:bound reduction} is to bound $\eta\bb{\sigma,\sigma_{0},\zeta}$ in terms of quantities that are known to the algorithm during execution, and this is utilized in \problemname~\pref{prob:localization}.
			\figurename~\ref{fig:bound tightness} demonstrates the tightness of the bound in \lemmaname~\ref{lem:bound reduction}, especially in the high SNR regime where $\sigma \approx \sigma_{0}$.
			\theoremname~\ref{thm:inner product bound} essentially utilizes error bounds on low-rank matrix completion and translates them into error bounds on the estimated singular vectors.
			Thereafter, it becomes conceptually straightforward to compute localization error bounds both numerically (by solving \problemname~\pref{prob:localization}) and analytically.
			Note that the dependence on the number of collected samples has been entirely captured in the quantity $\zeta$.
			This level of abstraction also allows us to compare localization performance for different decay profiles under a fixed sampling budget that is high enough for all the decay profiles in question.
			
			Since our localization algorithm is iterative in nature, to finish the proof of correctness we also need to show that it converges in a meaningful sense.
			The following theorem guarantees that the localized region shrinks \emph{geometrically} in each application of \algorithmname~\ref{alg:PAMC-UMR} until the localization boundaries are close enough to the true peak, provided that the observation noise is small enough for moderately good reconstruction in step~\sref{itm:reconstruction} and the target field admits a sufficiently sharp peak.

			\begin{figure}[t]
				\centering
				\makeatletter
					\if@twocolumn
						\includegraphics[width=0.7\figwidth]{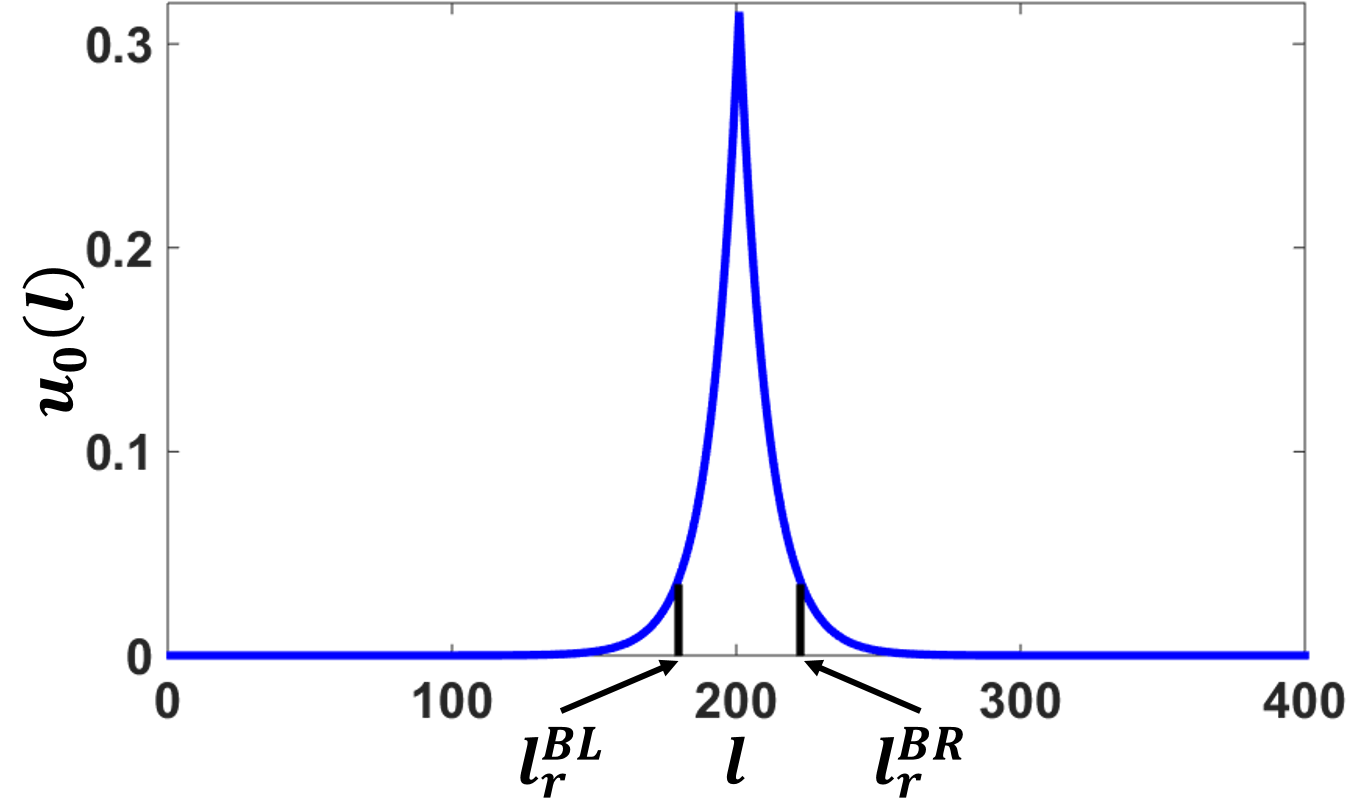}
					\else
						\includegraphics[width=0.9\figwidth]{GeoBBlims}
					\fi
				\makeatother
				\caption{Plot of an arbitrary non-negative unimodal vector $\vec{u}_{0} \in \setR^{401}$ with unit \ltwonorm, satisfying $\vec{u}_{0}\bb{l} \propto \exp\bb{-\abs{0.1\times l - 20.1}}$ over $1 \leq l \leq 401$.
				Choosing $\zeta' = 0.3\sqrt{2}$, calculations give $\lrBL = 176$ and $\lrBR = 226$.
				The threshold $\bb{\lrBR - \lrBL}$ is less than one-eighth of the length of $\vec{u}_{0}$.}
				\label{fig:BB-limits-geo-decay}
			\end{figure}

			\begin{theorem}
				\label{thm:convergence}
				Let $\mat{H} = \sigma_{0} \vec{u}_{0} \tpose{\vec{v}_{0}}$ denote the SVD of the positive matrix $\mat{H} \in \setRn{n}{n}$, where $\vec{u}_{0} \geq \vec{0}$ and $\vec{v}_{0} \geq \vec{0}$ are unimodal vectors (as described in \lemmaname~\ref{lem:peak-closeness}) with respective peaks at $l^{0}_{\text{r}}$ and $l^{0}_{\text{c}}$.
				Assume that the following are true.
				\begin{enumerate}
					\item	Step~\sref{itm:reconstruction} of \algorithmname~\ref{alg:PAMC-UMR} achieves a reconstruction error $\zeta$ upper bounded as $\zeta^{2}/\sigma^{2} < 7/16$.
					Let us define $\zeta' \triangleq 4\sqrt{1-\zeta^{2}/\sigma^{2}} - 3 > 0$.
					\item	$\onenorm{\vec{u}_{0}} = \rho_{\vec{u}}\sqrt{n}$ and $\onenorm{\vec{v}_{0}} = \rho_{\vec{v}}\sqrt{n}$ for some $\rho_{\vec{u}}, \rho_{\vec{v}} \leq \zeta'/2$.
					\item	$1 \leq \lcL \leq \lcBL$, $\lcBR \leq \lcR \leq n$, $1 \leq \lrL \leq \lrBL$ and $\lrBR \leq \lrR \leq n$, where
					\begin{subequations}
						\label{eqn:geo-shrink-bd}
						\makeatletter
							\if@twocolumn
								\begin{align}
									\lrBL & =	\argmaximizeol*{1 \leq j \leq n}{j}{\onenorm{\vec{u}_{0}\bb{1:j}} \leq \zeta'/\sqrt{2},}	\raisetag{0.7\baselineskip}	\label{eqn:lrBL-defn} \\
									\lcBL & =	\argmaximizeol*{1 \leq j \leq n}{j}{\onenorm{\vec{v}_{0}\bb{1:j}} \leq \zeta'/\sqrt{2},}	\\
									\lrBR & =	\argminimizeol*{1 \leq j \leq n}{j}{\onenorm{\vec{u}_{0}\bb{j:n}} \leq \zeta'/\sqrt{2},}	\\
									\lcBR & =	\argminimizeol*{1 \leq j \leq n}{j}{\onenorm{\vec{v}_{0}\bb{j:n}} \leq \zeta'/\sqrt{2}.}
								\end{align}
							\else
								\begin{align}
									\lrBL & =	\argmaximizeol{1 \leq j \leq n}{j}{\onenorm{\vec{u}_{0}\bb{1:j}} \leq \zeta'/\sqrt{2},}	\label{eqn:lrBL-defn} \\
									\lcBL & =	\argmaximizeol{1 \leq j \leq n}{j}{\onenorm{\vec{v}_{0}\bb{1:j}} \leq \zeta'/\sqrt{2},}	\\
									\lrBR & =	\argminimizeol{1 \leq j \leq n}{j}{\onenorm{\vec{u}_{0}\bb{j:n}} \leq \zeta'/\sqrt{2},}	\\
									\lcBR & =	\argminimizeol{1 \leq j \leq n}{j}{\onenorm{\vec{v}_{0}\bb{j:n}} \leq \zeta'/\sqrt{2}.}
								\end{align}
							\fi
						\makeatother
					\end{subequations}
				\end{enumerate}
				Then, \algorithmname~\ref{alg:PAMC-UMR} gives a localized region of size
				\begin{equation}
					\bb{\lcR-\lcL} \times \bb{\lrR-\lrL} < 16 \rho^{2}_{\vec{u}} \rho^{2}_{\vec{v}} n^{2}/\bb{\zeta'}^{4}.
					\label{eqn:loc-bound}
				\end{equation}
			\end{theorem}

			\begin{IEEEproof}
				\appendixname~\ref{sec:convergence proof}.
			\end{IEEEproof}

			Clearly, all assumptions in the above theorem are symmetric \wrt~vectors $\vec{u}_{0}$ and $\vec{v}_{0}$.
			We also make the following observations.
			\begin{enumerate}
				\item	All assumptions of \theoremname~\ref{thm:convergence} depend on the quantity $\zeta'$.
				If step~\sref{itm:reconstruction} of \algorithmname~\ref{alg:PAMC-UMR} achieves a better reconstruction, then $\zeta^{2}/\sigma^{2}$ is smaller and hence $\zeta'$ is larger.
				A larger value of $\zeta'$ means that the \lonenorm{} requirement on $\vec{u}_{0}$ is less stringent (allowing for a milder sharpness requirement on the peak).
				\item	If $\vec{u}_{0}$ showed a completely diffuse peak (all elements are of equal magnitude), then we would have $\vec{u}_{0} = \vec{1}/\sqrt{n}$ and $\onenorm{\vec{u}_{0}} = \sqrt{n}$, since $\twonorm{\vec{u}_{0}} = 1$ and $\vec{u}_{0} \geq \vec{0}$.
				Thus, the requirement of $\onenorm{\vec{u}_{0}} = \rho_{\vec{u}}\sqrt{n} \leq \zeta' \sqrt{n}/2$ as the criterion for sharpness of the peak is fairly modest, especially if $\rho_{\vec{u}} \leq \zeta'/2$ is a constant independent of $n$.
				\item	The conditions in \eqref{eqn:geo-shrink-bd} help describe the state of \algorithmname~\ref{alg:PAMC-UMR} at which one may expect geometric shrinkage of the localized region.
				In particular, geometric shrinkage continues only until the size of the localized region falls below $\bb{\lcBR - \lcBL} \times \bb{\lrBR - \lrBL}$.
				As illustrated in \figurename~\ref{fig:BB-limits-geo-decay}, if the peak in $\vec{u}_{0}$ (respectively $\vec{v}_{0}$) is sufficiently sharp, then this threshold $\bb{\lrBR - \lrBL}$ (respectively $\bb{\lcBR - \lcBL}$) for geometric shrinkage is fairly small.
			\end{enumerate}

			The proof of \theoremname~\ref{thm:convergence} relies on the following supporting lemmas, that also outline the high level proof strategy.
			One solution strategy for \problemname~\pref{prob:localization} involves solving a sequence of specific instances of \problemname~\pref{prob:peak-closeness}.
			\lemmaname~\ref{lem:peak-closeness} is essentially an implication of weak duality for \problemname~\pref{prob:peak-closeness}.
			Since \problemname~\pref{prob:peak-closeness} is stated as a feasibility problem, to simplify subsequent analysis, \lemmaname~\ref{lem:non-neg soln} states the equivalent optimization problem.
			Finally, \lemmaname~\ref{lem:cum-avg} helps to identify the dominant bound in \eqref{eqn:delta-bound} for the special case relevant to the proof of \theoremname~\ref{thm:convergence}.

			\begin{lemma}
				\label{lem:peak-closeness}
				Define the feasibility problem
				\makeatletter
					\if@twocolumn
						\find{\vec{z}}
						{\vec{z}\bb{j+1} \leq \vec{z}\bb{j}, \quad j = \lstar, \lstar + 1, \dotsc, n-1,
						\sep \vec{z}\bb{j-1} \leq \vec{z}\bb{j}, \quad j = 2, 3, \dotsc, \lstar,
						\sep \ip{\vec{z}}{\vec{v}} \geq \rho, \, \twonorm{\vec{z}} = 1, \, \vec{z} \geq \vec{0},}
						{\label{prob:peak-closeness}}
					\else
						\find{\vec{z}}
						{\vec{z}\bb{j+1} \leq \vec{z}\bb{j}, \quad j = \lstar, \lstar + 1, \dotsc, n-1,
						\sep \vec{z}\bb{j-1} \leq \vec{z}\bb{j}, \quad j = 2, 3, \dotsc, \lstar,
						\sep \ip{\vec{z}}{\vec{v}} \geq \rho,
						\sep \twonorm{\vec{z}} = 1,
						\sep \vec{z} \geq \vec{0},}
						{\label{prob:peak-closeness}}
					\fi
				\makeatother
				\wrt~$\vec{z} \in \setR^{n}$ and parametrized by $\lstar \in \cc{1,2,\dotsc,n}$, $0 \leq \rho \leq 1$ and $\vec{v} \in \setR^{n}_{+}$.
				Let $\twonorm{\vec{v}} = 1$ and suppose that $\vec{v}$ is unimodal with peak at $l^{0} \in \cc{\lstar, \lstar + 1, \dotsc, n}$, \ie~$\vec{v}\bb{j+1} \leq \vec{v}\bb{j}$ for every $l^{0} \leq j \leq n-1$ and $\vec{v}\bb{j-1} \leq \vec{v}\bb{j}$ for every $2 \leq j \leq l^{0}$.
				For \problemname~\pref{prob:peak-closeness} to be feasible, it is necessary that
				\makeatletter
					\if@twocolumn
						\begin{multline}
							\rho^{2} \leq \ip{\vec{1}}{\vec{v}}^{2} + \bb{\delta^{2} - 2\delta\ip{\vec{1}}{\vec{v}}}\bb{\kR+\kL+1}	\\
							{} + \delta^{2}\bb{\kR+\kL+1}^{2}
							\label{eqn:delta-l-inequality}
						\end{multline}
					\else
						\begin{equation}
							\rho^{2} \leq \ip{\vec{1}}{\vec{v}}^{2} + \bb{\delta^{2} - 2\delta\ip{\vec{1}}{\vec{v}}}\bb{\kR+\kL+1} + \delta^{2}\bb{\kR+\kL+1}^{2}
							\label{eqn:delta-l-inequality}
						\end{equation}
					\fi
				\makeatother
				holds for any integers $0 \leq \kL \leq l^{0}-\lstar$, $0 \leq \kR \leq n-l^{0}$, and any $\delta \in \setR_{+}$ satisfying
				\begin{equation}
					\delta \leq \frac{\ip{\vec{1}}{\vec{v}\bb{j+1:n}}}{\bb{l^{0}+\kR-j}},
					\quad	\forall \, j \in \cc{l^{0}-\kL-1,\dotsc, l^{0}+\kR-1}.
					\label{eqn:delta-bound}
				\end{equation}
			\end{lemma}

			\begin{IEEEproof}
				\appendixname~\ref{sec:peak-closeness proof}.
			\end{IEEEproof}

			\begin{lemma}
				\label{lem:non-neg soln}
				For $\vec{v} \in \setR^{n}_{+}$ and $\lstar \in \cc{1,2,\dotsc,n}$, the optimization problem
				\minimize{\vec{z}}
				{-\ip{\vec{z}}{\vec{v}}}
				{\vec{z}\bb{j+1} \leq \vec{z}\bb{j}, \quad j = \lstar, \lstar + 1, \dotsc, n-1,
				\sep \vec{z}\bb{j-1} \leq \vec{z}\bb{j}, \quad j = 2, 3, \dotsc, \lstar,
				\sep \twonorm{\vec{z}}^{2} = 1.}
				{\label{prob:peak-closeness-opt}}
				admits a solution $\vec{z}_{\opt}$ that satisfies $\vec{z}_{\opt} \geq \vec{0}$.
				Furthermore, $\vec{z}_{\opt}$ is feasible for \problemname~\pref{prob:peak-closeness} if and only if the optimal value of \problemname~\pref{prob:peak-closeness-opt} is no greater than $-\rho$.
			\end{lemma}

			\begin{IEEEproof}
				\appendixname~\ref{sec:non-neg soln proof}.
			\end{IEEEproof}

			\begin{lemma}
				\label{lem:cum-avg}
				If $\vec{v} \in \setR^{n}_{+}$ is a unimodal vector (as described in \lemmaname~\ref{lem:peak-closeness}) with peak at $l^{0} \in \cc{2,3,\dotsc,n}$, then $\ip{\vec{1}}{\vec{v}\bb{j+1:n}}/\bb{l^{0}-j}$ is a monotonically non-decreasing function of $j$ over $1 \leq j \leq l^{0}-1$.
			\end{lemma}
			
			\begin{IEEEproof}
				\appendixname~\ref{sec:cum-avg proof}.
			\end{IEEEproof}

		\subsection{Complexity Computations}
			\label{sec:tradeoff scaling laws}
			For a quantitative comparison of the trade-offs involved, we present the following analysis.
			In each round of sampling (each invocation of \algorithmname~\ref{alg:PAMC-UMR}), we collect $\BigOh{\nu n \log^{2} n}$ random samples on an $n \times n$ sub-matrix $\mat{H}$ formed by sampling $H\bb{\cdot}$ on a regular grid (using results from~\cite{gross2011recovering}, dependence of the number of samples on the coherence parameter $\nu$ has been factored in).
			Let us assume that sampling the field at the Nyquist rate would have required $m^{2} n^{2}$ samples, \ie~discretization of $H\bb{\cdot}$ into an $mn \times mn$ sized grid would allow for reconstruction of $H\bb{\cdot}$ using linear low-pass filtering.
			\theoremname~\ref{thm:convergence} guarantees a geometric reduction in the size of the search space as long as the search space is large in an appropriate sense, implying that the search space becomes small after at most $N_{\text{R}} = \BigOh{\log mn}$ sampling rounds.
			Assuming that the small search space is of size independent of $n$ and can be covered using a constant number of samples, the total number of samples collected equals
			\makeatletter
				\if@twocolumn
					\begin{equation}
						\begin{split}
							N_{\text{S}} & = N_{\text{R}} \cdot \BigOh{\nu n \log^{2} n}
							= \BigOh{\nu n \log^{2} n \log mn} \\
							& = \BigOh{\bb{1-\gamma\bb{n}} n \log^{2} n \log mn}.
						\end{split}
					\end{equation}
				\else
					\begin{equation}
						N_{\text{S}} = N_{\text{R}} \cdot \BigOh{\nu n \log^{2} n}
						= \BigOh{\nu n \log^{2} n \log mn} = \BigOh{\bb{1-\gamma\bb{n}} n \log^{2} n \log mn}.
					\end{equation}
				\fi
			\makeatother
			On the other hand, one-step naive matrix completion would have required $\BigOh{\nu mn \log^{2} mn}$ samples which is order-wise larger than $N_{\text{S}}$ by a factor of $\BigOh{\frac{m \log mn}{\log^{2} n}}$.

			Let us denote the total runtime by $N_{\text{T}}$.
			To compute this, we denote the run time of the $n \times n$ matrix completion problem from $\BigOh{\nu n \log^{2} n}$ random samples by $R_{1}\bb{n}$.
			We note that \algorithmname~\ref{alg:PAMC-UMR} involves solving \problemname~\pref{prob:stableMC} once and solving \problemname~\pref{prob:localization} four times.
			Clearly, \problemname~\pref{prob:stableMC} is a matrix completion problem and \problemname~\pref{prob:localization} can be solved by solving the $n$ distinct instances of the feasibility problem~\pref{prob:peak-closeness}[\lstar,\cdot,\cdot] for $\lstar \in \cc{1,2,\dotsc,n}$.
			By \lemmaname~\ref{lem:non-neg soln}, \problemname~\pref{prob:peak-closeness} is equivalent to \problemname~\pref{prob:peak-closeness-opt} which is a very simple convex quadratic program with non-negativity constraints in its dual form (see \eqref{eqn:peak-closeness-dual}).
			Denoting the running time for solving \problemname~\pref{prob:peak-closeness-opt} in its dual form by $R_{2}\bb{n}$, we have the complexity of \algorithmname~\ref{alg:PAMC-UMR} as $R_{1}\bb{n} + 4n R_{2}\bb{n}$ and the total running time as $N_{\text{R}}$ times the complexity of \algorithmname~\ref{alg:PAMC-UMR}, or equivalently, $N_{\text{T}} = \BigOh{\bb{R_{1}\bb{n} + n R_{2}\bb{n}} \log mn}$.
			In contrast, one-step naive matrix completion would have required $\BigOh{R_{1}\bb{mn}}$ running time; substantially larger than $\BigOh{R_{1}\bb{n} \log mn}$, if using general purpose Semidefinite Program Solvers like SeDuMi with CVX~\cite{gb08,cvx} that scale as $R_{1}\bb{p} = \BigOh{p^{3.5}}$.
			Thus, reconstructing the entire field turns out to be much worse from both sampling and computational viewpoints.

	\section{Baseline algorithms}
		\label{sec:methods}
		In this section, we introduce some baseline methods which shall serve as points of comparison for our \texttt{PAMCUR} algorithm.
		Each baseline algorithm is similar to \texttt{PAMCUR} in terms of the multi-resolution approach, \ie~each stage involves sampling on a subset of a $n \times n$ grid within some region of interest (ROI) with the sampling resolution getting finer with each progressive stage.
		However, the various algorithms differ in the localization strategies employed to shrink down the ROI in subsequent iterations.
		For all baseline algorithms presented below, we use a fixed scale $\kappa \leq 1$ in simulations for progressive reduction of the ROI with each stage (leading to a geometric reduction at a fixed rate) till the update in target localization across consecutive stages falls below a tolerance threshold.
		In contrast, \texttt{PAMCUR} (by design) chooses the reduction in ROI adaptively at each stage and thus does not correspond to a fixed value of $\kappa$.
		For the purpose of comparison, we have considered fixed sized grids at each stage and presented results for $n = 50$ and $n = 100$.
		To solve the low-rank matrix completion problem~\pref{prob:stableMC} we used the \emph{LMaFit} implementation~\cite{xu2012alternating,shen2012augmented}, while the unimodal regression was solved using the \emph{Pair Adjacent Violators} approach~\cite{ayer1955empiricaldistributionfunction,stout2008unimodal}.

		\subsection{Matrix Completion based variants (\texttt{MConly} and \texttt{MCuni})}
			These two algorithms are closely related to \texttt{PAMCUR} and differ only in which parts of \texttt{PAMCUR} they employ.
			The \texttt{MConly} algorithm, at each stage, performs a standard noisy low-rank matrix completion followed by peak detection along the horizontal and vertical directions as proposed in \cite{choudhary2014activetargetdetection}.
			The \texttt{MCuni} algorithm, at each stage, additionally finds the best unimodal fit (in the sense of \ltwonorm) to the estimated singular vectors $\vec{u}$ and $\vec{v}$ after the matrix completion step but before the peak detection step.
			This unimodal regression step provides robustness in the presence of spurious peaks and is posed analogous to \problemname~\pref{prob:peak-closeness}.

		\subsection{Surface Interpolation (\texttt{interp})}
			This method, at each stage, attempts to impute missing data after sampling by employing interpolation over a smooth surface using a nearest-neighbor approach.
			This is done by simply searching for the nearest sampled location and duplicating the measurement at that location.
			Such nearest-neighbor interpolation could lead to a noisy completed matrix, so we smooth it further by a moving average mask before executing peak detection on the imputed matrix.
			If implemented efficiently (using space partitioning methods like k-d trees), this method admits $\BigOh[]{n^3}$ running time on an $n \times n$ grid~\cite{lee1977Worstcaseanalysis}.

		\subsection{Mean-shift based Gradient Ascent (\texttt{MS})}
			Mean-shift (\texttt{MS}) is actually a popular algorithm used in pattern recognition for unsupervised clustering of data points in the feature space and we present below a suitable adaptation to perform gradient-ascent on the target field.
			Unlike the algorithms discussed so far, this method only exploits local information.
			The \texttt{MS} algorithm proceeds by collecting samples, approximating the local gradient from these samples and then performing a gradient ascent step to determine the next sampling neighborhood.
			Specifically, the gradient direction at a location $v_{k}$ is computed using a mean-shift update over a window of size $\omega$ according to a center of mass type computation ($p$ below denotes an arbitrary location)
			\begin{equation}
				\begin{split}
					v_{k+1}	& =	\frac{\sum_{p} p \cdot U\bb{p-v_{k}}}{\sum_{p} U\bb{p-v_{k}}},	\\
					U\bb{p-v}	& =	\begin{cases}
										1,	&	\norm{p-v}_{c} \leq \omega,	\\
										0,	&	\text{otherwise}.
									\end{cases}
				\end{split}
				\label{eq:MSC}
			\end{equation}
			Note that the new location $v_{k+1}$ is in the direction of one of the eight adjacent locations $v_{k}$ based on the direction of the gradient.

			Being local, this algorithm is quite susceptible to finding local peaks.
			To give this algorithm a fighting chance, within each stage, we shall allow it to execute a few times with random starting points so that multiple local peaks can be detected and the highest peak can be returned.
			The computational complexity is measured accordingly and (if implemented efficiently using space partitioning methods like k-d trees~\cite{lee1977Worstcaseanalysis}) equals $\BigOh{M n^3}$ where $M$ is the number of restarts and $n \times n$ is the grid size.
			The pseudo-code for this method is outlined as \algorithmname~\ref{algo:msc} in \appendixname~\ref{sec:MS algo pseudocode}.
			A larger $\omega$ provides robustness to local noise characteristics at the cost of increasing the number of collected samples and the risk of smoothing out peaks with small spreads.
			Thus, the number of samples acquired depends not only on the length of the gradient ascent trail but also on $\omega$.

	\section{Numerical Experiments: Synthetic Data} 
		\label{sec:synthetic simulations}
\begin{figure}[t]
				\centering
				\makeatletter
					\if@twocolumn
						\includegraphics[width=0.65\figwidth]{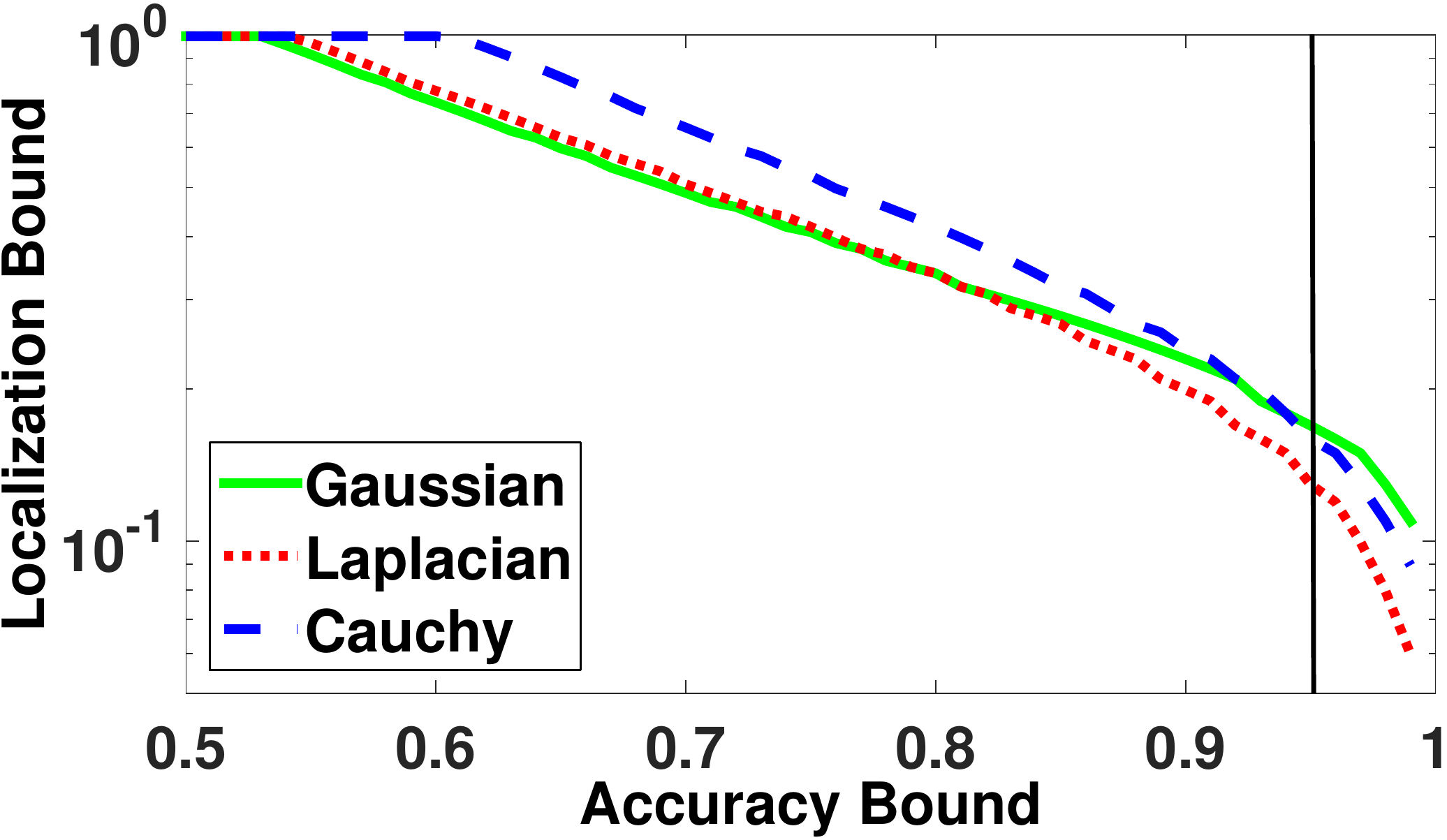}
					\else
						\includegraphics[width=0.95\figwidth]{Loc_Acc_Tradeoff_three}
					\fi
				\makeatother
				\caption{Trade-off showing the localization bound (normalized by size of search space) achievable for a given accuracy bound $\sqrt{1 - \zeta^{2}/\sigma^{2}}$ (and hence for a given sampling budget) for discretized versions of standard Gaussian, Laplacian and Cauchy fields.}
				\label{fig:localization accuracy tradeoff}
			\end{figure}

\begin{figure}
				\centering
				\makeatletter
					\if@twocolumn
						\subfloat[Laplacian]{\includegraphics[width=0.55\linewidth]{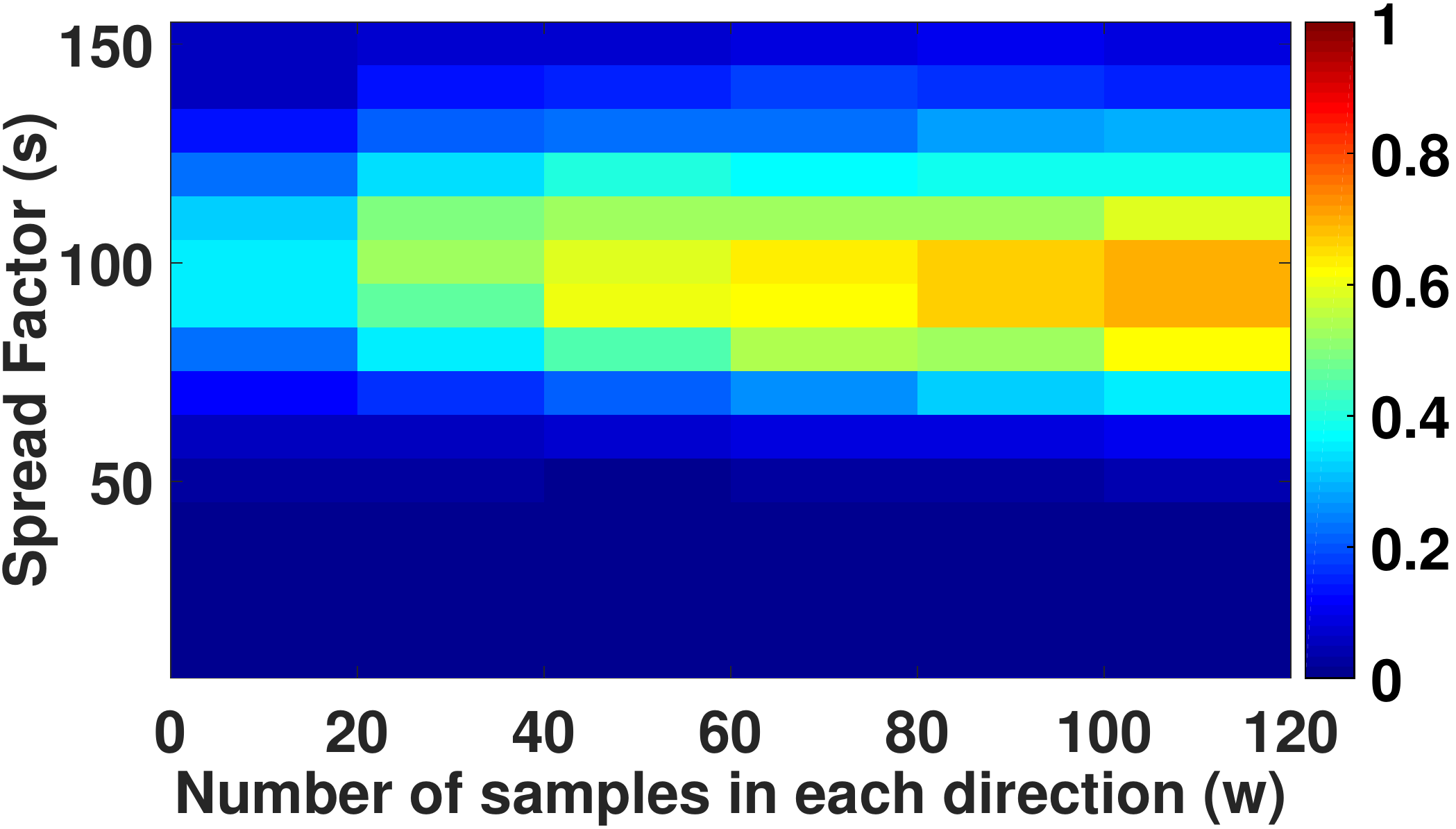}}	\\
						\subfloat[Gaussian]{\includegraphics[width=0.49\linewidth]{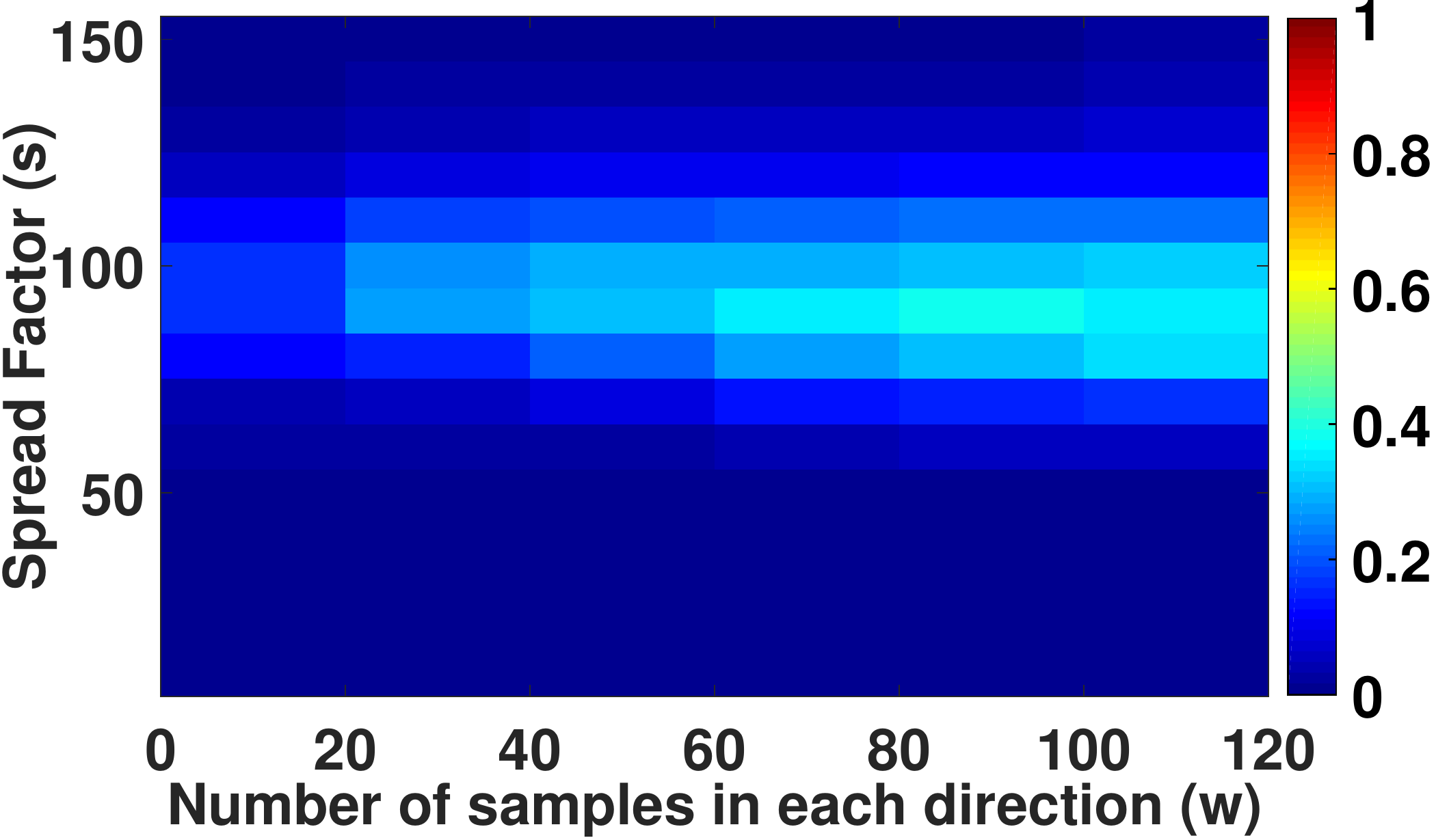}}
						\hfill
						\subfloat[Cauchy]{\includegraphics[width=0.49\linewidth]{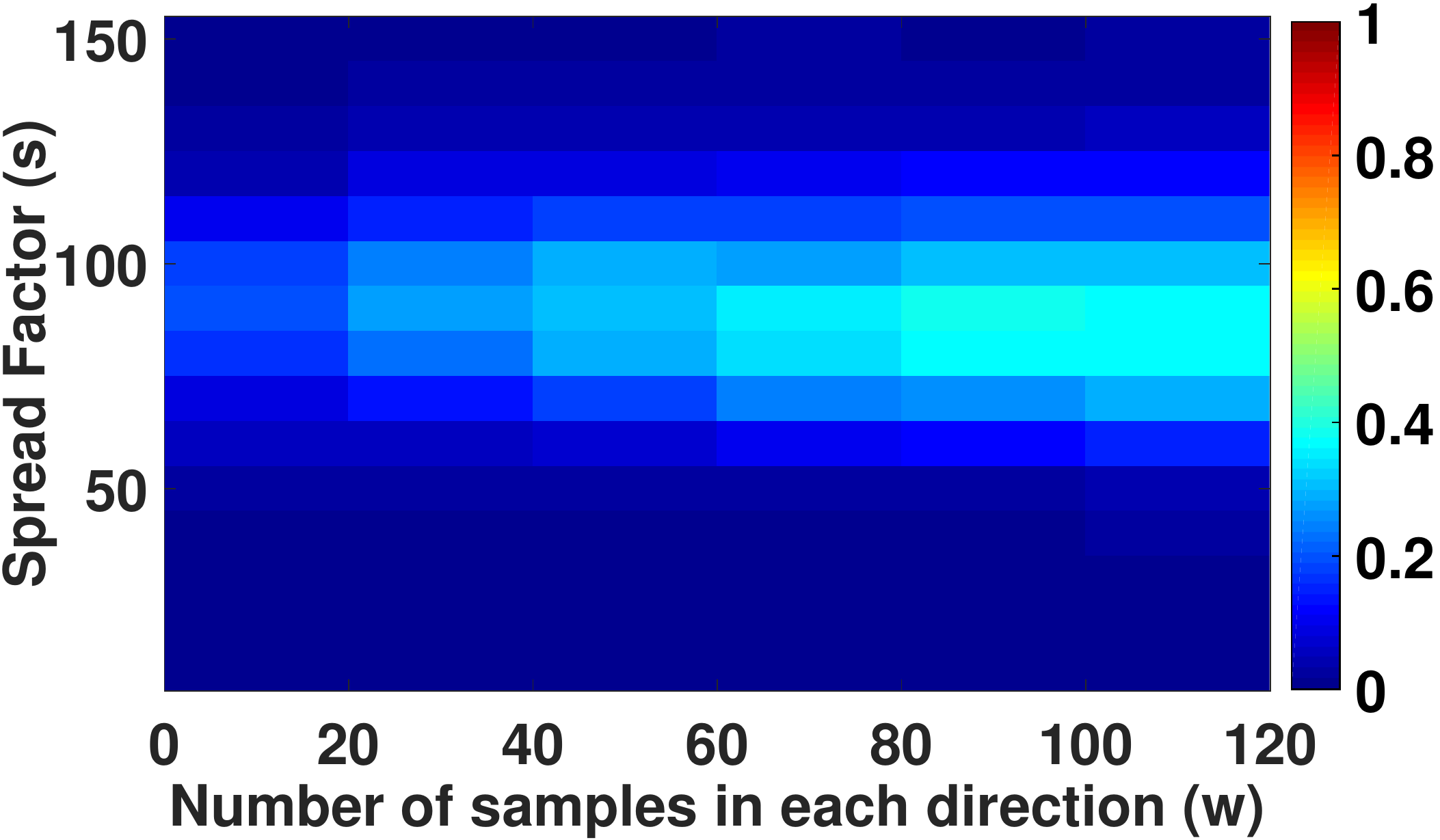}}
					\else
						\subfloat[Laplacian]{\includegraphics[width=0.65\figwidth]{det_laplace}}
						\hspace{0.01\figwidth}
						\subfloat[Gaussian]{\includegraphics[width=0.65\figwidth]{det_gaussian}}
						\hspace{0.01\figwidth}
						\subfloat[Cauchy]{\includegraphics[width=0.65\figwidth]{det_cauchy}}
					\fi
				\makeatother
				\caption{Variation of the probability of correct localization by \texttt{MConly} algorithm to within 4\% of the search space, averaged over 10 trials.
				Results are for three different decay profiles across a range of sampled window sizes and field spread factors with Gaussian distributed background noise.}
				\label{fig:simulation}
			\end{figure}

			It is instructive to study the localization bound vs accuracy trade-off for some known unimodal decay profiles from the exponential and power law families.
			\figurename~\ref{fig:localization accuracy tradeoff} shows the results for the standard Gaussian, Laplacian ($F\bb{x} = 0.5\exp\bb{-\abs{x}}$) and Cauchy ($F\bb{x} = 1/\bb{\pi + \pi x^{2}}$) fields, with $\vec{u}$ in \problemname~\pref{prob:localization} representing the discretization of a one dimensional continuous function.
			We see that for a given (sufficiently high) accuracy level $\sqrt{1-\zeta^{2}/\sigma^{2}}$ (which translates to a fixed sampling budget), the Laplacian field admits the best one-step localization bound.
			This is somewhat surprising at first sight since Gaussian fields are inherently far more localized than Laplacian fields.
			However, the same phenomenon was confirmed via actual simulation for these decay profiles across a range of window sizes and spread factors (see \figurename~\ref{fig:simulation}).
			Intuitively, good localization per sample requires the right balance of ``spread'' and support of the ``gradient'' of the field which seems to be better in case of Laplacian fields and hence they show the best localization performance for a given sampling budget.

	\section{Numerical Experiments: Elevation Dataset}
		\label{sec:real simulations}
		\subsection{3-D road network dataset}
			\begin{figure}[t]
				\centering
				\makeatletter
					\if@twocolumn
						\includegraphics[width=0.8\figwidth]{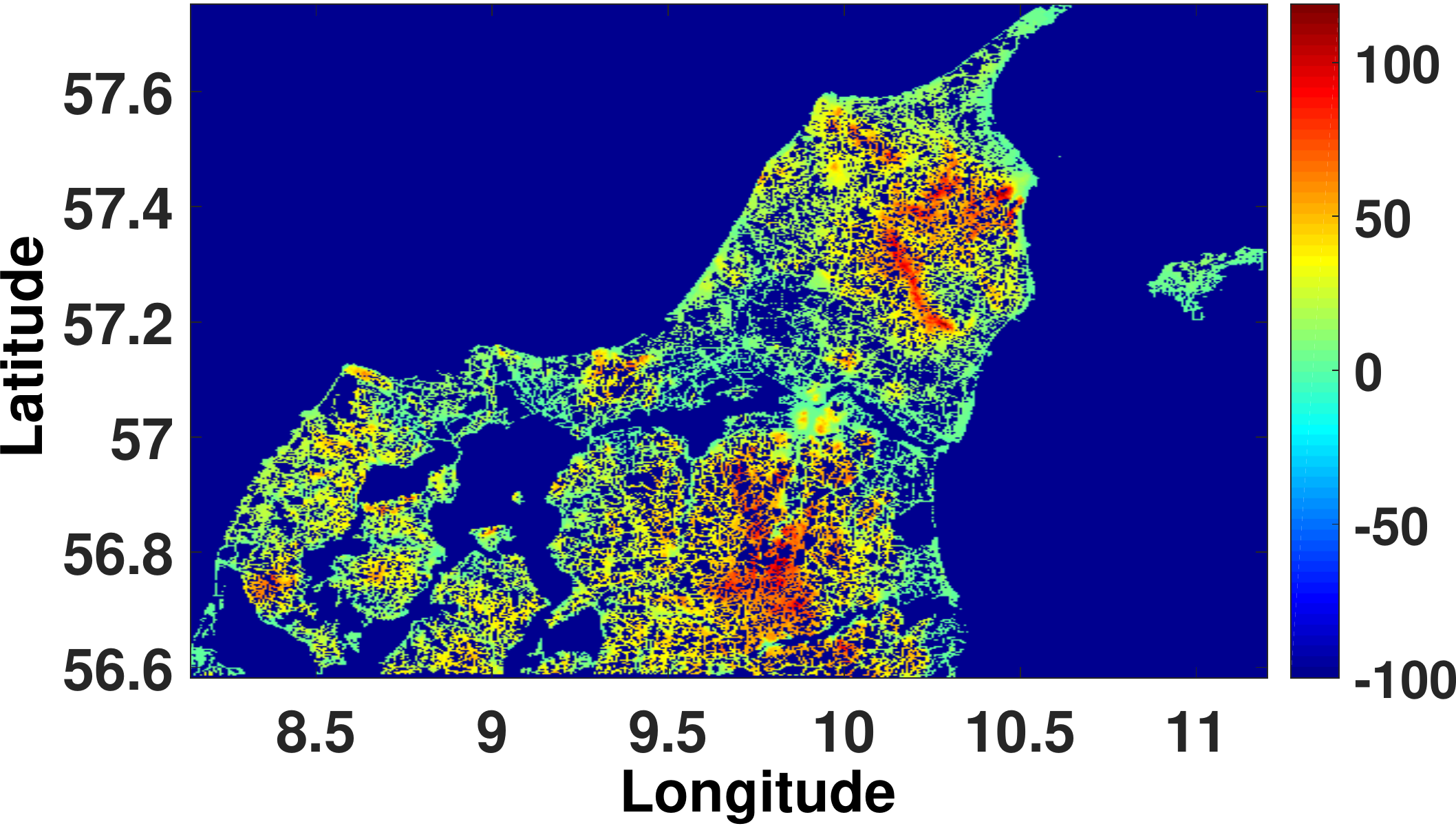}
					\else
						\includegraphics[width=0.95\figwidth]{global-map}
					\fi
				\makeatother
				\caption{A low resolution visualization of the elevations on the road network dataset.
				The regions where readings are not available are shown in blue.}
				\label{fig:global_map}
			\end{figure}
			For testing on real data, we use an altitude dataset for road networks in North Jutland, Denmark~\cite{kaul2013building}.
			The dataset covers a region of $185\times135~km^2$ and comprises of elevation measurements in metres at $434874$ locations sampled along physical roads.
			In the past, this dataset has been mostly used in unsupervised learning tasks or in applications that require accurate elevation information, \eg~eco-routing~\cite{guo2012ecomark}.
			For the purpose of evaluation, the objective is to locate the region with the highest elevation in the map.
			Two such regions are clearly visible from the elevation heat map as shown in \figurename~\ref{fig:global_map}.
			
			To simulate the uniform grids assumed by the algorithms under consideration, if the data at a sampled location is missing, we fill it in by selecting the nearest available sample.
			A fixed grid size of $n \times n$ is used for all experiments as mentioned earlier in \sectionname~\ref{sec:methods}.
			For all algorithms other than \texttt{MS}, we vary the number of collected samples by controlling the percentage of samples acquired on the $n \times n$ subgrid at each stage (denoted by the fraction $\alpha$).
			For the \texttt{MS} algorithm, the number of collected samples depends on the number of restarts $K$.

			\begin{figure}[t]
				\centering
				\makeatletter
					\if@twocolumn
						\subfloat[$n = 100$]{\includegraphics[width=0.7\figwidth]{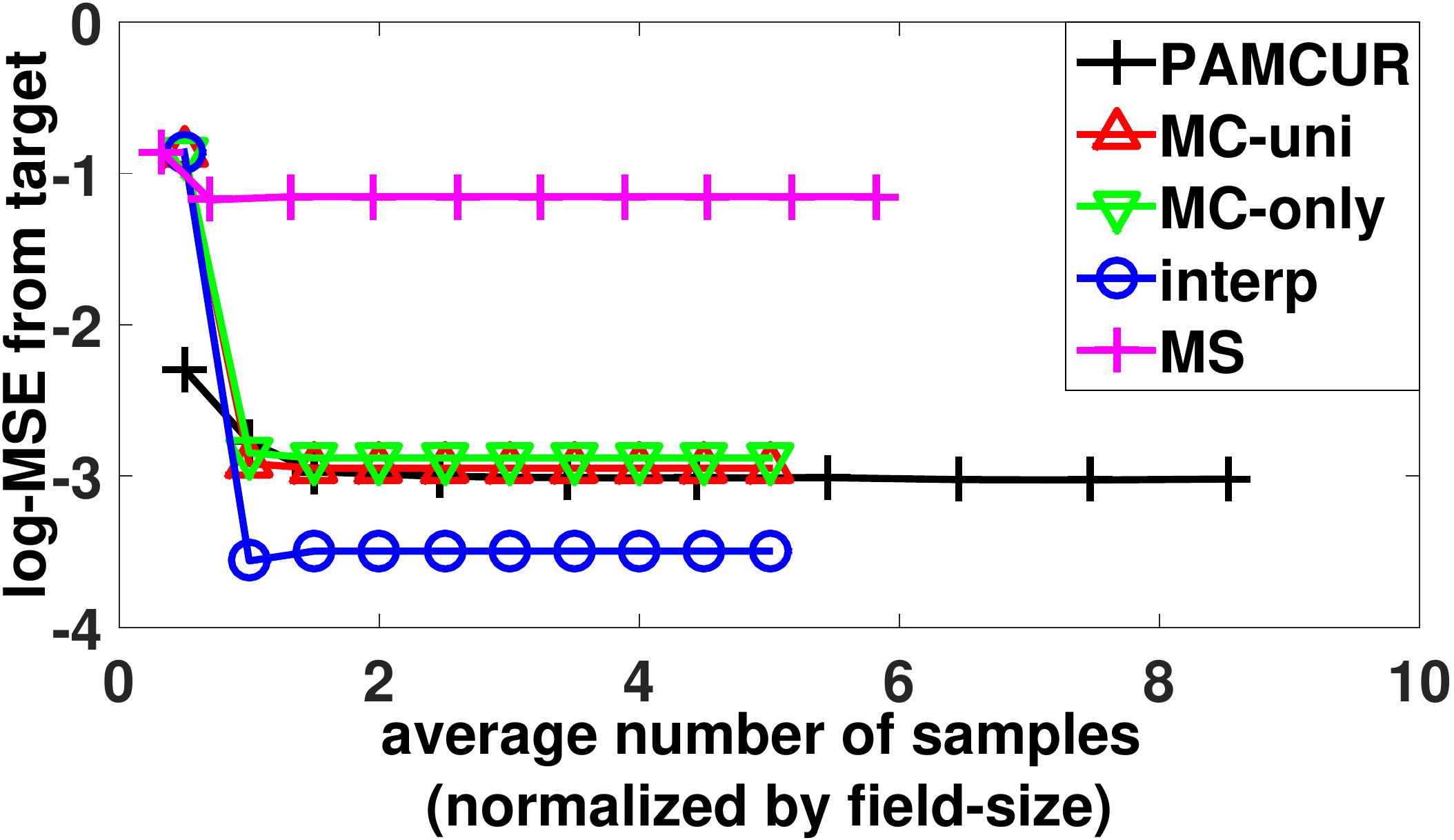}	\label{fig:100sim}}	\\
						\subfloat[$n = 50$]{\includegraphics[width=0.7\figwidth]{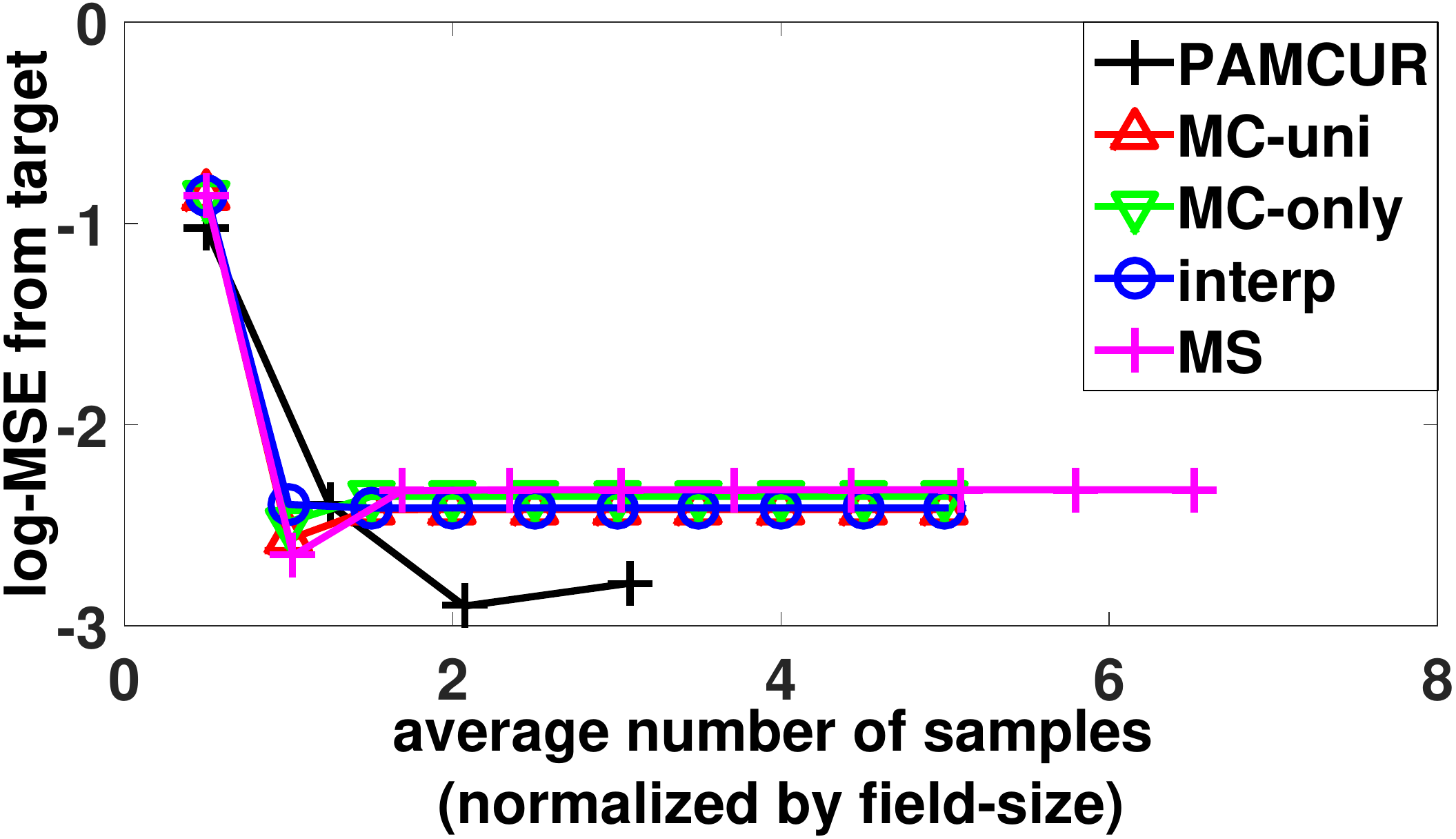}	\label{fig:50sim}}
					\else
						\subfloat[$n = 100$]{\includegraphics[width=0.95\figwidth]{field_100_periter}	\label{fig:100sim}}
						\hfill
						\subfloat[$n = 50$]{\includegraphics[width=0.95\figwidth]{field_50_periter}	\label{fig:50sim}}
					\fi
				\makeatother
				\caption{Trade-off between number of samples collected and the localization accuracy achieved for the \texttt{MConly}, \texttt{MCuni}, \texttt{interp}, \texttt{MS} and \texttt{PAMCUR} algorithms averaged over 500 runs for two different grid sizes, \viz~$n = 50$ and $n = 100$.}
				\label{fig:comparison_simulate}
			\end{figure}

		\subsection{Results}
			To evaluate each approach, we use the location of the highest peak (see the map in \figurename~\ref{fig:global_map}) as ground truth.
			Average distance of the localization from the highest peak is then measured as a metric for accuracy of the algorithm which is plotted in \figurename~\ref{fig:comparison_simulate} against the number of samples collected (normalized \wrt~the field size).
			We expect that as the number of collected samples is increased the accuracy of localization should improve for each algorithm, resulting in a trade-off curve.
			This could happen through better accuracy of the smoothing or the low-rank reconstruction process as the number of samples is increased.
			Generally speaking, we do observe this to be true in \figurename~\ref{fig:comparison_simulate} in terms of the absolute log-mean square error being lower for a grid size of $n = 100$ than for a grid size of $n = 50$.
			
			We further note that all the algorithms based on matrix completion, \viz~\texttt{MConly}, \texttt{MCuni} and \texttt{PAMCUR}, show very similar trade-off curves.
			The \texttt{PAMCUR} algorithm is somewhat more efficient at low sampling density ($n = 50$) owing to its built-in adaptive sampling strategy and hence yields better localization for the same number of collected samples.
			The \texttt{MS} algorithm, which only employs local search, performs poorly at higher sampling density ($n = 100$) which may be attributed to the formation of multiple noise induced local peaks.
			Finally, we note that the best localization accuracy trade-off on $n = 100$ sized grid is achieved by the \texttt{interp} algorithm, which is not at all surprising since an exhaustive search is performed by this algorithm on the completed 2D-grid and the higher sampling density ensures sufficient smoothness.
			For $n = 50$, smoothness of the completed 2D-grid seems to be inadequate for the \texttt{interp} algorithm to outperform other approaches.

			\begin{figure}[t]
				\centering
				\makeatletter
					\if@twocolumn
						\subfloat[localization area vs number of samples]{\includegraphics[width=0.7\figwidth]{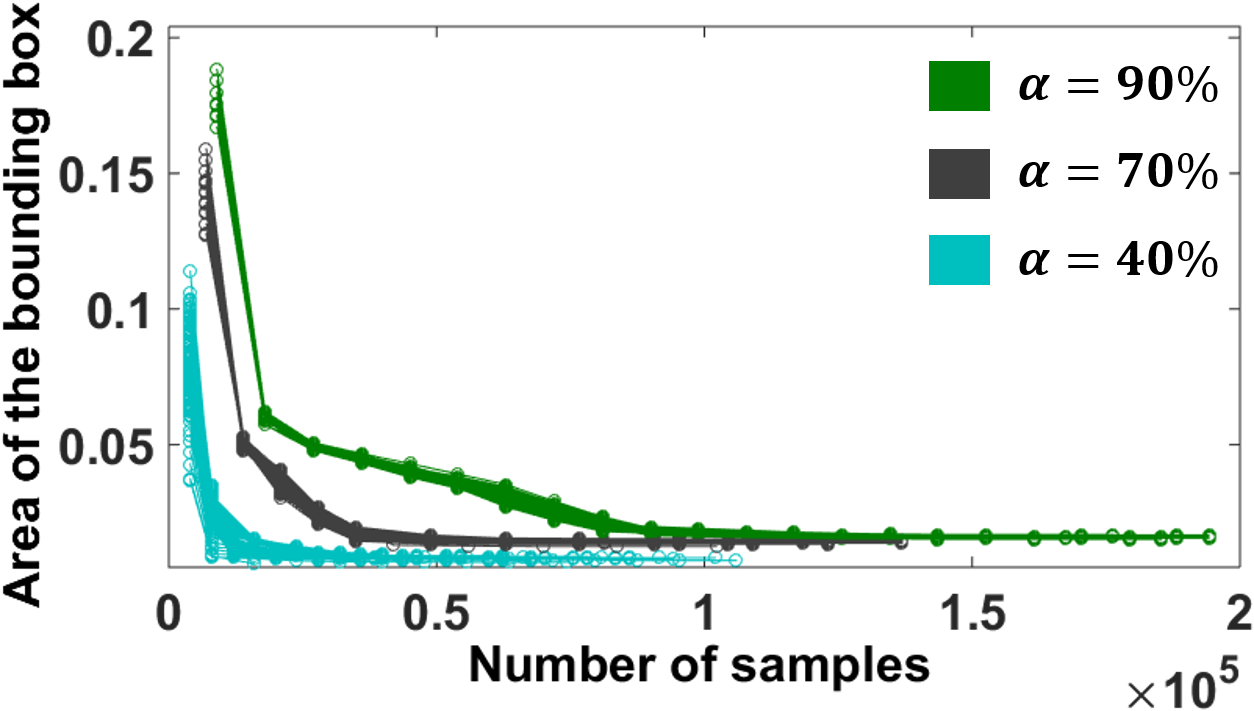}	\label{fig:areaBB-reduction}}	\\
						\subfloat[localization error vs number of samples]{\includegraphics[width=0.7\figwidth]{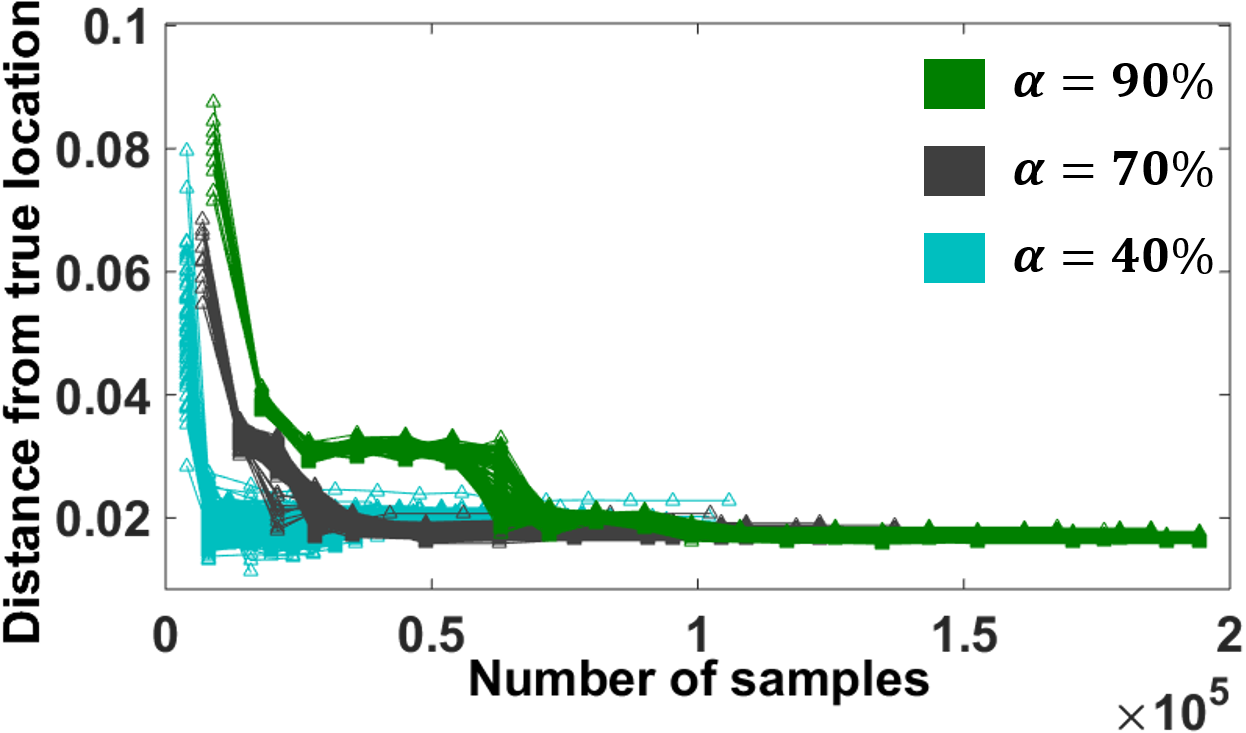}	\label{fig:locerr-reduction}}
					\else
						\subfloat[localization area vs number of samples]{\includegraphics[width=0.9\figwidth]{NS-areaBB-alpha-3}	\label{fig:areaBB-reduction}}
						\hspace{0.05\linewidth}
						\subfloat[localization error vs number of samples]{\includegraphics[width=0.9\figwidth]{NS-locerror-alpha-3}	\label{fig:locerr-reduction}}
					\fi
				\makeatother
				\caption{Localization error performance vs number of samples for \texttt{PAMCUR} algorithm on the elevation dataset in \figurename~\ref{fig:global_map}.
				Each color represents the result of 500 independent trials plotted individually.}
				\label{fig:performance curves}
			\end{figure}

			Further confirmation of the efficacy of \texttt{PAMCUR} for low sampling density is evident from~\figurename~\ref{fig:performance curves} where the decrease in
			\begin{enumerate}[a)]
				\item	the size of the localization area (\figurename~\ref{fig:areaBB-reduction}), and
				\item	the localization error (\figurename~\ref{fig:locerr-reduction})
			\end{enumerate}
			have been plotted against increase in the number of samples, for different intra-stage sampling fractions $\alpha$.
			It is clear that the trade-off vs number of samples is better for lower values of $\alpha$, subject to $\alpha$ being greater than the information theoretic lower limit for \texttt{PAMCUR}.
			This lower limit seems to be somewhere between $\alpha = 0.3$ and $\alpha = 0.4$ since we did not get interpretable results for the former while the latter gave algorithmic convergence.
			Finally, we make the pleasing observation that the initial rate of error reduction in the peak estimate as well as the rate of shrinkage of the localization bounding box is at a geometric rate and occurs with high probability over the realization of the sampling locations.
			This is in agreement with our theoretical result in \theoremname~\ref{thm:convergence}.

	\section{Conclusions}
		\label{sec:conclusions}
		In this paper, the problem of target localization from incomplete samples of the target field was examined with the goal of reducing the number of samples necessary to solve the problem by utilizing the structural properties of the target field.
		An algorithm (\texttt{PAMCUR}) was presented that exploited separability and unimodality of the decaying field around the target to use a low-rank matrix completion based approach coupled with unimodal regression at multiple resolutions, and a theoretical trade-off analysis between sampling density, noise level and convergence rate of localization was developed.
		Knowledge of exact decay profiles was shown to be unnecessary.
		It was demonstrated (somewhat surprisingly) that Laplacian fields achieve better localization vs accuracy trade-off under a fixed sampling budget, as compared to Gaussian or Cauchy fields.
		Numerical experiments and comparisons on synthetic and real datasets were performed to test the efficacy and robustness of the presented approach, and the results demonstrated the advantages of the \texttt{PAMCUR} algorithm (for low sampling density) over other approaches based on mean-shift clustering, surface interpolation and naive low-rank matrix completion with peak detection.

	\bibliographystyle{IEEEtran}
	\bibliography{IEEEabrv,Bib-Files/UWA,Bib-Files/strings,Bib-Files/refs,Bib-Files/ownpub,Bib-Files/PaperList,Bib-Files/LitSearch}

	\appendices

	\section{Proof of \theoremname~\ref{thm:inner product bound}}
		\label{sec:inner product bound proof}
		We let $\mat{P}_{\vec{u}} = \vec{u} \tpose{\vec{u}}$ and $\mat{P}_{\vec{v}} = \vec{v} \tpose{\vec{v}}$ respectively denote the matrices projecting onto the vectors $\vec{u}$ and $\vec{v}$, and let $\mat{P}_{\vec{u}^{\perp}} = \eye - \mat{P}_{\vec{u}}$ and $\mat{P}_{\vec{v}^{\perp}} = \eye - \mat{P}_{\vec{v}}$ denote the projection matrices onto the respective orthogonal complement spaces.
		We have,
		\makeatletter
			\if@twocolumn
				\begin{subequations}
					\label{eqn:projected error representation}
					\begin{align}
						\MoveEqLeft[1] \fronorm{\mat{Z}}^{2} = \fronorm{\bb{\mat{P}_{\vec{u}} + \mat{P}_{\vec{u}^{\perp}}} \mat{Z} \bb{\mat{P}_{\vec{v}} + \mat{P}_{\vec{v}^{\perp}}}}^{2} \\
						& = \fronorm{\mat{P}_{\vec{u}} \mat{Z} \mat{P}_{\vec{v}} + \mat{P}_{\vec{u}} \mat{Z} \mat{P}_{\vec{v}^{\perp}} + \mat{P}_{\vec{u}^{\perp}} \mat{Z} \mat{P}_{\vec{v}} + \mat{P}_{\vec{u}^{\perp}} \mat{Z} \mat{P}_{\vec{v}^{\perp}}}^{2} \label{eqn:orthogonal representation} \\
						& = \fronorm{\mat{P}_{\vec{u}} \mat{Z} \mat{P}_{\vec{v}}}^{2} + \fronorm{\mat{P}_{\vec{u}} \mat{Z} \mat{P}_{\vec{v}^{\perp}}}^{2} \notag \\
						& \quad + \fronorm{\mat{P}_{\vec{u}^{\perp}} \mat{Z} \mat{P}_{\vec{v}}}^{2} + \fronorm{\mat{P}_{\vec{u}^{\perp}} \mat{Z} \mat{P}_{\vec{v}^{\perp}}}^{2} \label{eqn:norm decomposition}\\
						& \geq \fronorm{\mat{P}_{\vec{u}} \mat{Z} \mat{P}_{\vec{v}}}^{2} + \fronorm{\mat{P}_{\vec{u}} \mat{Z} \mat{P}_{\vec{v}^{\perp}}}^{2} + \fronorm{\mat{P}_{\vec{u}^{\perp}} \mat{Z} \mat{P}_{\vec{v}}}^{2}. \label{eqn:sum of orthogonal error terms}
					\end{align}
				\end{subequations}
			\else
				\begin{subequations}
					\label{eqn:projected error representation}
					\begin{align}
						\fronorm{\mat{Z}}^{2} & =	\fronorm{\bb{\mat{P}_{\vec{u}} + \mat{P}_{\vec{u}^{\perp}}} \mat{Z} \bb{\mat{P}_{\vec{v}} + \mat{P}_{\vec{v}^{\perp}}}}^{2}
						=	\fronorm{\mat{P}_{\vec{u}} \mat{Z} \mat{P}_{\vec{v}} + \mat{P}_{\vec{u}} \mat{Z} \mat{P}_{\vec{v}^{\perp}} + \mat{P}_{\vec{u}^{\perp}} \mat{Z} \mat{P}_{\vec{v}} + \mat{P}_{\vec{u}^{\perp}} \mat{Z} \mat{P}_{\vec{v}^{\perp}}}^{2}	\label{eqn:orthogonal representation} \\
						& = \fronorm{\mat{P}_{\vec{u}} \mat{Z} \mat{P}_{\vec{v}}}^{2} + \fronorm{\mat{P}_{\vec{u}} \mat{Z} \mat{P}_{\vec{v}^{\perp}}}^{2} + \fronorm{\mat{P}_{\vec{u}^{\perp}} \mat{Z} \mat{P}_{\vec{v}}}^{2} + \fronorm{\mat{P}_{\vec{u}^{\perp}} \mat{Z} \mat{P}_{\vec{v}^{\perp}}}^{2}	\label{eqn:norm decomposition} \\
						& \geq \fronorm{\mat{P}_{\vec{u}} \mat{Z} \mat{P}_{\vec{v}}}^{2} + \fronorm{\mat{P}_{\vec{u}} \mat{Z} \mat{P}_{\vec{v}^{\perp}}}^{2} + \fronorm{\mat{P}_{\vec{u}^{\perp}} \mat{Z} \mat{P}_{\vec{v}}}^{2}. \label{eqn:sum of orthogonal error terms}
					\end{align}
				\end{subequations}
			\fi
		\makeatother
		where \eqref{eqn:norm decomposition} follows from \eqref{eqn:orthogonal representation} since each term within the $\fronorm{\cdot}^{2}$ expression in \eqref{eqn:orthogonal representation} is orthogonal to the other three terms \wrt~the standard trace inner product over the vector space of $n \times n$ real matrices.
		Furthermore, we neglect the last term in \eqref{eqn:norm decomposition} to arrive at \eqref{eqn:sum of orthogonal error terms} since we anticipate it to be small in the high SNR regime.
		This is because the dominant singular vectors $\bb{\vec{u}, \vec{v}}$ of $\widehat{\mat{H}}$ should be a good approximation of the true singular vectors $\bb{\vec{u}_{0}, \vec{v}_{0}}$ at high SNR so that projecting $\vec{u}_{0}$ (respectively $\vec{v}_{0}$) on to the orthogonal complement space $\vec{u}^{\perp}$ (respectively $\vec{v}^{\perp}$) should incur only a small amount of energy.
		We can evaluate each of the terms on the \rhs~of \eqref{eqn:sum of orthogonal error terms} as below.
		\makeatletter
			\if@twocolumn
				\begin{subequations}
					\label{eqn:individual projected error norms}
					\begin{gather}
						\begin{split}
							\fronorm{\mat{P}_{\vec{u}} \mat{Z} \mat{P}_{\vec{v}}}
							& = \fronorm{\bb{\sigma_{0} \ip{\vec{u}_{0}}{\vec{u}} \ip{\vec{v}_{0}}{\vec{v}} - \sigma}
							\vec{u} \tpose{\vec{v}} + \mat{0}}	\\
							& = \abs{\sigma_{0} \ip{\vec{u}_{0}}{\vec{u}} \ip{\vec{v}_{0}}{\vec{v}} - \sigma}.
						\end{split}	\\
						\begin{split}
							\MoveEqLeft \fronorm{\mat{P}_{\vec{u}} \mat{Z} \mat{P}_{\vec{v}^{\perp}}}
							= \fronorm{\sigma_{0} \ip{\vec{u}_{0}}{\vec{u}} \vec{u} \tpose{\vec{v}_{0}} \mat{P}_{\vec{v}^{\perp}} + \mat{0}}	\\
							& = \abs{\ip{\vec{u}_{0}}{\vec{u}}} \sigma_{0} \fronorm{\tpose{\vec{v}_{0}} \mat{P}_{\vec{v}^{\perp}}}
							= \abs{\ip{\vec{u}_{0}}{\vec{u}}} \sigma_{0} \fronorm{\mat{P}_{\vec{v}^{\perp}}\bb{\vec{v}_{0}}}	\\
							& = \abs{\ip{\vec{u}_{0}}{\vec{u}}} \sigma_{0} \sqrt{1 - \abs{\ip{\vec{v}_{0}}{\vec{v}}}^{2}}.
						\end{split}	\raisetag{\baselineskip}	\\
						\begin{split}
							\MoveEqLeft \fronorm{\mat{P}_{\vec{u}^{\perp}} \mat{Z} \mat{P}_{\vec{v}}}
							= \fronorm{\sigma_{0} \mat{P}_{\vec{u}^{\perp}}\bb{\vec{u}_{0}} \ip{\vec{v}_{0}}{\vec{v}} \tpose{\vec{v}} + \mat{0}}	\\
							& = \abs{\ip{\vec{v}_{0}}{\vec{v}}} \sigma_{0} \fronorm{\mat{P}_{\vec{u}^{\perp}}\bb{\vec{u}_{0}}}
							= \abs{\ip{\vec{v}_{0}}{\vec{v}}} \sigma_{0} \sqrt{1 - \abs{\ip{\vec{u}_{0}}{\vec{u}}}^{2}}.
						\end{split}	\raisetag{2.3\baselineskip}
					\end{gather}
				\end{subequations}
			\else
				\begin{subequations}
					\label{eqn:individual projected error norms}
					\begin{gather}
						\fronorm{\mat{P}_{\vec{u}} \mat{Z} \mat{P}_{\vec{v}}}
						= \fronorm{\bb{\sigma_{0} \ip{\vec{u}_{0}}{\vec{u}} \ip{\vec{v}_{0}}{\vec{v}} - \sigma}
						\vec{u} \tpose{\vec{v}} + \mat{0}}
						= \abs{\sigma_{0} \ip{\vec{u}_{0}}{\vec{u}} \ip{\vec{v}_{0}}{\vec{v}} - \sigma}.	\\
						\begin{split}
							\fronorm{\mat{P}_{\vec{u}} \mat{Z} \mat{P}_{\vec{v}^{\perp}}}
							& =	\fronorm{\sigma_{0} \ip{\vec{u}_{0}}{\vec{u}} \vec{u} \tpose{\vec{v}_{0}} \mat{P}_{\vec{v}^{\perp}} + \mat{0}}
							= \abs{\ip{\vec{u}_{0}}{\vec{u}}} \sigma_{0} \fronorm{\tpose{\vec{v}_{0}} \mat{P}_{\vec{v}^{\perp}}}	\\
							& = \abs{\ip{\vec{u}_{0}}{\vec{u}}} \sigma_{0} \fronorm{\mat{P}_{\vec{v}^{\perp}}\bb{\vec{v}_{0}}}
							= \abs{\ip{\vec{u}_{0}}{\vec{u}}} \sigma_{0} \sqrt{1 - \abs{\ip{\vec{v}_{0}}{\vec{v}}}^{2}}.
						\end{split}	\\
						\begin{split}
							\fronorm{\mat{P}_{\vec{u}^{\perp}} \mat{Z} \mat{P}_{\vec{v}}}
							= \fronorm{\sigma_{0} \mat{P}_{\vec{u}^{\perp}}\bb{\vec{u}_{0}} \ip{\vec{v}_{0}}{\vec{v}} \tpose{\vec{v}} + \mat{0}}
							& = \abs{\ip{\vec{v}_{0}}{\vec{v}}} \sigma_{0} \fronorm{\mat{P}_{\vec{u}^{\perp}}\bb{\vec{u}_{0}}}
							= \abs{\ip{\vec{v}_{0}}{\vec{v}}} \sigma_{0} \sqrt{1 - \abs{\ip{\vec{u}_{0}}{\vec{u}}}^{2}}.
						\end{split}
					\end{gather}
				\end{subequations}
			\fi
		\makeatother
		For brevity of notation we let $\alpha = \ip{\vec{u}_{0}}{\vec{u}} \in \BB{-1,1}$ and $\beta = \ip{\vec{v}_{0}}{\vec{v}} \in \BB{-1,1}$.
		From the assumptions of the theorem, $\fronorm{\mat{Z}} \leq \zeta$ and combining this with \eqref{eqn:projected error representation} and \eqref{eqn:individual projected error norms} implies
		\makeatletter
			\if@twocolumn
				\begin{subequations}
					\begin{align}
						\zeta^{2} & \geq \bb{\sigma_{0}\alpha\beta - \sigma}^{2} + \sigma_{0}^{2} \alpha^{2} \bb{1 - \beta^{2}} + \sigma_{0}^{2} \beta^{2} \bb{1 - \alpha^{2}} \\
						& = \sigma^{2} - 2\sigma\sigma_{0}\alpha\beta - \sigma_{0}^{2} \alpha^{2} \beta^{2} + \sigma_{0}^{2} \bb{\alpha^{2} + \beta^{2}} \label{eqn:interim 1} \\
						& \geq \sigma^{2} - 2\sigma\sigma_{0}\alpha\beta - \sigma_{0}^{2} \alpha^{2} \beta^{2} + 2\sigma_{0}^{2}\abs{\alpha\beta}, \label{eqn:interim 2}
					\end{align}
				\end{subequations}
			\else
				\begin{subequations}
					\begin{align}
						\zeta^{2} \geq \bb{\sigma_{0}\alpha\beta - \sigma}^{2} + \sigma_{0}^{2} \alpha^{2} \bb{1 - \beta^{2}} + \sigma_{0}^{2} \beta^{2} \bb{1 - \alpha^{2}}
						& =	\sigma^{2} - 2\sigma\sigma_{0}\alpha\beta - \sigma_{0}^{2} \alpha^{2} \beta^{2} + \sigma_{0}^{2} \bb{\alpha^{2} + \beta^{2}}	\label{eqn:interim 1} \\
						& \geq \sigma^{2} - 2\sigma\sigma_{0}\alpha\beta - \sigma_{0}^{2} \alpha^{2} \beta^{2} + 2\sigma_{0}^{2}\abs{\alpha\beta},	\label{eqn:interim 2}
					\end{align}
				\end{subequations}
			\fi
		\makeatother
		where \eqref{eqn:interim 2} was obtained from \eqref{eqn:interim 1} using the relation $\alpha^{2} + \beta^{2} \geq 2\abs{\alpha\beta}$.
		Because the signs of $\vec{u}$ and $\vec{v}$ can be switched globally without changing the estimate $\sigma \vec{u} \tpose{\vec{v}}$, \WLOG~we assume $\alpha = \ip{\vec{u}}{\vec{u}_{0}} \gneq 0$.
		We have
		\begin{equation}
			\sigma_{0}^{2} \alpha^{2} \beta^{2} + 2\sigma_{0}\alpha\bb{\sigma\beta - \sigma_{0}\abs{\beta}} + \zeta^{2}-\sigma^{2} \geq 0,
			\label{eqn:alpha-beta-ineq}
		\end{equation}
		which is a quadratic inequality in $\alpha\beta$.
		If \eqref{eqn:alpha-beta-ineq} were satisfied with equality, then the corresponding quadratic equation \wrt~the variable $\sigma_{0} \alpha$ would have roots in the set $\cc{\beta^{-2}\BB{\bb{\sigma_{0}\abs{\beta} - \sigma\beta} \pm \sqrt{\bb{\sigma_{0}\abs{\beta} - \sigma\beta}^{2} - \beta^{2}\bb{\zeta^{2} - \sigma^{2}}}}}$, by the quadratic formula.
		Since $\zeta \leq \sigma$ from the premise of the theorem, the two roots are of opposite signs.
		Further, $\sigma_{0}\alpha \gneq 0$ by assumption and therefore, to satisfy \eqref{eqn:alpha-beta-ineq}, $\sigma_{0}\alpha$ must be greater than or equal to the larger root.
		Thus, we have
		\makeatletter
			\if@twocolumn
				\begin{equation}
					\begin{split}
						\sigma_{0}\alpha & \geq \beta^{-2}\BB{\bb{\sigma_{0}\abs{\beta} - \sigma\beta} + \sqrt{\bb{\sigma_{0}\abs{\beta} - \sigma\beta}^{2} - \beta^{2}\bb{\zeta^{2} - \sigma^{2}}}}	\\
						& >	\beta^{-2}\bb{\sigma_{0}\abs{\beta} - \sigma\beta}.	\raisetag{0.7\baselineskip}
						\label{eqn:alpha LB}
					\end{split}
				\end{equation}
			\else
				\begin{equation}
					\sigma_{0}\alpha \geq \beta^{-2}\BB{\bb{\sigma_{0}\abs{\beta} - \sigma\beta} + \sqrt{\bb{\sigma_{0}\abs{\beta} - \sigma\beta}^{2} - \beta^{2}\bb{\zeta^{2} - \sigma^{2}}}}
					>	\beta^{-2}\bb{\sigma_{0}\abs{\beta} - \sigma\beta}.
					\label{eqn:alpha LB}
				\end{equation}
			\fi
		\makeatother
		Since $\alpha \leq 1$ and $\abs{\beta} \leq 1$, assuming $\beta < 0$ leads to \eqref{eqn:alpha LB} implying that
		\makeatletter
			\if@twocolumn
				\begin{equation}
					\begin{split}
						\sigma_{0} & \geq \sigma_{0}\alpha > \beta^{-2}\bb{\sigma_{0}\abs{\beta} - \sigma\beta}
						=	\beta^{-2}\bb{\sigma_{0}\abs{\beta} + \sigma\abs{\beta}}	\\
						& =	\bb{\sigma_{0} + \sigma}/\abs{\beta} \geq \sigma_{0} + \sigma,
					\end{split}
				\end{equation}
			\else
				\begin{equation}
					\sigma_{0} \geq \sigma_{0}\alpha > \beta^{-2}\bb{\sigma_{0}\abs{\beta} - \sigma\beta}
					=	\beta^{-2}\bb{\sigma_{0}\abs{\beta} + \sigma\abs{\beta}}
					=	\bb{\sigma_{0} + \sigma}/\abs{\beta} \geq \sigma_{0} + \sigma,
				\end{equation}
			\fi
		\makeatother
		which is a clear contradiction.
		Hence, $\beta > 0$ and the left inequality in \eqref{eqn:alpha LB} yields the joint bound
		\makeatletter
			\if@twocolumn
				\begin{equation}
					\begin{split}
						\alpha\beta & \geq \frac{\inv{\beta}}{\sigma_{0}}\BB{\bb{\sigma_{0}\beta - \sigma\beta} + \sqrt{\bb{\sigma_{0}\beta - \sigma\beta}^{2} - \beta^{2}\bb{\zeta^{2} - \sigma^{2}}}}	\\
						& \geq	\bb{1-\frac{\sigma}{\sigma_{0}}} + \sqrt{\bb{1-\frac{\sigma}{\sigma_{0}}}^{2} + \bb{\frac{\sigma}{\sigma_{0}}}^{2} - \bb{\frac{\zeta}{\sigma_{0}}}^{2}},
					\end{split}
				\end{equation}
			\else
				\begin{equation}
					\alpha\beta \geq \frac{\inv{\beta}}{\sigma_{0}}\BB{\bb{\sigma_{0}\beta - \sigma\beta} + \sqrt{\bb{\sigma_{0}\beta - \sigma\beta}^{2} - \beta^{2}\bb{\zeta^{2} - \sigma^{2}}}}
					\geq	\bb{1-\frac{\sigma}{\sigma_{0}}} + \sqrt{\bb{1-\frac{\sigma}{\sigma_{0}}}^{2} + \bb{\frac{\sigma}{\sigma_{0}}}^{2} - \bb{\frac{\zeta}{\sigma_{0}}}^{2}},
				\end{equation}
			\fi
		\makeatother
		thus proving the theorem.

	\section{Proof of \theoremname~\ref{thm:convergence}}
		\label{sec:convergence proof}
		\begin{figure}[t]
			\centering
			\makeatletter
				\if@twocolumn
					\includegraphics[width=0.4\figwidth]{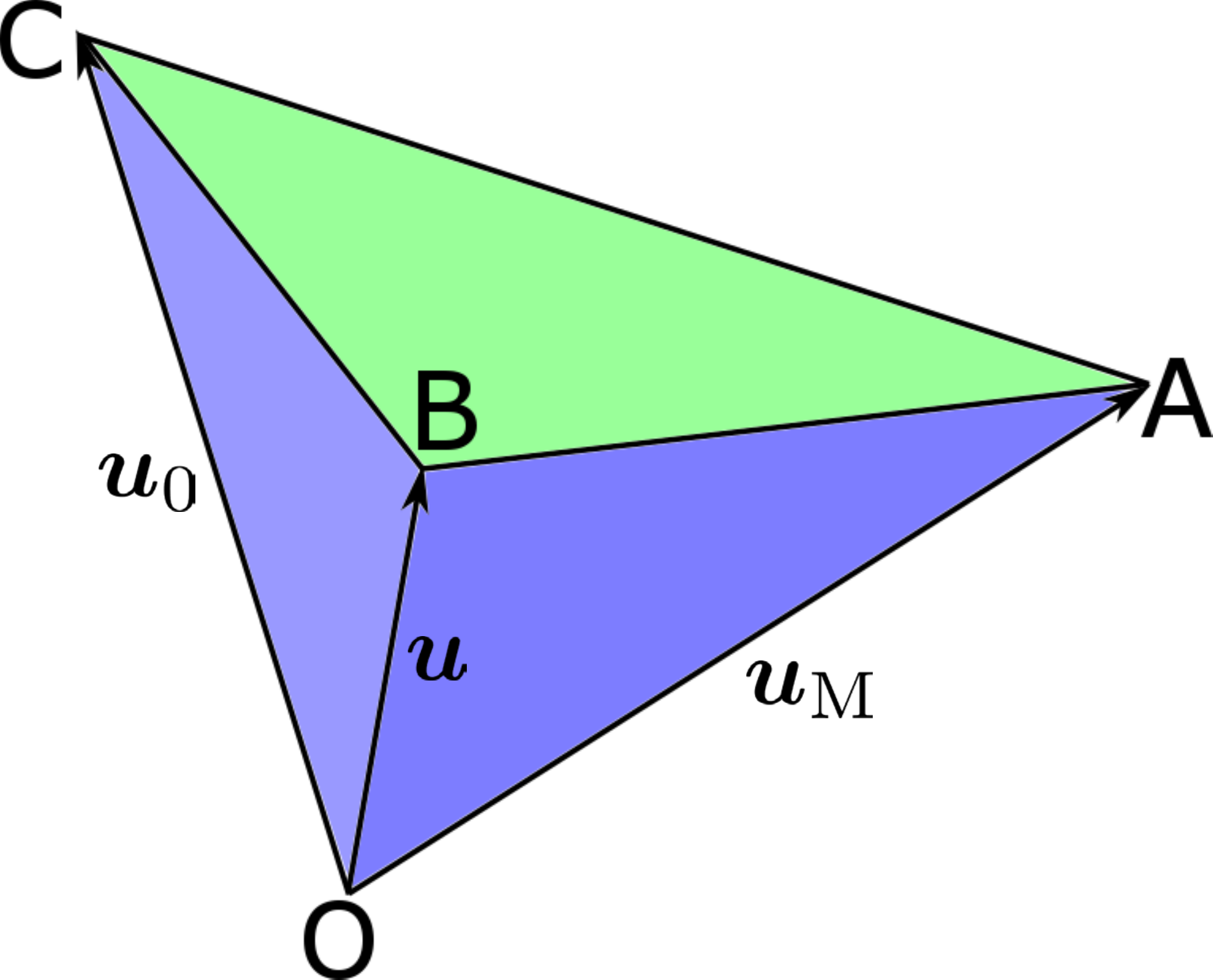}
				\else
					\includegraphics[width=0.6\figwidth]{Inner-Product-Tetrahedron}
				\fi
			\makeatother
			\caption{A three dimensional embedding of vectors $\vec{u}_{\text{M}}$, $\vec{u}$ and $\vec{u}_{0}$ for aiding visualization in proof of \theoremname~\ref{thm:convergence}.}
			\label{fig:OABC-tetrahedron}
		\end{figure}
		We shall use $\zeta$, $\sigma$, $\vec{u}$ and $\vec{v}$ as defined in the steps~\sref{itm:SVD} and~\sref{itm:localization LB} of \algorithmname~\ref{alg:PAMC-UMR}, and $\eta\bb{\sigma,\sigma_{0},\zeta}$ as defined in \theoremname~\ref{thm:inner product bound}.
		The proof proceeds by separately bounding $\bb{\lcR-\lcL}$ and $\bb{\lrR-\lrL}$.
		We shall only derive the bound on $\bb{\lrR-\lrL}$ since both bounds follow from the same sequence of steps.
		
		\theoremname~\ref{thm:inner product bound} and \lemmaname~\ref{lem:bound reduction} together imply that $\ip{\vec{u}_{0}}{\vec{u}} \ip{\vec{v}_{0}}{\vec{v}} \geq \eta\bb{\sigma,\sigma_{0},\zeta} \geq \sqrt{1-\zeta^{2}/\sigma^{2}}$.
		Since $\abs{\ip{\vec{v}_{0}}{\vec{v}}} \leq \twonorm{\vec{v}_{0}} \twonorm{\vec{v}} \leq 1$ (respectively $\abs{\ip{\vec{u}_{0}}{\vec{u}}} \leq \twonorm{\vec{u}_{0}} \twonorm{\vec{u}} \leq 1$) by the Cauchy-Schwartz inequality, we have $\abs{\ip{\vec{u}_{0}}{\vec{u}}} \geq \sqrt{1-\zeta^{2}/\sigma^{2}}$ (respectively $\abs{\ip{\vec{v}_{0}}{\vec{v}}} \geq \sqrt{1-\zeta^{2}/\sigma^{2}}$).
		We assume \WLOG~that $\ip{\vec{u}_{0}}{\vec{u}} > 0$ implying that $\ip{\vec{u}_{0}}{\vec{u}} \geq \sqrt{1-\zeta^{2}/\sigma^{2}}$ and that \problemname~\pref{prob:localization}[n,\zeta,\sigma,\vec{u}] is feasible at step~\sref{itm:localization LB} of \algorithmname~\ref{alg:PAMC-UMR}
		(otherwise $\ip{\vec{u}_{0}}{\vec{u}} < 0$ implying that $\ip{\vec{u}_{0}}{-\vec{u}} \geq \sqrt{1-\zeta^{2}/\sigma^{2}}$ and that \problemname~\pref{prob:localization}[n,\zeta,\sigma,-\vec{u}] is feasible).
		Let $\bb{\zrL,\lrL}$ denote a solution to \problemname~\pref{prob:localization}[n,\zeta,\sigma,\vec{u}].
		It is clear from the constraints in \problemname~\pref{prob:localization} that $\zrL$ is a unimodal vector and satisfies $\ip{\zrL}{\vec{u}} \geq \twonorm{\zrL}\sqrt{1-\zeta^{2}/\sigma^{2}}$.
		For brevity of notation, we set $\vec{u}_{\text{M}} = \zrL/\twonorm{\zrL}$ and get $\ip{\vec{u}_{\text{M}}}{\vec{u}} \geq \sqrt{1-\zeta^{2}/\sigma^{2}}$.
		The proof proceeds by bounding $\ip{\vec{u}_{\text{M}}}{\vec{u}_{0}}$ using the bounds on $\ip{\vec{u}_{\text{M}}}{\vec{u}}$ and $\ip{\vec{u}_{0}}{\vec{u}}$.
		
		Let points A, B and C respectively represent the vectors $\vec{u}_{\text{M}}$, $\vec{u}$ and $\vec{u}_{0}$ in $n$-dimensional space with O as origin (see \figurename~\ref{fig:OABC-tetrahedron} as an aid to visualization).
		Therefore, OA, OB and OC are all unit length line segments and the inner products $\ip{\vec{u}_{\text{M}}}{\vec{u}}$, $\ip{\vec{u}_{0}}{\vec{u}}$ and $\ip{\vec{u}_{\text{M}}}{\vec{u}_{0}}$ are respectively equal to $\cos\angle\text{AOB}$, $\cos\angle\text{BOC}$ and $\cos\angle\text{COA}$.
		Using the cosine rule from elementary trigonometry on triangle COA, we have
		\makeatletter
			\if@twocolumn
				\begin{multline}
					\text{AC} =	\sqrt{\text{OA}^{2} + \text{OC}^{2} - 2\cdot\text{OA}\cdot\text{OC}\cdot\cos\angle\text{COA}}	\\
					=	\sqrt{2 - 2\cdot\ip{\vec{u}_{\text{M}}}{\vec{u}_{0}}}.
				\end{multline}
			\else
				\begin{equation}
					\text{AC} =	\sqrt{\text{OA}^{2} + \text{OC}^{2} - 2\cdot\text{OA}\cdot\text{OC}\cdot\cos\angle\text{COA}}
					=	\sqrt{2 - 2\cdot\ip{\vec{u}_{\text{M}}}{\vec{u}_{0}}}.
				\end{equation}
			\fi
		\makeatother
		Similarly, using the cosine rule on triangles AOB and BOC respectively gives $\text{AB} = \sqrt{2 - 2\cdot\ip{\vec{u}_{\text{M}}}{\vec{u}}}$ and $\text{BC} = \sqrt{2 - 2\cdot\ip{\vec{u}_{0}}{\vec{u}}}$.
		By the triangle inequality, we have $\text{AC} \leq \text{AB} + \text{BC}$ leading to
		\begin{subequations}
			\makeatletter
				\if@twocolumn
					\begin{align}
						\MoveEqLeft[1]	\sqrt{2 - 2\cdot\ip{\vec{u}_{\text{M}}}{\vec{u}_{0}}}	\leq
						\sqrt{2 - 2\cdot\ip{\vec{u}_{\text{M}}}{\vec{u}}} + \sqrt{2 - 2\cdot\ip{\vec{u}_{0}}{\vec{u}}}	\notag \\
						& \implies	\sqrt{1 - \ip{\vec{u}_{\text{M}}}{\vec{u}_{0}}}	\leq
						\sqrt{1 - \ip{\vec{u}_{\text{M}}}{\vec{u}}} + \sqrt{1 - \ip{\vec{u}_{0}}{\vec{u}}}	\notag \\
						& \implies	\sqrt{1 - \ip{\vec{u}_{\text{M}}}{\vec{u}_{0}}}
						\leq	\sqrt{1 - \sqrt{1-\zeta^{2}/\sigma^{2}}}	\notag \\
						& \phantom{{} \implies	\sqrt{1 - \ip{\vec{u}_{\text{M}}}{\vec{u}_{0}}}}
						\qquad {} + \sqrt{1 - \sqrt{1-\zeta^{2}/\sigma^{2}}}	\label{eqn:ip bound used} \\
						& \implies	1 - \ip{\vec{u}_{\text{M}}}{\vec{u}_{0}}
							\leq	4\cdot\bb{1 - \sqrt{1-\zeta^{2}/\sigma^{2}}}	\notag \\
						& \implies	\ip{\vec{u}_{\text{M}}}{\vec{u}_{0}}
						\geq	1 - 4\cdot\bb{1 - \sqrt{1-\zeta^{2}/\sigma^{2}}}	\notag \\
						& \phantom{{} \implies	\ip{\vec{u}_{\text{M}}}{\vec{u}_{0}}}
						=	4\sqrt{1-\zeta^{2}/\sigma^{2}} - 3	=	\zeta'	\label{eqn:ip-bound}
					\end{align}
				\else
					\begin{align}
						\MoveEqLeft	\sqrt{2 - 2\cdot\ip{\vec{u}_{\text{M}}}{\vec{u}_{0}}}	\leq
						\sqrt{2 - 2\cdot\ip{\vec{u}_{\text{M}}}{\vec{u}}} + \sqrt{2 - 2\cdot\ip{\vec{u}_{0}}{\vec{u}}}	\notag \\
						& \implies	\sqrt{1 - \ip{\vec{u}_{\text{M}}}{\vec{u}_{0}}}	\leq
						\sqrt{1 - \ip{\vec{u}_{\text{M}}}{\vec{u}}} + \sqrt{1 - \ip{\vec{u}_{0}}{\vec{u}}}	\notag \\
						& \implies	\sqrt{1 - \ip{\vec{u}_{\text{M}}}{\vec{u}_{0}}}	\leq
						\sqrt{1 - \sqrt{1-\zeta^{2}/\sigma^{2}}} + \sqrt{1 - \sqrt{1-\zeta^{2}/\sigma^{2}}}	\label{eqn:ip bound used} \\
						& \implies	1 - \ip{\vec{u}_{\text{M}}}{\vec{u}_{0}}
								\leq	4\cdot\bb{1 - \sqrt{1-\zeta^{2}/\sigma^{2}}}	\notag \\
						& \implies	\ip{\vec{u}_{\text{M}}}{\vec{u}_{0}}
								\geq	1 - 4\cdot\bb{1 - \sqrt{1-\zeta^{2}/\sigma^{2}}}
								=	4\sqrt{1-\zeta^{2}/\sigma^{2}} - 3
								\triangleq	\zeta'	\label{eqn:ip-bound}
					\end{align}
				\fi
			\makeatother
		\end{subequations}
		where \eqref{eqn:ip bound used} uses the lower bounds on the inner products $\ip{\vec{u}_{\text{M}}}{\vec{u}}$ and $\ip{\vec{u}_{0}}{\vec{u}}$.
		For the \rhs~of \eqref{eqn:ip-bound} to be a useful bound, we need it to be positive, or equivalently, $\sqrt{1-\zeta^{2}/\sigma^{2}} > 3/4 \iff \zeta^{2}/\sigma^{2} < 7/16$, which is assumed in the premise of this theorem.

		Let us refer to \problemsname~\pref{prob:peak-closeness} and~\pref{prob:peak-closeness-opt} as \problemsname~\pref{prob:peak-closeness}[\lstar,\rho,\vec{v}] and~\pref{prob:peak-closeness-opt}[\lstar,\vec{v}] to make the dependence on the parameters $\lstar$, $\rho$ and $\vec{v}$ explicit.
		From the premise of the theorem, $\vec{u}_{0} \geq \vec{0}$ is a unimodal vector with its peak at the index $l^{0}_{\text{r}}$.
		Recall that $\vec{u}_{\text{M}}$ is also a unimodal vector with its peak at index $\lrL$ and suppose \WLOG~that $\lrL \leq l^{0}_{\text{r}}$.
		It is clear that $\vec{u}_{\text{M}}$ is feasible for \problemname~\pref{prob:peak-closeness-opt}[\lrL,\vec{u}_{0}] and \eqref{eqn:ip-bound} implies that $-\ip{\vec{u}_{\text{M}}}{\vec{u}_{0}} \leq -\zeta'$.
		By \lemmaname~\ref{lem:non-neg soln}, we can assume that $\vec{u}_{\text{M}} \geq \vec{0}$ and that $\vec{u}_{\text{M}}$ is feasible for \problemname~\pref{prob:peak-closeness}[\lrL,\zeta',\vec{u}_{0}].
		Next, using \lemmaname~\ref{lem:peak-closeness}, we get the bound in \eqref{eqn:delta-l-inequality} provided that the restrictions on $\kL$, $\kR$ and $\delta$ are satisfied.
		In the notation for \problemname~\pref{prob:peak-closeness}[\lrL,\zeta',\vec{u}_{0}], \eqref{eqn:delta-l-inequality} says that
		\makeatletter
			\if@twocolumn
				\begin{multline}
					\bb{\zeta'}^{2} \leq	\ip{\vec{1}}{\vec{u}_{0}}^{2} + \bb{\delta^{2} - 2\delta\ip{\vec{1}}{\vec{u}_{0}}} \cdot \bb{\kL+\kR+1}	\\
					{} + \delta^{2}\bb{\kL+\kR+1}^{2} \triangleq	h\bb{\delta}
					\label{eqn:delta-l-bound-mod}
				\end{multline}
			\else
				\begin{equation}
					\bb{\zeta'}^{2} \leq	\ip{\vec{1}}{\vec{u}_{0}}^{2} + \bb{\delta^{2} - 2\delta\ip{\vec{1}}{\vec{u}_{0}}} \cdot \bb{\kL+\kR+1} + \delta^{2}\bb{\kL+\kR+1}^{2} \triangleq	h\bb{\delta}
					\label{eqn:delta-l-bound-mod}
				\end{equation}
			\fi
		\makeatother
		is true for any integers $0 \leq \kL \leq l^{0}_{\text{r}} - \lrL$, $0 \leq \kR \leq n - l^{0}_{\text{r}}$, and any $\delta \in \setR_{+}$ satisfying
		\makeatletter
			\if@twocolumn
				\begin{equation}
					\delta \leq \frac{\ip{\vec{1}}{\vec{u}_{0}\bb{j+1:n}}}{\bb{l^{0}_{\text{r}}+\kR-j}},
					\forall \, j \in \cc{l^{0}_{\text{r}}-\kL-1,\dotsc, l^{0}_{\text{r}}+\kR-1}.
					\label{eqn:delta-bound-mod}
				\end{equation}
			\else
				\begin{equation}
					\delta \leq \frac{\ip{\vec{1}}{\vec{u}_{0}\bb{j+1:n}}}{\bb{l^{0}_{\text{r}}+\kR-j}},
					\quad	\forall \, j \in \cc{l^{0}_{\text{r}}-\kL-1,\dotsc, l^{0}_{\text{r}}+\kR-1}.
					\label{eqn:delta-bound-mod}
				\end{equation}
			\fi
		\makeatother
		We will use $\kL = l^{0}_{\text{r}} - \lrL$ and $\kR = 0$.
		Invoking \lemmaname~\ref{lem:cum-avg} on $\vec{u}_{0}$ implies that $\ip{\vec{1}}{\vec{u}_{0}\bb{j+1:n}}/\bb{l^{0}_{\text{r}}-j}$ is monotonically non-decreasing in $j$ over $1 \leq j \leq l^{0}_{\text{r}}-1$.
		Hence, the dominating bound in \eqref{eqn:delta-bound-mod} is obtained for $j = l^{0}_{\text{r}}-\kL-1 = \lrL-1$.
		This gives the largest permissible value of $\delta$ as $\delta_{\ast} = \ip{\vec{1}}{\vec{u}_{0}\bb{\lrL:n}}/\bb{l^{0}_{\text{r}}-\lrL+1}$, leading to
		\makeatletter
			\if@twocolumn
				\begin{equation}
					\begin{split}
						h\bb{\delta_{\ast}}
						& =	\ip{\vec{1}}{\vec{u}_{0}}^{2} + \bb{\delta_{\ast}^{2} - 2\delta_{\ast}\ip{\vec{1}}{\vec{u}_{0}}}\bb{l^{0}_{\text{r}}-\lrL+1}	\\
						& \phantom{{} = {}} + \delta_{\ast}^{2}\bb{l^{0}_{\text{r}}-\lrL+1}^{2}	\\
						& =	\ip{\vec{1}}{\vec{u}_{0}}^{2} - 2\ip{\vec{1}}{\vec{u}_{0}}\delta_{\ast}\bb{l^{0}_{\text{r}}-\lrL+1}	\\
						& \phantom{{} = {}} + \delta_{\ast}^{2}\bb{l^{0}_{\text{r}}-\lrL+2}\bb{l^{0}_{\text{r}}-\lrL+1}	\\
						& =	\ip{\vec{1}}{\vec{u}_{0}}^{2} - 2\ip{\vec{1}}{\vec{u}_{0}}\ip{\vec{1}}{\vec{u}_{0}\bb{\lrL:n}}	\\
						& \phantom{{} = {}} + \frac{l^{0}_{\text{r}}-\lrL+2}{l^{0}_{\text{r}}-\lrL+1} \ip{\vec{1}}{\vec{u}_{0}\bb{\lrL:n}}^{2}	\\
						& =	\bb{\ip{\vec{1}}{\vec{u}_{0}} - \ip{\vec{1}}{\vec{u}_{0}\bb{\lrL:n}}}^{2} + \frac{\ip{\vec{1}}{\vec{u}_{0}\bb{\lrL:n}}^{2}}{l^{0}_{\text{r}}-\lrL+1}	\\
						& <	\ip{\vec{1}}{\vec{u}_{0}\bb{1:\lrL-1}}^{2} + \frac{\ip{\vec{1}}{\vec{u}_{0}}^{2}}{l^{0}_{\text{r}}-\lrL+1}	\\
						& \leq	\frac{1}{2}\bb{\zeta'}^{2} + \frac{\rho_{\vec{u}}^{2} n}{l^{0}_{\text{r}}-\lrL+1}
					\end{split}	\raisetag{\baselineskip}
					\label{eqn:h-delta-star}
				\end{equation}
			\else
				\begin{equation}
					\begin{split}
						h\bb{\delta_{\ast}} & =	\ip{\vec{1}}{\vec{u}_{0}}^{2} + \bb{\delta_{\ast}^{2} - 2\delta_{\ast}\ip{\vec{1}}{\vec{u}_{0}}}\bb{l^{0}_{\text{r}}-\lrL+1} + \delta_{\ast}^{2}\bb{l^{0}_{\text{r}}-\lrL+1}^{2}	\\
						& =	\ip{\vec{1}}{\vec{u}_{0}}^{2} - 2\ip{\vec{1}}{\vec{u}_{0}}\delta_{\ast}\bb{l^{0}_{\text{r}}-\lrL+1} + \delta_{\ast}^{2}\bb{l^{0}_{\text{r}}-\lrL+2}\bb{l^{0}_{\text{r}}-\lrL+1}	\\
						& =	\ip{\vec{1}}{\vec{u}_{0}}^{2} - 2\ip{\vec{1}}{\vec{u}_{0}}\ip{\vec{1}}{\vec{u}_{0}\bb{\lrL:n}} + \frac{l^{0}_{\text{r}}-\lrL+2}{l^{0}_{\text{r}}-\lrL+1} \ip{\vec{1}}{\vec{u}_{0}\bb{\lrL:n}}^{2}	\\
						& =	\bb{\ip{\vec{1}}{\vec{u}_{0}} - \ip{\vec{1}}{\vec{u}_{0}\bb{\lrL:n}}}^{2} + \frac{\ip{\vec{1}}{\vec{u}_{0}\bb{\lrL:n}}^{2}}{l^{0}_{\text{r}}-\lrL+1}	\\
						& <	\ip{\vec{1}}{\vec{u}_{0}\bb{1:\lrL-1}}^{2} + \frac{\ip{\vec{1}}{\vec{u}_{0}}^{2}}{l^{0}_{\text{r}}-\lrL+1}	\\
						& \leq	\frac{1}{2}\bb{\zeta'}^{2} + \frac{\rho_{\vec{u}}^{2} n}{l^{0}_{\text{r}}-\lrL+1}
					\end{split}
					\label{eqn:h-delta-star}
				\end{equation}
			\fi
		\makeatother
		where the last inequality follows from the premise $\ip{\vec{1}}{\vec{u}_{0}} = \onenorm{\vec{u}_{0}} \leq \rho_{\vec{u}}\sqrt{n}$ and using $\lrL-1 < \lrL \leq \lrBL$ with \eqref{eqn:lrBL-defn}.
		Using \eqref{eqn:h-delta-star} in \eqref{eqn:delta-l-bound-mod} gives
		\begin{equation}
			\begin{split}
				\MoveEqLeft	\bb{\zeta'}^{2} \leq h\bb{\delta_{\ast}}
				<	\frac{1}{2}\bb{\zeta'}^{2} + \frac{\rho_{\vec{u}}^{2} n}{l^{0}_{\text{r}}-\lrL+1}	\\
				& \implies	l^{0}_{\text{r}}-\lrL < \frac{2 \rho_{\vec{u}}^{2} n}{\bb{\zeta'}^{2}} - 1
			\end{split}
			\label{eqn:lrL-guarantee}
		\end{equation}

		Note that step~\sref{itm:localization UB} of \algorithmname~\ref{alg:PAMC-UMR} essentially uses \problemname~\pref{prob:localization}[n,\zeta,\sigma,\vec{u}] to find $\lrR$; just like step~\sref{itm:localization LB} except that we are now looking for the maximum index $l$ permitting \problemname~\pref{prob:localization}[n,\zeta,\sigma,\vec{u}] to be feasible.
		Hence, all arguments in the preceding three paragraphs are still valid with $\vec{u}_{\text{M}}$ now representing a unimodal vector with its peak at index $\lrR$ and $\lrR \geq l^{0}_{\text{r}}$.
		Analogous to \eqref{eqn:lrL-guarantee}, this leads to the bound
		\begin{equation}
			\lrR - l^{0}_{\text{r}} < \frac{2 \rho_{\vec{u}}^{2} n}{\bb{\zeta'}^{2}} - 1
		\end{equation}
		which, when added to \eqref{eqn:lrL-guarantee} gives
		\begin{equation}
			\lrR-\lrL < \frac{4 \rho_{\vec{u}}^{2} n}{\bb{\zeta'}^{2}} - 2
			< \frac{4 \rho_{\vec{u}}^{2} n}{\bb{\zeta'}^{2}}.
			\label{eqn:u-loc-bound}
		\end{equation}
		All arguments in this proof \wrt~$\vec{u}_{0}$ can be duplicated \wrt~$\vec{v}_{0}$, starting at the second paragraph from $\abs{\ip{\vec{v}_{0}}{\vec{v}}} \geq \sqrt{1-\zeta^{2}/\sigma^{2}}$.
		Hence, analogous to \eqref{eqn:u-loc-bound}, we can derive an upper bound on $\lcR-\lcL$ which when multiplied with \eqref{eqn:u-loc-bound} gives \eqref{eqn:loc-bound} and completes the proof.

	\section{Proof of \lemmaname~\ref{lem:peak-closeness}}
		\label{sec:peak-closeness proof}
		We shall reason about the feasibility of \problemname~\pref{prob:peak-closeness} by studying the closely related optimization problem~\pref{prob:peak-closeness-opt}.
		Using \lemmaname~\ref{lem:non-neg soln}, solutions to \problemname~\pref{prob:peak-closeness-opt} can be translated to and from \problemname~\pref{prob:peak-closeness}.
		Thus, it suffices to show that under the assumptions of this lemma, it is necessary for the inequality in \eqref{eqn:delta-l-inequality} to hold if the optimal value of \problemname~\pref{prob:peak-closeness-opt} is not to exceed $-\rho$.
		
		The \emph{unimodality constraints} (first two constraints) in \problemname~\pref{prob:peak-closeness-opt} can be written more compactly as a linear inequality constraint $\mat{A}\vec{z} \leq \vec{0}$ where $\mat{A} \in \setRn{\bb{n-1}}{n}$ is a bidiagonal matrix with non-zero elements
		\begin{equation}
			\mat{A}\bb{j,j:j+1} =	\begin{cases}
										\bb{1, -1},	&	1 \leq j \leq \lstar-1,	\\
										\bb{-1, 1},	&	\lstar \leq j \leq n-1.
									\end{cases}
		\end{equation}
		The Lagrangian for \problemname~\pref{prob:peak-closeness-opt} is
		\makeatletter
			\if@twocolumn
				\begin{equation}
					\begin{split}
						\mathcal{L}\bb{\vec{z};\vec{\lambda},\mu}
						& =	-\ip{\vec{z}}{\vec{v}} + \tpose{\vec{\lambda}}\mat{A}\vec{z} + \mu\bb{\twonorm{\vec{z}}^{2} - 1}	\\
						& =	\tpose{\vec{z}}\bb{\tpose{\mat{A}}\vec{\lambda} - \vec{v}} + \mu\tpose{\vec{z}}\vec{z} - \mu
					\end{split}
				\end{equation}
			\else
				\begin{equation}
					\mathcal{L}\bb{\vec{z};\vec{\lambda},\mu}
					=	-\ip{\vec{z}}{\vec{v}} + \tpose{\vec{\lambda}}\mat{A}\vec{z} + \mu\bb{\twonorm{\vec{z}}^{2} - 1}
					=	\tpose{\vec{z}}\bb{\tpose{\mat{A}}\vec{\lambda} - \vec{v}} + \mu\tpose{\vec{z}}\vec{z} - \mu
				\end{equation}
			\fi
		\makeatother
		and its partial first and second derivatives \wrt~$\vec{z}$ are $\diffp{}{{\vec{z}}}\mathcal{L}\bb{\vec{z};\vec{\lambda},\mu} = \tpose{\mat{A}}\vec{\lambda} - \vec{v} + 2\mu\vec{z}$ and $\diffp[2]{}{{\vec{z}}}\mathcal{L}\bb{\vec{z};\vec{\lambda},\mu} = 2\mu\eye$ respectively.
		Clearly, $\mathcal{L}\bb{\vec{z};\vec{\lambda},\mu}$ is minimized \wrt~$\vec{z}$ at $\vec{z} = \bb{\vec{v} - \tpose{\mat{A}}\vec{\lambda}}/\bb{2\mu}$ for $\mu > 0$, implying that the Lagrangian dual function is
		\makeatletter
			\if@twocolumn
				\begin{equation}
					\begin{split}
						g\bb{\vec{\lambda},\mu}
						& = \inf_{\vec{z}} \mathcal{L}\bb{\vec{z};\vec{\lambda},\mu}	\\
						& =	-\frac{1}{2\mu}\twonorm{\vec{v} - \tpose{\mat{A}}\vec{\lambda}}^{2} + \mu\bb{\frac{1}{4\mu^{2}}\twonorm{\vec{v} - \tpose{\mat{A}}\vec{\lambda}}^{2} - 1}	\\
						& =	-\frac{1}{4\mu}\twonorm{\vec{v} - \tpose{\mat{A}}\vec{\lambda}}^{2} - \mu.
					\end{split}	\raisetag{\baselineskip}
					\label{eqn:dual-fn}
				\end{equation}
			\else
				\begin{equation}
					g\bb{\vec{\lambda},\mu} = \inf_{\vec{z}} \mathcal{L}\bb{\vec{z};\vec{\lambda},\mu}
					=	-\frac{1}{2\mu}\twonorm{\vec{v} - \tpose{\mat{A}}\vec{\lambda}}^{2} + \mu\bb{\frac{1}{4\mu^{2}}\twonorm{\vec{v} - \tpose{\mat{A}}\vec{\lambda}}^{2} - 1}
					=	-\frac{1}{4\mu}\twonorm{\vec{v} - \tpose{\mat{A}}\vec{\lambda}}^{2} - \mu.
					\label{eqn:dual-fn}
				\end{equation}
			\fi
		\makeatother
		We further have $\diffp{}{\mu} g\bb{\vec{\lambda},\mu} = \inv{\bb{4\mu^{2}}}\twonorm{\vec{v} - \tpose{\mat{A}}\vec{\lambda}}^{2} - 1$ so that $g\bb{\vec{\lambda},\mu}$ is maximized \wrt~$\mu$ for $\mu > 0$ at $\mu = \twonorm{\vec{v} - \tpose{\mat{A}}\vec{\lambda}}/2$.
		Let $\vec{z}_{\opt}$ be a solution to \problemname~\pref{prob:peak-closeness-opt}.
		From \eqref{eqn:dual-fn}, we have
		\begin{equation}
			\sup_{\mu} g\bb{\vec{\lambda},\mu} =	-\frac{1}{4\mu}\bb{2\mu}^{2}-\mu =	-2\mu
			=	-\twonorm{\vec{v} - \tpose{\mat{A}}\vec{\lambda}}
		\end{equation}
		and using weak duality theory for \problemname~\pref{prob:peak-closeness-opt} gives
		\makeatletter
			\if@twocolumn
				\begin{equation}
					\begin{split}
						-\ip{\vec{z}_{\opt}}{\vec{v}}
						& \geq	\sup_{\vec{\lambda} \geq \vec{0}} \, \sup_{\mu} \, g\bb{\vec{\lambda},\mu}
						=	\sup_{\vec{\lambda} \geq \vec{0}} \, -\twonorm{\vec{v} -\tpose{\mat{A}}\vec{\lambda}}	\\
						& =	-\inf_{\vec{\lambda} \geq \vec{0}} \, \twonorm{\vec{v} -\tpose{\mat{A}}\vec{\lambda}}.
					\end{split}
					\label{eqn:peak-closeness-dual}
				\end{equation}
			\else
				\begin{equation}
					-\ip{\vec{z}_{\opt}}{\vec{v}} \geq	\sup_{\vec{\lambda} \geq \vec{0}} \, \sup_{\mu} \, g\bb{\vec{\lambda},\mu}
					=	\sup_{\vec{\lambda} \geq \vec{0}} \, -\twonorm{\vec{v} -\tpose{\mat{A}}\vec{\lambda}}
					=	-\inf_{\vec{\lambda} \geq \vec{0}} \, \twonorm{\vec{v} -\tpose{\mat{A}}\vec{\lambda}}.
					\label{eqn:peak-closeness-dual}
				\end{equation}
			\fi
		\makeatother
		\lemmaname~\ref{lem:non-neg soln} says that for feasibility of \problemname~\pref{prob:peak-closeness}, we must have $-\ip{\vec{z}_{\opt}}{\vec{v}} \leq -\rho$ which implies that $\rho \leq \twonorm{\vec{v} -\tpose{\mat{A}}\vec{\lambda}}$ for every $\vec{\lambda} \in \setR^{n}_{+}$.
		Next, we make an appropriate choice of $\vec{\lambda} \geq \vec{0}$ to get the inequality in \eqref{eqn:delta-l-inequality}.

		Let $0 \leq \kL \leq l^{0}-\lstar$ and $0 \leq \kR \leq n-l^{0}$ denote integers and let $\delta > 0$ be a real number, all chosen arbitrarily.
		The vector $\tpose{\vec{w}} \triangleq \tpose{\vec{\lambda}}\mat{A}$ can be expressed piecewise as
		\begin{equation}
			\vec{w}\bb{j}
			=	\begin{cases}
					\vec{\lambda}\bb{1}	&	,\quad	j = 1,	\\
					\vec{\lambda}\bb{j} - \vec{\lambda}\bb{j-1}	&	,\quad	2 \leq j \leq \lstar-1,	\\
					-\vec{\lambda}\bb{\lstar} - \vec{\lambda}\bb{\lstar-1}	&	,\quad	j = \lstar,	\\
					-\vec{\lambda}\bb{j} + \vec{\lambda}\bb{j-1}	&	,\quad	\lstar+1 \leq j \leq n-1,	\\
					\vec{\lambda}\bb{n-1}	&	,\quad	j = n.
				\end{cases}
			\label{eqn:w-expression}
		\end{equation}
		We select $\vec{\lambda}$ such that $\vec{w}$ satisfies
		\begin{equation}
			\vec{w}\bb{j} =	\begin{cases}
								\vec{v}\bb{j}	&	,\quad	1 \leq j \leq l^{0}-\kL-1, \, j \neq \lstar,	\\
								\vec{v}\bb{j} - \delta	&	,\quad	l^{0}-\kL \leq j \leq l^{0}+\kR,	\\
								\vec{v}\bb{j}	&	,\quad	l^{0}+\kR+1 \leq j \leq n.
							\end{cases}
			\label{eqn:w-assignment}
		\end{equation}
		We do not explicitly set $\vec{w}\bb{\lstar}$ but require it to satisfy the consistency of assignments using \eqref{eqn:w-expression} and \eqref{eqn:w-assignment}.
		We solve for $\vec{\lambda}\bb{1:\lstar-1}$ recursively element-wise starting from $\vec{\lambda}\bb{1}$ and for $\vec{\lambda}\bb{\lstar:n-1}$ recursively element-wise starting from $\vec{\lambda}\bb{n-1}$.
		\makeatletter
			\if@twocolumn
				Letting $c_{1}\bb{\delta} \triangleq -\bb{\kR+\kL+1}\delta$ and $c_{2}\bb{j,\delta} \triangleq -\bb{l^{0}+\kR-j}\delta$, we get
				\begin{equation}
					\vec{\lambda}\bb{j}
					=	\begin{cases}
							\displaystyle	\sum_{k=1}^{j} \vec{v}\bb{k}	&	,\quad	1 \leq j \leq \lstar-1,	\\
							\displaystyle	c_{1}\bb{\delta} + \sum_{k=j+1}^{n} \vec{v}\bb{k}	&	,\quad	\lstar \leq j \leq l^{0}-\kL-2,	\\
							\displaystyle	c_{2}\bb{j,\delta} + \sum_{k=j+1}^{n} \vec{v}\bb{k}	&	,\quad	\text{otherwise},	\\
							\displaystyle	\sum_{k=j+1}^{n} \vec{v}\bb{k}	&	,\quad	l^{0}+\kR \leq j \leq n-1,
						\end{cases}
					\label{eqn:lambda-assignment}
				\end{equation}
			\else
				We get
				\begin{equation}
					\vec{\lambda}\bb{j}
					=	\begin{cases}
							\displaystyle	\sum_{k=1}^{j} \vec{v}\bb{k}	&	,\quad	1 \leq j \leq \lstar-1,	\\
							\displaystyle	-\bb{\kR+\kL+1}\delta + \sum_{k=j+1}^{n} \vec{v}\bb{k}	&	,\quad	\lstar \leq j \leq l^{0}-\kL-2,	\\
							\displaystyle	-\bb{l^{0}+\kR-j}\delta + \sum_{k=j+1}^{n} \vec{v}\bb{k}	&	,\quad	l^{0}-\kL-1 \leq j \leq l^{0}+\kR-1,	\\
							\displaystyle	\sum_{k=j+1}^{n} \vec{v}\bb{k}	&	,\quad	l^{0}+\kR \leq j \leq n-1,
						\end{cases}
					\label{eqn:lambda-assignment}
				\end{equation}
			\fi
		\makeatother
		and for consistency, we have
		\begin{equation}
			\vec{w}\bb{\lstar} = -\vec{\lambda}\bb{\lstar} - \vec{\lambda}\bb{\lstar-1}
			=	\bb{\kR+\kL+1}\delta + \vec{v}\bb{\lstar} - \ip{\vec{1}}{\vec{v}}.
			\label{eqn:w-consistency}
		\end{equation}
		Since $\vec{\lambda} \geq \vec{0}$ is needed, we must ensure in \eqref{eqn:lambda-assignment} that $-\bb{l^{0}+\kR-j}\delta + \ip{\vec{1}}{\vec{v}\bb{j+1:n}} \geq 0$ for every $l^{0}-\kL-1 \leq j \leq l^{0}+\kR-1$, or equivalently, \eqref{eqn:delta-bound} should hold to guarantee $\vec{\lambda}\bb{l^{0}-\kL-1:l^{0}+\kR-1} \geq \vec{0}$.
		Since $\vec{v} \geq \vec{0}$, we already have $\vec{\lambda}\bb{1:\lstar-1} \geq \vec{0}$ and $\vec{\lambda}\bb{l^{0}+\kR:n-1} \geq \vec{0}$ in \eqref{eqn:lambda-assignment}.
		Further, $\vec{\lambda}\bb{\lstar:l^{0}-\kL-2} \geq \vec{0}$ follows from \eqref{eqn:delta-bound} with $j=l^{0}-\kL-1$ and the simple observation that $\vec{\lambda}\bb{\lstar} \geq \vec{\lambda}\bb{\lstar+1} \geq \dots \geq \vec{\lambda}\bb{l^{0}-\kL-1} \geq 0$.
		With $\vec{w} = \tpose{\mat{A}}\vec{\lambda}$ satisfying \eqref{eqn:w-assignment} and \eqref{eqn:w-consistency}, we get
		\makeatletter
			\if@twocolumn
				\begin{equation}
					\begin{split}
						\rho^{2}	& \leq \twonorm{\vec{v} -\tpose{\mat{A}}\vec{\lambda}}^{2}	\\
						& =	\bb[\big]{\vec{v}\bb{\lstar} - \vec{w}\bb{\lstar}}^{2}	\\
						& \phantom{{} = {}} + \twonorm[\big]{\vec{v}\bb{l^{0}-\kL : l^{0}+\kR} - \vec{w}\bb{l^{0}-\kL : l^{0}+\kR}}^{2}	\\
						& =	\bb[\big]{\bb{\kR+\kL+1}\delta - \ip{\vec{1}}{\vec{v}}}^{2} + \delta^{2}\bb{l^{0}+\kR - l^{0}+\kL+1}	\\
						& =	\ip{\vec{1}}{\vec{v}}^{2} - 2\delta\ip{\vec{1}}{\vec{v}}\bb{\kR+\kL+1} + \delta^{2}\bb{\kR+\kL+1}^{2}	\\
						& \phantom{{} = {}} + \delta^{2}\bb{\kR+\kL+1}	\\
						& =	\ip{\vec{1}}{\vec{v}}^{2} + \bb{\delta^{2} - 2\delta\ip{\vec{1}}{\vec{v}}}\cdot\bb{\kR+\kL+1}	\\
						& \phantom{{} = {}} + \delta^{2}\bb{\kR+\kL+1}^{2},
					\end{split}
				\end{equation}
			\else
				\begin{equation}
					\begin{split}
						\rho^{2} \leq \twonorm{\vec{v} -\tpose{\mat{A}}\vec{\lambda}}^{2}
						& =	\bb[\big]{\vec{v}\bb{\lstar} - \vec{w}\bb{\lstar}}^{2} + \twonorm[\big]{\vec{v}\bb{l^{0}-\kL : l^{0}+\kR} - \vec{w}\bb{l^{0}-\kL : l^{0}+\kR}}^{2}	\\
						& =	\bb[\big]{\bb{\kR+\kL+1}\delta - \ip{\vec{1}}{\vec{v}}}^{2} + \delta^{2}\cdot\bb{l^{0}+\kR - l^{0}+\kL+1}	\\
						& =	\ip{\vec{1}}{\vec{v}}^{2} - 2\delta\ip{\vec{1}}{\vec{v}}\bb{\kR+\kL+1} + \delta^{2}\bb{\kR+\kL+1}^{2} + \delta^{2}\bb{\kR+\kL+1}	\\
						& =	\ip{\vec{1}}{\vec{v}}^{2} + \bb{\delta^{2} - 2\delta\ip{\vec{1}}{\vec{v}}}\cdot\bb{\kR+\kL+1} + \delta^{2}\bb{\kR+\kL+1}^{2},
					\end{split}
				\end{equation}
			\fi
		\makeatother
		completing the proof.

	\section{Proof of \lemmaname~\ref{lem:cum-avg}}
		\label{sec:cum-avg proof}
		We will use mathematical induction.
		Let $f\bb{j} \triangleq \ip{\vec{1}}{\vec{v}\bb{j+1:n}}/\bb{l^{0}-j}$ be defined over $1 \leq j \leq l^{0}-1$.
		We have $\ip{\vec{1}}{\vec{v}\bb{l^{0}:n}} \geq \vec{v}\bb{l^{0}}$ from element-wise non-negativity of $\vec{v}$, and $\vec{v}\bb{l^{0}} \geq \vec{v}\bb{l^{0}-1}$ from unimodality of $\vec{v}$.
		This leads to the induction basis
		\makeatletter
			\if@twocolumn
				\begin{equation}
					\begin{split}
						f\bb{l^{0} - 2} - f\bb{l^{0} - 1}
						& =	\frac{\ip{\vec{1}}{\vec{v}\bb{l^{0}-1:n}}}{2} - \ip{\vec{1}}{\vec{v}\bb{l^{0}:n}}	\\
						& =	\frac{\vec{v}\bb{l^{0}-1} - \ip{\vec{1}}{\vec{v}\bb{l^{0}:n}}}{2}	\\
						& \leq	\frac{\vec{v}\bb{l^{0}-1} - \vec{v}\bb{l^{0}}}{2}	\leq 0.
					\end{split}	\raisetag{\baselineskip}
				\end{equation}
			\else
				\begin{equation}
					\begin{split}
						f\bb{l^{0} - 2} - f\bb{l^{0} - 1}
						& =	\frac{\ip{\vec{1}}{\vec{v}\bb{l^{0}-1:n}}}{2} - \ip{\vec{1}}{\vec{v}\bb{l^{0}:n}}	\\
						& =	\frac{\vec{v}\bb{l^{0}-1} - \ip{\vec{1}}{\vec{v}\bb{l^{0}:n}}}{2}
						\leq	\frac{\vec{v}\bb{l^{0}-1} - \vec{v}\bb{l^{0}}}{2}	\leq 0.
					\end{split}
				\end{equation}
			\fi
		\makeatother
		For the inductive step, we have
		\makeatletter
			\if@twocolumn
				\begin{equation}
					\begin{split}
						f\bb{j-1}	& = \frac{\ip{\vec{1}}{\vec{v}\bb{j:n}}}{\bb{l^{0}-j+1}}	\\
						& =	\frac{\vec{v}\bb{j}}{\bb{l^{0}-j+1}}
							+ \frac{\bb{l^{0}-j}}{\bb{l^{0}-j+1}} \frac{\ip{\vec{1}}{\vec{v}\bb{j+1:n}}}{\bb{l^{0}-j}}	\\
						& =	\frac{\vec{v}\bb{j}}{\bb{l^{0}-j+1}} + \frac{\bb{l^{0}-j}}{\bb{l^{0}-j+1}} f\bb{j},
					\end{split}	\raisetag{\baselineskip}
				\end{equation}
			\else
				\begin{equation}
					f\bb{j-1}	= \frac{\ip{\vec{1}}{\vec{v}\bb{j:n}}}{\bb{l^{0}-j+1}}
					= \frac{\vec{v}\bb{j}}{\bb{l^{0}-j+1}}
					+ \frac{\bb{l^{0}-j}}{\bb{l^{0}-j+1}} \frac{\ip{\vec{1}}{\vec{v}\bb{j+1:n}}}{\bb{l^{0}-j}}
					= \frac{\vec{v}\bb{j}}{\bb{l^{0}-j+1}} + \frac{\bb{l^{0}-j}}{\bb{l^{0}-j+1}} f\bb{j},
				\end{equation}
			\fi
		\makeatother
		implying that $f\bb{j-1}$ is a convex combination of $\vec{v}\bb{j}$ and $f\bb{j}$.
		If $\vec{v}\bb{j} \leq f\bb{j}$ is true, then we would immediately have $f\bb{j-1} \leq f\bb{j}$ since $f\bb{j-1}$ must lie on the real line between $\vec{v}\bb{j}$ and $f\bb{j}$.
		From unimodality of $\vec{v}$, we have $\vec{v}\bb{l^{0}} \geq \vec{v}\bb{l^{0}-1} \geq \dots \geq \vec{v}\bb{j}$ and therefore
		\makeatletter
			\if@twocolumn
				\begin{equation}
					\begin{split}
						\vec{v}\bb{j}	& = \frac{\sum_{k=j+1}^{l^{0}} \vec{v}\bb{j}}{l^{0}-j}
						\leq	\frac{\sum_{k=j+1}^{l^{0}} \vec{v}\bb{k}}{l^{0}-j}	\\
						& =	\frac{\ip{\vec{1}}{\vec{v}\bb{j+1:l^{0}}}}{l^{0}-j}
						\leq	\frac{\ip{\vec{1}}{\vec{v}\bb{j+1:n}}}{l^{0}-j}	= f\bb{j},
					\end{split}
				\end{equation}
			\else
				\begin{equation}
					\vec{v}\bb{j}	= \frac{\sum_{k=j+1}^{l^{0}} \vec{v}\bb{j}}{l^{0}-j}
					\leq	\frac{\sum_{k=j+1}^{l^{0}} \vec{v}\bb{k}}{l^{0}-j}
					= \frac{\ip{\vec{1}}{\vec{v}\bb{j+1:l^{0}}}}{l^{0}-j}
					\leq	\frac{\ip{\vec{1}}{\vec{v}\bb{j+1:n}}}{l^{0}-j}	= f\bb{j},
				\end{equation}
			\fi
		\makeatother
		completing the proof.

	\makeatletter
		\if@twocolumn
			\newpage
		\else
		\fi
	\makeatother

	\section{Mean-shift based gradient ascent (\texttt{MS}) algorithm}
		\label{sec:MS algo pseudocode}
		\begin{algorithm}
			\caption{Modified mean-shift gradient ascent algorithm for sampling and peak detection on the current grid}
			\KwIn{$n \times n$ grid selected at the current stage}	\\
			\KwOut{Location of peaks $\mu_1,\mu_2,\ldots$ where the number of peaks is initially unknown}	\\
			\textbf{Steps:}
			\begin{algorithmic}[1]
				\State Set the maximum number of iterations to $M$, and the size of window to $\omega$
				\State Cluster index $K\leftarrow 1$
				\For {$iter ~\leftarrow 1$ \KwTo $M$}
					\State Randomly select an initial location $X$ on the map.
					\While{$X$ hasn't already been visited}
						\LineComment{assuming we're going to find a new peak}
						\State Mark $X$ as visited and belonging to cluster $K$
						\State Compute the new center of mass $X_c$ within a window of size $\omega$ around $X$.
						\State Compute the direction from $X$ \KwTo $X_c$ and quantize the angle into 8 equi-partitioned bins in $\setA$
						\State Update the position of $X$ to one of the 8 adjacent positions based on the quantized direction 
					\EndWhile
					\State Mark all points in this trail leading up to $X$ as belonging to the same cluster as $X$
					\If{$X$ belongs to the new cluster $K$}
						\LineComment{new peak found}
						\State Append $X$ as $\mu_K$ to the set of peaks
						\State $K\leftarrow K+1$
					\EndIf
				\EndFor
				\State Return the highest peak among $\mu_1,\mu_2,\ldots,\mu_K$
			\end{algorithmic}
			\label{algo:msc}
		\end{algorithm}

	\section{Proof of \lemmaname~\ref{lem:non-neg soln}}
		\label{sec:non-neg soln proof}
		Assuming $\vec{z}_{\opt} \geq \vec{0}$, the second part of the lemma follows on observing that
		\begin{enumerate}
			\item	the constraint $\twonorm{\vec{z}_{\opt}}^{2} = 1$ is equivalent to the constraint $\twonorm{\vec{z}_{\opt}} = 1$,
			\item	the optimal value of \problemname~\pref{prob:peak-closeness-opt} is $-\ip{\vec{z}_{\opt}}{\vec{v}}$ and $-\ip{\vec{z}_{\opt}}{\vec{v}} \leq -\rho \iff \ip{\vec{z}_{\opt}}{\vec{v}} \geq \rho$, and
			\item	the remaining constraints in \problemname~\pref{prob:peak-closeness} are also present in \problemname~\pref{prob:peak-closeness-opt} and $\vec{z}_{\opt} \geq \vec{0}$.
		\end{enumerate}
		To show the first part of the lemma, we start from an arbitrary solution $\vec{z}_{\ast}$ for \problemname~\pref{prob:peak-closeness-opt} and transform it into a vector in $\setR^{n}_{+}$ that is feasible for \problemname~\pref{prob:peak-closeness-opt} and gives the same or a better value of the objective function than $\vec{z}_{\ast}$.
		If $\vec{z}_{\ast} \geq \vec{0}$ then no transformation is necessary.
		Otherwise, we invoke the following sequence of arguments.
		For brevity, we refer to the first two constraints in \problemname~\pref{prob:peak-closeness-opt} as the unimodality constraints.
		\begin{enumerate}
			\item	If $\vec{z}_{\ast}\bb{\lstar} < 0$, then form a vector $\vec{z}' \in \setR^{n}$ that agrees with $\vec{z}_{\ast}$ on the indices $\cc{1,\dotsc,n}\setminus\cc{\lstar}$ and $\vec{z}'\bb{\lstar} = \abs{\vec{z}_{\ast}\bb{\lstar}} > \vec{z}_{\ast}\bb{\lstar}$.
			Clearly, $\twonorm{\vec{z}'}^{2} = \twonorm{\vec{z}_{\ast}}^{2} = 1$ and $\vec{z}'$ satisfies the unimodality constraints since $\lstar$ is still the index of the largest element and the other elements are same as those in $\vec{z}_{\ast}$.
			Further, $\vec{v}\bb{\lstar} \geq 0$ implies that
			\makeatletter
				\if@twocolumn
					\begin{equation}
						\begin{split}
							\MoveEqLeft	\bb{-\ip{\vec{z}'}{\vec{v}}} - \bb{-\ip{\vec{z}_{\ast}}{\vec{v}}}
							=	\ip{\vec{z}_{\ast}}{\vec{v}} - \ip{\vec{z}'}{\vec{v}}	\\
							& =	\vec{z}_{\ast}\bb{\lstar}\vec{v}\bb{\lstar} - \vec{z}'\bb{\lstar}\vec{v}\bb{\lstar}	\\
							& =	\bb{\vec{z}_{\ast}\bb{\lstar} - \abs{\vec{z}_{\ast}\bb{\lstar}}} \vec{v}\bb{\lstar}
							\leq	0.
						\end{split}
					\end{equation}
				\else
					\begin{equation}
						\bb{-\ip{\vec{z}'}{\vec{v}}} - \bb{-\ip{\vec{z}_{\ast}}{\vec{v}}}
						=	\ip{\vec{z}_{\ast}}{\vec{v}} - \ip{\vec{z}'}{\vec{v}}
						=	\vec{z}_{\ast}\bb{\lstar}\vec{v}\bb{\lstar} - \vec{z}'\bb{\lstar}\vec{v}\bb{\lstar}
						=	\bb{\vec{z}_{\ast}\bb{\lstar} - \abs{\vec{z}_{\ast}\bb{\lstar}}} \vec{v}\bb{\lstar}
						\leq	0.
					\end{equation}
				\fi
			\makeatother
			Thus, $\vec{z}'$ is feasible for \problemname~\pref{prob:peak-closeness-opt} and is optimal \wrt~the value of the objective function since $-\ip{\vec{z}'}{\vec{v}} \leq -\ip{\vec{z}_{\ast}}{\vec{v}}$.
			Therefore, \WLOG~we subsequently assume $\vec{z}_{\ast}\bb{\lstar} \geq 0$.
			
			\item	If $\vec{z}_{\ast}\bb{\lstar} = 0$ then $\vec{z}_{\ast} \leq \vec{0}$ implying that the optimal value of \problemname~\pref{prob:peak-closeness-opt} is $-\ip{\vec{z}_{\ast}}{\vec{v}} = \ip{-\vec{z}_{\ast}}{\vec{v}} \geq 0$ since $\vec{v} \geq \vec{0}$.
			Define the vector $\vec{z}' \in \setR^{n}$ such that $\vec{z}'\bb{\lstar} = 1$ and $\vec{z}'\bb{\cc{1,\dotsc,n}\setminus\cc{\lstar}} = \vec{0}$.
			Then, $\vec{z}'$ has unit length and trivially satisfies the unimodality constraints with peak at index $\lstar$, making it feasible for \problemname~\pref{prob:peak-closeness-opt}.
			Furthermore, $-\ip{\vec{z}'}{\vec{v}} = -\vec{v}\bb{\lstar} \leq 0$ implies that $\vec{z}'$ achieves an objective function value that is at least as good as $\vec{z}_{\ast}$.
			Therefore, \WLOG~we may subsequently assume $\vec{z}_{\ast}\bb{\lstar} > 0$.
			
			\item	Let $\Lambda \subseteq \cc{1,2,\dotsc,n} \setminus \cc{\lstar}$ denote the set of indices on which $\vec{z}_{\ast}$ is negative.
			Since $\vec{z}_{\ast}$ is monotonically non-decreasing on the index set $\cc{1,2,\dotsc,\lstar-1}$, $\Lambda_{1} \triangleq \Lambda \bigcap \cc{1,2,\dotsc,\lstar-1}$ is either empty or is the contiguous set $\bb{1,2,\dotsc,\card{\Lambda_{1}}}$.
			By an analogous reasoning, $\Lambda_{2} \triangleq \Lambda \bigcap \cc{\lstar+1,\lstar+2,\dotsc,n}$ is either empty or is the contiguous set $\cc{n-\card{\Lambda_{2}}+1, n-\card{\Lambda_{2}}+1, \dotsc, n}$.
			Consider a vector $\vec{z}' \in \setR^{n}$ such that $\vec{z}'\bb{\Lambda} = \vec{0}$ and $\vec{z}'\bb{\Lambda^{\comp}} = \vec{z}_{\ast}\bb{\Lambda^{\comp}}/\twonorm{\vec{z}_{\ast}\bb{\Lambda^{\comp}}}$.
			Since $\lstar \in \Lambda^{\comp}$ and $\vec{z}_{\ast}\bb{\lstar} > 0$, $\twonorm{\vec{z}_{\ast}\bb{\Lambda^{\comp}}} > 0$ and $\vec{z}'\bb{\Lambda^{\comp}}$ is well defined.
			Clearly, $\twonorm{\vec{z}'}^{2} = 1$ and $\vec{z}'$ satisfies the unimodality constraints because
			\begin{enumerate}
				\item	on index subset $\Lambda^{\comp}$, $\vec{z}'$ is a positively rescaled version of $\vec{z}_{\ast}$ and thus honors the element-wise inequality constraints,
				\item	on index subset $\Lambda$, $\vec{z}'$ coincides with a zero vector and trivially satisfies the unimodality constraints with equality, and
				\item	the boundary cases $0 = \vec{z}'\bb{\card{\Lambda_{1}}} \leq \vec{z}'\bb{\card{\Lambda_{1}} + 1}$ and $\vec{z}'\bb{n - \card{\Lambda_{2}}} \geq \vec{z}'\bb{n - \card{\Lambda_{2}} + 1} = 0$ are also satisfied since $\vec{z}'\bb{\card{\Lambda_{1}} + 1}$ and $\vec{z}'\bb{n - \card{\Lambda_{2}}}$ are non-negative by definition of $\Lambda_{1}$ and $\Lambda_{2}$.
			\end{enumerate}
			Further, using the non-negativity of $\vec{v}$,
			\makeatletter
				\if@twocolumn
					\begin{equation}
						\begin{split}
							\MoveEqLeft[0.5]	\bb{-\ip{\vec{z}'}{\vec{v}}} - \bb{-\ip{\vec{z}_{\ast}}{\vec{v}}}
							=	\ip{\vec{z}_{\ast}}{\vec{v}} - \ip{\vec{z}'}{\vec{v}}	\\
							& =	\ip{\vec{z}_{\ast}\bb{\Lambda}}{\vec{v}\bb{\Lambda}} + \ip{\vec{z}_{\ast}\bb{\Lambda^{\comp}}}{\vec{v}\bb{\Lambda^{\comp}}} - \ip{\vec{z}'\bb{\Lambda^{\comp}}}{\vec{v}\bb{\Lambda^{\comp}}}	\\
							& =	-\ip{-\vec{z}_{\ast}\bb{\Lambda}}{\vec{v}\bb{\Lambda}} + \ip{\vec{z}_{\ast}\bb{\Lambda^{\comp}}}{\vec{v}\bb{\Lambda^{\comp}}}	\\
							& \phantom{{} =	-\ip{-\vec{z}_{\ast}\bb{\Lambda}}{\vec{v}\bb{\Lambda}}} -\frac{1}{\twonorm{\vec{z}_{\ast}\bb{\Lambda^{\comp}}}}\ip{\vec{z}_{\ast}\bb{\Lambda^{\comp}}}{\vec{v}\bb{\Lambda^{\comp}}}	\\
							& \leq	0 + \frac{\twonorm{\vec{z}_{\ast}\bb{\Lambda^{\comp}}} - 1}{\twonorm{\vec{z}_{\ast}\bb{\Lambda^{\comp}}}}
							\ip{\vec{z}_{\ast}\bb{\Lambda^{\comp}}}{\vec{v}\bb{\Lambda^{\comp}}}
							\leq	0,
						\end{split}
						\raisetag{1.2\baselineskip}
					\end{equation}
				\else
					\begin{equation}
						\begin{split}
							\MoveEqLeft	\bb{-\ip{\vec{z}'}{\vec{v}}} - \bb{-\ip{\vec{z}_{\ast}}{\vec{v}}}
							=	\ip{\vec{z}_{\ast}}{\vec{v}} - \ip{\vec{z}'}{\vec{v}}
							=	\ip{\vec{z}_{\ast}\bb{\Lambda}}{\vec{v}\bb{\Lambda}} + \ip{\vec{z}_{\ast}\bb{\Lambda^{\comp}}}{\vec{v}\bb{\Lambda^{\comp}}} - \ip{\vec{z}'\bb{\Lambda^{\comp}}}{\vec{v}\bb{\Lambda^{\comp}}}	\\
							& =	-\ip{-\vec{z}_{\ast}\bb{\Lambda}}{\vec{v}\bb{\Lambda}} + \ip{\vec{z}_{\ast}\bb{\Lambda^{\comp}}}{\vec{v}\bb{\Lambda^{\comp}}} - \frac{1}{\twonorm{\vec{z}_{\ast}\bb{\Lambda^{\comp}}}}\ip{\vec{z}_{\ast}\bb{\Lambda^{\comp}}}{\vec{v}\bb{\Lambda^{\comp}}}	\\
							& \leq	0 + \frac{\twonorm{\vec{z}_{\ast}\bb{\Lambda^{\comp}}} - 1}{\twonorm{\vec{z}_{\ast}\bb{\Lambda^{\comp}}}}
							\ip{\vec{z}_{\ast}\bb{\Lambda^{\comp}}}{\vec{v}\bb{\Lambda^{\comp}}}
							\leq	0,
						\end{split}
					\end{equation}
				\fi
			\makeatother
			where the first inequality is because $-\vec{z}_{\ast}\bb{\Lambda} > \vec{0} \implies \ip{-\vec{z}_{\ast}\bb{\Lambda}}{\vec{v}\bb{\Lambda}} \geq 0$ and the second inequality is because $0 < \twonorm{\vec{z}_{\ast}\bb{\Lambda^{\comp}}} \leq \twonorm{\vec{z}_{\ast}} = 1$ and $\vec{z}_{\ast}\bb{\Lambda^{\comp}} \geq \vec{0} \implies \ip{\vec{z}_{\ast}\bb{\Lambda^{\comp}}}{\vec{v}\bb{\Lambda^{\comp}}} \geq 0$.
			Thus, $\vec{z}'$ is feasible for \problemname~\pref{prob:peak-closeness-opt}, satisfies $\vec{z}' \geq \vec{0}$ and is optimal \wrt~the value of the objective function, since $-\ip{\vec{z}'}{\vec{v}} \leq -\ip{\vec{z}_{\ast}}{\vec{v}}$.
			Therefore, \WLOG~we can assume $\vec{z}_{\ast} \geq \vec{0}$, completing the proof.
		\end{enumerate}

	\section{Proof of \lemmaname~\ref{lem:bound reduction}}
		\label{sec:bound reduction proof}
		Setting $x = \zeta/\sigma_{0}$ and $y = \sigma/\sigma_{0}$ we have $0 \leq \zeta \leq \sigma \leq \sigma_{0} \iff 0 \leq x \leq y \leq 1$.
		We also have $\sqrt{1-\zeta^{2}/\sigma^{2}} = \sqrt{1-x^{2}/y^{2}}$ and $\eta\bb{\sigma,\sigma_{0},\zeta} = \bb{1-y} + \sqrt{\bb{1-y}^{2} + y^{2} - x^{2}}$ from \eqref{eqn:eta-defn}.
		To prove the result, it thus suffices to show that
		\makeatletter
			\if@twocolumn
				\begin{equation}
					\begin{split}
						f\bb{x,y} & \triangleq \eta\bb{\sigma,\sigma_{0},\zeta} - \sqrt{1-\zeta^{2}/\sigma^{2}}	\\
						& =	\bb{1-y} + \sqrt{\bb{1-y}^{2} + y^{2} - x^{2}} - \sqrt{1-x^{2}/y^{2}}
					\end{split}
				\end{equation}
			\else
				\begin{equation}
					f\bb{x,y} \triangleq \eta\bb{\sigma,\sigma_{0},\zeta} - \sqrt{1-\zeta^{2}/\sigma^{2}}
					=	\bb{1-y} + \sqrt{\bb{1-y}^{2} + y^{2} - x^{2}} - \sqrt{1-x^{2}/y^{2}}
				\end{equation}
			\fi
		\makeatother
		is non-negative over the domain $0 \leq x \leq y \leq 1$.
		We have
		\makeatletter
			\if@twocolumn
				\begin{equation}
					\begin{split}
						f\bb{0,y} & =	\bb{1-y} + \sqrt{\bb{1-y}^{2} + y^{2}} - \sqrt{1}	\\
						& =	\sqrt{\bb{1-y}^{2} + y^{2}} - y \geq 0.
					\end{split}
				\end{equation}
			\else
				\begin{equation}
					f\bb{0,y} =	\bb{1-y} + \sqrt{\bb{1-y}^{2} + y^{2}} - \sqrt{1}
					=	\sqrt{\bb{1-y}^{2} + y^{2}} - y \geq 0.
				\end{equation}
			\fi
		\makeatother
		We further have
		\makeatletter
			\if@twocolumn
				\begin{equation}
					\begin{split}
						\diffp{}{x} f\bb{x,y}
						& =	\diffp{}{x} \bb{\sqrt{\bb{1-y}^{2} + y^{2} - x^{2}} - \sqrt{1-x^{2}/y^{2}}}	\\
						& =	\frac{-x}{\sqrt{\bb{1-y}^{2} + y^{2} - x^{2}}} - \frac{-x/y^{2}}{\sqrt{1-x^{2}/y^{2}}}	\\
						& =	\frac{-x}{\sqrt{\bb{1-y}^{2} + y^{2} - x^{2}}} + \frac{x}{y\sqrt{y^{2} - x^{2}}}.
					\end{split}
					\label{eqn:diff-derivative}
				\end{equation}
			\else
				\begin{equation}
					\begin{split}
						\diffp{}{x} f\bb{x,y}
						& =	\diffp{}{x} \bb{\sqrt{\bb{1-y}^{2} + y^{2} - x^{2}} - \sqrt{1-x^{2}/y^{2}}}	\\
						& =	\frac{-x}{\sqrt{\bb{1-y}^{2} + y^{2} - x^{2}}} - \frac{-x/y^{2}}{\sqrt{1-x^{2}/y^{2}}}
						=	\frac{-x}{\sqrt{\bb{1-y}^{2} + y^{2} - x^{2}}} + \frac{x}{y\sqrt{y^{2} - x^{2}}}.
					\end{split}
					\label{eqn:diff-derivative}
				\end{equation}
			\fi
		\makeatother
		Since
		\makeatletter
			\if@twocolumn
				\begin{subequations}
					\begin{align}
						\MoveEqLeft \diffp{}{x} f\bb{x,y} \geq 0	\\
						&	\iff \frac{x}{y\sqrt{y^{2} - x^{2}}} \geq \frac{x}{\sqrt{\bb{1-y}^{2} + y^{2} - x^{2}}}	\\
						&	\iff \frac{1}{y\sqrt{y^{2} - x^{2}}}	\geq \frac{1}{\sqrt{\bb{1-y}^{2} + y^{2} - x^{2}}}	\\
						&	\iff y\sqrt{y^{2} - x^{2}}	\leq \sqrt{\bb{1-y}^{2} + y^{2} - x^{2}}	\\
						&	\iff y^{2}\bb{y^{2} - x^{2}}	\leq \bb{1-y}^{2} + y^{2} - x^{2}	\\
						&	\iff 0 \leq \bb{1-y}^{2} + \bb{1 - y^{2}}\bb{y^{2} - x^{2}}
					\end{align}
				\end{subequations}
			\else
				\begin{subequations}
					\begin{align}
						\diffp{}{x} f\bb{x,y} \geq 0
						&	\iff \frac{x}{y\sqrt{y^{2} - x^{2}}} \geq \frac{x}{\sqrt{\bb{1-y}^{2} + y^{2} - x^{2}}}
							\iff \frac{1}{y\sqrt{y^{2} - x^{2}}}	\geq \frac{1}{\sqrt{\bb{1-y}^{2} + y^{2} - x^{2}}}	\\
						&	\iff y\sqrt{y^{2} - x^{2}}	\leq \sqrt{\bb{1-y}^{2} + y^{2} - x^{2}}
							\iff y^{2}\bb{y^{2} - x^{2}}	\leq \bb{1-y}^{2} + y^{2} - x^{2}	\\
						&	\iff 0 \leq \bb{1-y}^{2} + \bb{1 - y^{2}}\bb{y^{2} - x^{2}}
					\end{align}
				\end{subequations}
			\fi
		\makeatother
		is true over $0 \leq x \leq y \leq 1$, it follows that $\diffp{}{x} f\bb{x,y} \geq 0$ over $0 \leq x \leq y \leq 1$.
		Thus, for every $y \in \BB{0,1}$, $f\bb{x,y}$ is increasing \wrt~$x$ over $0 \leq x \leq y$ and $f\bb{0,y}$ is non-negative.
		Therefore, $f\bb{x,y} \geq f\bb{0,y} \geq 0$ over $0 \leq x \leq y \leq 1$, completing the proof.

	\section{Relaxing Positivity and Sampling Grid Assumptions in \algorithmname~\ref{alg:PAMC-UMR}}
		\label{sec:relax-positivity}
		For a non-square sampling grid of size $\nr \times \nc$, the sample complexity bound to guarantee success of low-rank matrix completion \whp~changes~\cite{gross2011recovering} to $\card{\mathcal{V}'} = \BigOh{\bb{\nr+\nc} \log^{2} \max\cc{\nr,\nc}}$ and steps~\sref{itm:random sampling} and~\sref{itm:reconstruction} in \algorithmname~\ref{alg:PAMC-UMR} should be adjusted accordingly.
		Steps~\sref{itm:localization LB} and~\sref{itm:localization UB} should be changed to operate on \problemname~\pref{prob:localization}[\nr,\zeta,\sigma,\vec{u}] to give $\lrL,\lrR \in \cc{1,2,\dotsc,\nr}$, and analogously, step~\sref{itm:localization second set} should operate on \problemname~\pref{prob:localization}[\nc,\zeta,\sigma,\vec{v}] to give $\lcL,\lcR \in \cc{1,2,\dotsc,\nc}$.
		The functional forms for $\zeta$ and $C\bb{q,n}$ should change according to~\cite{candes2010noisematrixcompletion}, however \theoremname~\ref{thm:inner product bound} is valid as is.
		\theoremname~\ref{thm:convergence} undergoes only a small change with $n$ being replaced by $\nr$ in all assumptions pertaining to $\vec{u}_{0}$ and $n$ being replaced by $\nc$ for all assumptions about $\vec{v}_{0}$, implying that $n^{2}$ in the localization bound \eqref{eqn:loc-bound} is replaced by $\nr \cdot \nc$.

		To relax the positivity assumption on $H\bb{\cdot}$, we note that \algorithmname~\ref{alg:PAMC-UMR} works in exactly the same way even if only the weaker condition of $\abs{\vec{u}_{0}} \geq \vec{0}$ and $\abs{\vec{v}_{0}} \geq \vec{0}$ being unimodal is satisfied.
		This is apparent from examining the statements for \theoremname~\ref{thm:convergence} and \lemmasname~\ref{lem:peak-closeness} and~\ref{lem:non-neg soln}.
		The properties of positivity and unimodality of $\vec{u}_{0}$ and $\vec{v}_{0}$ are not used by \theoremname~\ref{thm:inner product bound}, and by examining the proof of \theoremname~\ref{thm:convergence} we see that these properties of $\vec{u}_{0}$ and $\vec{v}_{0}$ are relevant only in the steps~\sref{itm:localization LB} through~\sref{itm:localization second set} of \algorithmname~\ref{alg:PAMC-UMR} through the use of \problemname~\pref{prob:localization}.
		We make the following claim without proof (note that the absolute value operator $\abs{\cdot}$ is understood to act element-wise on vectors).

		\begin{corollary}
			\label{cor:convergence-non-positive}
			Consider a modification of \algorithmname~\ref{alg:PAMC-UMR} with all instances of \problemname~\pref{prob:localization}[n,\zeta,\sigma,\vec{u}] replaced by \problemname~\pref{prob:localization}[n,\zeta,\sigma,\abs{\vec{u}}] and all instances of \problemname~\pref{prob:localization}[n,\zeta,\sigma,\vec{v}] replaced by \problemname~\pref{prob:localization}[n,\zeta,\sigma,\abs{\vec{v}}].
			The conclusion of \theoremname~\ref{thm:convergence} holds for this modified algorithm under the weaker assumption of $\abs{\vec{u}_{0}}$ and $\abs{\vec{v}_{0}}$ being unimodal vectors with respective peaks at $l^{0}_{\text{r}}$ and $l^{0}_{\text{c}}$, where $\mat{H} = \sigma_{0} \vec{u}_{0} \tpose{\vec{v}_{0}}$ is the SVD of the not necessarily positive matrix $\mat{H} \in \setRn{n}{n}$, provided that all other assumptions of \theoremname~\ref{thm:convergence} remain unchanged.
		\end{corollary}

	\section{Coherence Computation}
		\label{sec:coherence computation}
		We follow the definitions laid out in \cite{gross2011recovering}.
		Let $\mat{H} = \sigma_{0} \vec{u}_{0} \tpose{\vec{v}_{0}}$ denote the SVD of $\mat{H}$ and let $\nu > 0$ denote the coherence parameter defined as the minimum value of $\nu_{0}$ satisfying the bounds
		\makeatletter
			\if@twocolumn
				\begin{subequations}
					\label{eqn:coherence defn}
					\begin{align}
						\max_{i,j} \, \abs{\ip{\vec{u}_{0}}{\vec{e}_{i}}}^{2} + \abs{\ip{\vec{v}_{0}}{\vec{e}_{j}}}^{2}
						& \leq \frac{2\nu_{0}}{\sqrt{n}},	\\
						\max_{i,j} \, \abs{\ip{\vec{u}_{0}}{\vec{e}_{i}}}^{2} \cdot \abs{\ip{\vec{v}_{0}}{\vec{e}_{j}}}^{2} & \leq \frac{\nu_{0}}{n}.
					\end{align}
				\end{subequations}
			\else
				\begin{equation}
					\max_{i,j} \, \abs{\ip{\vec{u}_{0}}{\vec{e}_{i}}}^{2} + \abs{\ip{\vec{v}_{0}}{\vec{e}_{j}}}^{2}
					\leq \frac{2\nu_{0}}{\sqrt{n}}	\quad	\text{and}	\quad
					\max_{i,j} \, \abs{\ip{\vec{u}_{0}}{\vec{e}_{i}}}^{2} \cdot \abs{\ip{\vec{v}_{0}}{\vec{e}_{j}}}^{2} \leq \frac{\nu_{0}}{n}.
					\label{eqn:coherence defn}
				\end{equation}
			\fi
		\makeatother
		Since $\mat{H}$ is formed by discretization of the function $H\bb{\vec{y}} = F\bb{\yc} G\bb{\yr}$, for high enough resolution of discretization, we can write
		\makeatletter
			\if@twocolumn
				\begin{equation}
					\begin{split}
						\MoveEqLeft[0]	\mu\bb{\vec{u}_{0}} \triangleq
						\frac{\max\limits_{i} \, \abs{\ip{\vec{u}_{0}}{\vec{e}_{i}}}^{2}}{\twonorm{\vec{u}_{0}}^{2}}
						\approx \frac{\max\limits_{i} \, \abs{\int_{i/\sqrt{n}}^{\bb{i+1}/\sqrt{n}} F\bb{\yc} d\yc}^{2}}{\int_{-1/2}^{1/2} F^{2}\bb{\yc} d\yc \int_{i/\sqrt{n}}^{\bb{i+1}/\sqrt{n}} d\yc}	\\
						& = \frac{\max\limits_{i} \, \abs{\sqrt{n} \int_{i/\sqrt{n}}^{\bb{i+1}/\sqrt{n}} F\bb{\yc} d\yc}^{2}}{\sqrt{n} \int_{-1/2}^{1/2} F^{2}\bb{\yc} d\yc}
						\approx \frac{\esssup\limits_{\yc} F^{2}\bb{\yc}}{\sqrt{n} \int_{-1/2}^{1/2} F^{2}\bb{\yc} d\yc}.
					\end{split}
					\label{eqn:cont appx coh}
				\end{equation}
			\else
				\begin{equation}
					\begin{split}
						\mu\bb{\vec{u}_{0}} & \triangleq
						\frac{\displaystyle \max_{i} \, \abs{\ip{\vec{u}_{0}}{\vec{e}_{i}}}^{2}}{\twonorm{\vec{u}_{0}}^{2}}
						\approx \frac{\displaystyle \max_{i} \, \abs{\int_{i/\sqrt{n}}^{\bb{i+1}/\sqrt{n}} F\bb{\yc} d\yc}^{2}}{\displaystyle \int_{-1/2}^{1/2} F^{2}\bb{\yc} d\yc \int_{i/\sqrt{n}}^{\bb{i+1}/\sqrt{n}} d\yc}	\\
						& =	\frac{\displaystyle \max_{i} \, \abs{\sqrt{n} \int_{i/\sqrt{n}}^{\bb{i+1}/\sqrt{n}} F\bb{\yc} d\yc}^{2}}{\displaystyle \sqrt{n} \int_{-1/2}^{1/2} F^{2}\bb{\yc} d\yc}
						\approx \frac{\displaystyle \esssup_{\yc} F^{2}\bb{\yc}}{\displaystyle \sqrt{n} \int_{-1/2}^{1/2} F^{2}\bb{\yc} d\yc}.
					\end{split}
					\label{eqn:cont appx coh}
				\end{equation}
			\fi
		\makeatother
		where we have assumed $-1/2 \leq \yc \leq 1/2$.
		If the function $F\bb{\cdot}$ is highly localized within $\BB{-1/2, 1/2}$ then,
		\begin{equation}
			\mu\bb{\vec{u}_{0}}	= \frac{\displaystyle \bb{1-\gamma\bb{n}} \esssup_{\yc} F^{2}\bb{\yc}}{\displaystyle \sqrt{n} \int_{-\infty}^{\infty} F^{2}\bb{\yc} d\yc}.
			\label{eqn:cont appx coh infty}
		\end{equation}
		where the approximation factor $\bb{1-\gamma\bb{n}}$ encapsulates all of the foregoing approximations.
		Similarly,
		\begin{equation}
			\mu\bb{\vec{v}_{0}} \triangleq
			\frac{\displaystyle \max_{j} \, \abs{\ip{\vec{v}_{0}}{\vec{e}_{j}}}^{2}}{\twonorm{\vec{v}_{0}}^{2}}
			= \frac{\displaystyle \bb{1-\gamma'\bb{n}} \esssup_{\yr} G^{2}\bb{\yr}}{\displaystyle \sqrt{n} \int_{-\infty}^{\infty} G^{2}\bb{\yr} d\yr}.
			\label{eqn:cont appx coh second}
		\end{equation}
		Equation \eqref{eqn:coherence defn} implies that
		\begin{equation}
			\nu = \max \cc{\frac{\sqrt{n}}{2}\bb[\big]{\mu\bb{\vec{u}_{0}} + \mu\bb{\vec{v}_{0}}}, n\mu\bb{\vec{u}_{0}}\mu\bb{\vec{v}_{0}}}.
			\label{eqn:coherence parameter}
		\end{equation}
		Barring the approximation factors of $\bb{1-\gamma\bb{n}}$ and $\bb{1 - \gamma'\bb{n}}$, it is clear that $\sqrt{n}\mu\bb{\vec{u}_{0}}$ and $\sqrt{n}\mu\bb{\vec{v}_{0}}$ are independent of $n$ as long as the approximations in \eqref{eqn:cont appx coh}, \eqref{eqn:cont appx coh infty} and \eqref{eqn:cont appx coh second} are valid.
		In particular, the coherence parameter $\nu$ is unchanged by sub-sampling on a $\sqrt{g} \times \sqrt{g}$ uniform grid as long as $\gamma\bb{n} \approx \gamma\bb{g}$ and $\gamma'\bb{n} \approx \gamma'$.

		\subsection{Exponential Fields}
			Let $H\bb{\vec{y}} = H_{0}\exp\bb{-\ac\abs{\yc}^{\pc} -\ar\abs{\yr}^{\pr}}$ with $F\bb{\yc} = \sqrt{H_{0}}\exp\bb{-\ac\abs{\yc}^{\pc}}$ and $G\bb{\yr} = \sqrt{H_{0}}\exp\bb{-\ar\abs{\yr}^{\pr}}$.
			We have
			\makeatletter
				\if@twocolumn
					\begin{subequations}
						\begin{align}
							\mu\bb{\vec{u}_{0}}	& = \frac{\displaystyle \esssup_{\yc} \exp\bb{-2\ac\abs{\yc}^{\pc}}}{\displaystyle \sqrt{n} \int_{-\infty}^{\infty} \exp\bb{-2\ac\abs{\yc}^{\pc}} d\yc}	\\
							& = \frac{1}{2\sqrt{n}} \inv{\bb{\int_{0}^{\infty} \exp\bb{-2\ac\yc^{\pc}} d\yc}}	\label{eqn:exp expr} \\
							& = \frac{1}{2\sqrt{n}} \inv{\bb{\frac{1}{\pc\bb{2\ac}^{1/\pc}} \int_{0}^{\infty} t^{\frac{1}{\pc} - 1} \exp\bb{-t} dt}}	\label{eqn:gamma expr} \\
							& = \frac{\bb{2\ac}^{1/\pc}}{\sqrt{n}\bb{2/\pc}\Gamma\bb{1/\pc}},	\label{eqn:by gamma defn}
						\end{align}
					\end{subequations}
				\else
					\begin{subequations}
						\begin{align}
							\mu\bb{\vec{u}_{0}}	& = \frac{\displaystyle \esssup_{\yc} \exp\bb{-2\ac\abs{\yc}^{\pc}}}{\displaystyle \sqrt{n} \int_{-\infty}^{\infty} \exp\bb{-2\ac\abs{\yc}^{\pc}} d\yc}
							=	\frac{1}{2\sqrt{n}} \inv{\bb{\int_{0}^{\infty} \exp\bb{-2\ac\yc^{\pc}} d\yc}}	\label{eqn:exp expr} \\
							& = \frac{1}{2\sqrt{n}} \inv{\bb{\frac{1}{\pc\bb{2\ac}^{1/\pc}} \int_{0}^{\infty} t^{\frac{1}{\pc} - 1} \exp\bb{-t} dt}}	\label{eqn:gamma expr} \\
							& = \frac{\bb{2\ac}^{1/\pc}}{\sqrt{n}\bb{2/\pc}\Gamma\bb{1/\pc}},	\label{eqn:by gamma defn}
						\end{align}
					\end{subequations}
				\fi
			\makeatother
			where \eqref{eqn:gamma expr} is obtained from \eqref{eqn:exp expr} by the change of variables $t = 2\ac\yc^{\pc}$, and \eqref{eqn:by gamma defn} uses the definition of the $\Gamma$-function.
			Similarly,
			\begin{equation}
				\mu\bb{\vec{v}_{0}}	= \frac{\bb{2\ar}^{1/\pr}}{\sqrt{n}\bb{2/\pr}\Gamma\bb{1/\pr}}
			\end{equation}
			and the coherence parameter is determined as in \eqref{eqn:coherence parameter}.

		\subsection{Power Law Fields}
			Let $H\bb{\vec{y}} = H_{0} \inv{\bb{\ac + \abs{\yc}^{\pc}}} \inv{\bb{\ar + \abs{\yr}^{\pr}}}$ with $F\bb{\yc} = \sqrt{H_{0}}\inv{\bb{\ac + \abs{\yc}^{\pc}}}$ and $G\bb{\yr} = \sqrt{H_{0}}\inv{\bb{\ar + \abs{\yr}^{\pr}}}$.
			We have
			\makeatletter
				\if@twocolumn
					\begin{subequations}
						\begin{align}
							\MoveEqLeft[1]	\mu\bb{\vec{u}_{0}} = \frac{\displaystyle \esssup_{\yc} \bb{\ac + \abs{\yc}^{\pc}}^{-2}}{\displaystyle \sqrt{n} \int_{-\infty}^{\infty} \bb{\ac + \abs{\yc}^{\pc}}^{-2} d\yc}	\\
							& = \frac{1}{2\ac^{2}\sqrt{n}} \inv{\bb{\int_{0}^{\infty} \bb{\ac + \yc^{\pc}}^{-2} d\yc}}	\label{eqn:poly expr} \\
							& = \begin{cases}
									\frac{1}{2\ac^{2}\sqrt{n}} \inv{\bb{\frac{\bb{\pi/\pc}\bb{1-1/\pc}}{\ac^{2-1/\pc} \sin\bb{\pi/\pc}}}}	& ,\quad \pc \in \bb{\frac{1}{2},1} \bigcup \bb{1,\infty} \\
									\frac{1}{2\ac^{2}\sqrt{n}} \inv{\bb{\frac{1}{\ac}}} & ,\quad \pc = 1
								\end{cases}
							\label{eqn:evaluated poly expr} \\
							& = \begin{cases}
									\frac{\pc^{2} \sin\bb{\pi/\pc}}{2\sqrt{n}\pi\bb{\pc-1}\ac^{1/\pc}} & ,\quad \pc \in \bb{\frac{1}{2},1} \bigcup \bb{1,\infty} \\
									\inv{\bb{2\ac\sqrt{n}}} & ,\quad \pc = 1
								\end{cases}
						\end{align}
					\end{subequations}
				\else
					\begin{subequations}
						\begin{align}
							\mu\bb{\vec{u}_{0}}	& = \frac{\displaystyle \esssup_{\yc} \bb{\ac + \abs{\yc}^{\pc}}^{-2}}{\displaystyle \sqrt{n} \int_{-\infty}^{\infty} \bb{\ac + \abs{\yc}^{\pc}}^{-2} d\yc}
							=	\frac{1}{2\ac^{2}\sqrt{n}} \inv{\bb{\int_{0}^{\infty} \bb{\ac + \yc^{\pc}}^{-2} d\yc}}	\label{eqn:poly expr} \\
							& =	\begin{cases}
									\displaystyle	\frac{1}{2\ac^{2}\sqrt{n}} \inv{\bb{\frac{\bb{\pi/\pc}\bb{1-1/\pc}}{\ac^{2-1/\pc} \sin\bb{\pi/\pc}}}}	& ,\quad \pc \in \bb{\frac{1}{2},1} \bigcup \bb{1,\infty} \\
									\displaystyle	\frac{1}{2\ac^{2}\sqrt{n}} \inv{\bb{\frac{1}{\ac}}} & ,\quad \pc = 1
								\end{cases}
								\label{eqn:evaluated poly expr} \\
							& = \begin{cases}
									\displaystyle	\frac{\pc^{2} \sin\bb{\pi/\pc}}{2\sqrt{n}\pi\bb{\pc-1}\ac^{1/\pc}} & ,\quad \pc \in \bb{\frac{1}{2},1} \bigcup \bb{1,\infty} \\
									\inv{\bb{2\ac\sqrt{n}}} & ,\quad \pc = 1
								\end{cases}
						\end{align}
					\end{subequations}
				\fi
			\makeatother
			where \eqref{eqn:evaluated poly expr} was obtained from \eqref{eqn:poly expr} by considering the following cases.
			For $\pc = 1$, we have
			\makeatletter
				\if@twocolumn
					\begin{equation}
						\begin{split}
							\int_{0}^{\infty} \bb{\ac + \yc^{\pc}}^{-2} d\yc
							& = \int_{0}^{\infty} \bb{\ac + \yc}^{-2} d\bb{\ac + \yc} \\
							& = \mleft. -\bb{\ac + \yc}^{-1} \mright\rvert_{\yc = 0}^{\infty}
							= \inv{\ac}.
						\end{split}
					\end{equation}
				\else
					\begin{equation}
						\int_{0}^{\infty} \bb{\ac + \yc^{\pc}}^{-2} d\yc
						= \int_{0}^{\infty} \bb{\ac + \yc}^{-2} d\bb{\ac + \yc}
						= \mleft. -\bb{\ac + \yc}^{-1} \mright\rvert_{\yc = 0}^{\infty}
						= \inv{\ac}.
					\end{equation}
				\fi
			\makeatother
			For $\pc > 1$, we invoke the definite integral formula
			\makeatletter
				\if@twocolumn
					\begin{equation}
						\begin{split}
							\MoveEqLeft \int_{0}^{\infty} \frac{t^{m} dt}{\bb{t^{n} + a^{n}}^{r}}	\\
							& = \frac{\bb{-1}^{r-1} \pi a^{m+1-nr} \cdot \Gamma\BB{\bb{m+1}/n}}{n\sin\BB{\bb{m+1}\pi/n} \cdot \bb{r-1}! \cdot \Gamma\BB{\bb{m+1}/n - r + 1}}
						\end{split}
					\end{equation}
				\else
					\begin{equation}
						\int_{0}^{\infty} \frac{t^{m} dt}{\bb{t^{n} + a^{n}}^{r}}
						=	\frac{\bb{-1}^{r-1} \pi a^{m+1-nr} \cdot \Gamma\BB{\bb{m+1}/n}}{n\sin\BB{\bb{m+1}\pi/n} \cdot \bb{r-1}! \cdot \Gamma\BB{\bb{m+1}/n - r + 1}}
					\end{equation}
				\fi
			\makeatother
			from \cite{spiegel2008handbookformulas}, valid in the range $n\bb{r-2} < m+1 < nr$, with the values $m=0,n=\pc,a=\ac^{1/\pc},r=2$.
			To see that the range criterion is satisfied, we observe that $n\bb{r-2} < m+1 < nr$ reduces to $0 < 1 < 2\pc$ which is true for $\pc > 1/2$.
			Similarly,
			\makeatletter
				\if@twocolumn
					\begin{equation}
						\mu\bb{\vec{v}_{0}}
						=   \begin{cases}
								\frac{\pr^{2} \sin\bb{\pi/\pr}}{2\sqrt{n}\pi\bb{\pr-1}\ar^{1/\pr}} & ,\quad \pr \in \bb{\frac{1}{2},1} \bigcup \bb{1,\infty} \\
								\inv{\bb{2\ar\sqrt{n}}} & ,\quad \pr = 1
							\end{cases}
					\end{equation}
				\else
					\begin{equation}
						\mu\bb{\vec{v}_{0}}
						=   \begin{cases}
								\displaystyle	\frac{\pr^{2} \sin\bb{\pi/\pr}}{2\sqrt{n}\pi\bb{\pr-1}\ar^{1/\pr}} & ,\quad \pr \in \bb{\frac{1}{2},1} \bigcup \bb{1,\infty}	\\
								\inv{\bb{2\ar\sqrt{n}}} & ,\quad \pr = 1
							\end{cases}
					\end{equation}
				\fi
			\makeatother
			and the coherence parameter is determined as in \eqref{eqn:coherence parameter}.

\end{document}